\documentclass[a4paper,eqsecnum,nofootinbib]{revtex4}
\usepackage{slashed,amsmath,amssymb,wrapfig} 
\usepackage{color,epsfig}
\usepackage{hyperref}

\newcommand{\nn}{\nonumber}
\newcommand{\beq} {\begin{equation}}
\newcommand{\eeq} {\end{equation}}
\newcommand{\beqa} {\begin{eqnarray}}
\newcommand{\eeqa} {\end{eqnarray}}

\newcommand{\cf}{{\it cf.}}
\newcommand{\ie}{{\it i.e.}}

\newcommand{\eg}{{\it e.g.}}

\newcommand{\as}{{\alpha_s}}
\newcommand{\lqcd}{\Lambda_{QCD}}

\newcommand{\la}{\Lambda}
\newcommand{\eps}{\epsilon}
\newcommand{\ieps}{i\varepsilon}
\newcommand{\vphi}{\varphi}

\newcommand{\order}[1]{${\cal O}\left(#1 \right)$}
\newcommand{\morder}[1]{{\cal O}\left(#1 \right)}
\newcommand{\eq}[1]{(\ref{#1})}
\newcommand{\fig}[1]{Fig.~\ref{#1}}
\newcommand{\lsim}{\lesssim}
\newcommand{\gsim}{\gtrsim}

\newcommand{\inv}[1]{\frac{1}{#1}}
\newcommand{\halft}{{\textstyle \frac{1}{2}}}
\newcommand{\halfs}{{\scriptstyle \frac{1}{2}}}
\newcommand{\quart}{{\textstyle \frac{1}{4}}}
\newcommand{\sfrac}[2]{{\textstyle\frac{#1}{#2}}}
\newcommand{\ket}[1]{\left\vert{#1}\right\rangle}
\newcommand{\bra}[1]{\langle{#1}\vert}

\newcommand{\com}[2]{\left[{#1},{#2}\right]}
\newcommand{\comb}[2]{\big[{#1},{#2}\big]}
\newcommand{\acom}[2]{\left\{{#1},{#2}\right\}}
\newcommand{\acomb}[2]{\big\{{#1},{#2}\big\}}
\newcommand{\tr}{\mathrm{Tr}\,}

\newcommand{\bs}[1]{\boldsymbol{#1}}

\newcommand{\psl}{{\slashed{p}}}
\newcommand{\Psl}{{\slashed{P}}}
\newcommand{\Pisl}{{\slashed{\Pi}}}
\newcommand{\ksl}{{\slashed{k}}}

\newcommand{\Vsl}{{/\hspace{-1.7mm}V}}

\newcommand{\mB}{\mathcal{B}}

\newcommand{\mH}{\mathcal{H}}
\newcommand{\mL}{\mathcal{L}}
\newcommand{\mM}{\mathcal{M}}

\newcommand{\xv}{{\bs{x}}} 
\newcommand{\yv}{{\bs{y}}}
\newcommand{\zv}{{\bs{z}}}
\newcommand{\pv}{{\bs{p}}}
\newcommand{\kv}{{\bs{k}}}
\newcommand{\qv}{{\bs{q}}}

\newcommand{\Pv}{{\bs{P}}}
\newcommand{\Av}{{\bs{A}}}

\newcommand{\gv}{\bs{\gamma}}
\newcommand{\delv}{\bs{\delta}}
\newcommand{\gz}{\gamma^0}
\newcommand{\go}{\gamma^1}
\newcommand{\gf}{\gamma^0\gamma^1}
\newcommand{\nv}{\bs{\nabla}}
\newcommand{\sv}{\bs{\sigma}}

\newcommand{\moe}{{m}}

\newcommand{\kum}{{\,{_1}F_1}}
\newcommand{\xbj}{{x_{Bj}}}

\newcommand{\rar}{\rightarrow}
\newcommand{\lar}{\leftarrow}

\newcommand{\rder}{{\buildrel\rar\over{\partial}}}
\newcommand{\lder}{{\buildrel\lar\over{\partial}}}

\newcommand{\rnab}{{\buildrel\rar\over{\nv}}}
\newcommand{\lnab}{{\buildrel\lar\over{\nv}}}

\newcommand{\Pxi}{P+\xi E}
\newcommand{\dpi}{\Pi}
\newcommand{\dsi}{\sigma}

\begin{document}
{\par\raggedleft \texttt{\vspace{-.3cm} 20 February 2014}\par}

\title{Bound states -- from QED to QCD\footnote{Based on lectures presented at the {\it ``Mini-school on theoretical methods in particle physics''} at the Higgs Centre for Theoretical Physics, University of Edinburgh on 30 September to 4 October 2013.}}

\author{Paul Hoyer}
\affiliation{Department of Physics and Helsinki Institute of Physics\\ POB 64, FIN-00014 University of Helsinki, Finland}

\begin{abstract} 

These lectures are divided into two parts. In Part 1 I discuss bound state topics at the level of a basic course in field theory: The derivation of the Schr\"odinger and Dirac equations from the QED Lagrangian, by summing Feynman diagrams and in a Hamiltonian framework. Less well known topics include the equal-time wave function of Positronium in motion and the properties of the Dirac wave function for a linear potential. The presentation emphasizes physical aspects and provides the framework for Part 2, which discusses the derivation of relativistic bound states at Born level in QED and QCD. A central aspect is the maintenance of Poincar\'e invariance. The transformation of the wave function under boosts is studied in detail in $D=1+1$ dimensions, and its generalization to $D=3+1$ is indicated. Solving Gauss' law for $A^0$ with a non-vanishing boundary condition leads to a linear potential for QCD mesons, and an analogous confining potential for baryons. 

\end{abstract}


\maketitle

\vspace{-.5cm}

\tableofcontents

\parindent 0cm
\vspace{-.2cm}

\section{Introduction}

The aim of these lectures is to review and develop the field theory description of bound states at a basic level. Bound state calculations differ from those of scattering amplitudes, yet are typically not discussed in modern textbooks. Sophisticated calculations of atoms at high orders in $\alpha$ provide precision tests of QED. Here we are mainly concerned with the principles and practice of bound states at lowest order.

The established framework for bound states in QED may be useful also for QCD hadrons. Heavy quarkonia are often referred to as the ``Positronium of QCD'', based on the astonishing similarity of their spectra. With this in mind we emphasize a physical understanding of bound state calculations in QED.

Part 1 (Sections \ref{possection} and \ref{diracbound}) shows how lowest order approximations to bound states, at the level of the Schr\"odinger and Dirac equations, are derived from the QED Lagrangian. One possibility is to sum an infinite set of Feynman diagrams. We discuss why QED perturbation theory diverges, giving rise to bound state poles in scattering amplitudes. Feynman diagrams may be evaluated in any frame, allowing to consider the wave function of atoms in motion. The Positronium (equal-time) wave function turns out to transform not only by Lorentz contracting, as would be suggested by classical relativity.

The Schr\"odinger and Dirac equations can also be obtained by constructing eigenstates of the QED Hamiltonian. This clarifies the multi-particle nature of Dirac states, and motivates the interpretation of the norm of the Dirac wave function as an {\it inclusive} particle density.

Part 2 (Sections \ref{ffstatesec} and \ref{3dpotential}) contains research-level material. Based on the experience in Part 1 we define ``Born level'' bound states as eigenstates of the field theory Hamiltonian, with a gauge field that satisfies the equations of motions at lowest order in the coupling. In this way we obtain bound states of any CM momentum, with a non-trivial, exact Poincar\'e symmetry. We study the properties of these states in some detail in $D=1+1$ dimensions, including their parton distributions. The states turn out to have a parton-hadron duality similar to what is observed for hadrons. 

Both the Hamiltonian and the equations of motion are fixed by the field theory action. The bound states are thus almost uniquely determined, raising the question of how confinement can be described in $D=3+1$ dimensions. In the present framework the only possibility is to consider a homogeneous, \order{g^0} solution of Gauss' law. For neutral states this leads to an exactly linear potential, analogous to the potential in $D=1+1$.

Parts 1 and 2 are summarized and discussed in Sections \ref{disc1} and \ref{disc2}, respectively. A derivation of bound states with scalar (rather than fermion) constituents in $D=1+1$ is given in appendix \ref{sqedapp}. Appendix \ref{NRlimit} shows how the relativistic wave functions reduce to Schr\"odinger ones in the non-relativistic limit.

\vspace{1cm}

\centerline{\Large PART 1: Basics of bound states}

\section{Positronium\label{possection}}

\subsection{Divergence of the perturbative expansion\label{divsum}}

On general grounds we know that bound states appear as poles in scattering amplitudes. The poles are on the real axis of the complex energy plane for stable bound states (like protons) and below the real axis in case of unstable states. It is perhaps worthwhile to illustrate this using the free scalar propagator
\beq\label{scalarprop}
D(p^0,\pv) = \frac{i}{p^2-m^2+\ieps}=\frac{i}{(p^0-E_p+\ieps)(p^0+E_p-\ieps)}
\eeq
where $E_p = \sqrt{\pv^2+m^2}$. Fourier transforming $p^0 \to t$,
\beq\label{ft}
D(t,\pv) \equiv \int \frac{d p^0}{2\pi} D(p^0,\pv) \exp(-ip^0 t)
       = \inv{2E_p}\Big[\theta(t) e^{-iE_p t} + \theta(-t) e^{iE_p t} \Big]
\eeq
In the reverse transformation $t \to p^0$ the poles of \eq{scalarprop} in $p^0$ are created by the infinite range of the $t$-integration. 

\begin{wrapfigure}{r}{0.5\textwidth}
  \vspace{-20pt}
  \begin{center}
    \includegraphics[width=0.48\textwidth]{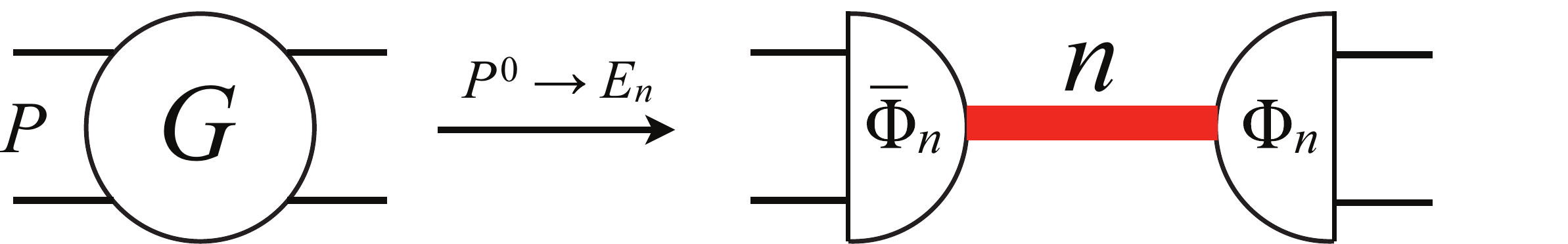}
  \end{center}
  \vspace{-18pt}
  \caption{Bound states appear as poles in scattering amplitudes. Unitarity requires that the residue factorizes into a product of incoming and outgoing wavefunctions.}
\label{pole}
\end{wrapfigure}

Bound states are by definition stationary in time,
\beq\label{Seq}
H\ket{P,t} = P^0 \ket{P,t} \ \ \ \Longrightarrow\ \ \ket{P,t} = e^{-iP^0t}\ket{P,0}
\eeq
where $P$ is the 4-momentum and $P^0=\sqrt{\Pv^2+M^2}$. The bound state contribution to a completeness sum in an amplitude $\bra{f,t_f}i,t_i\rangle$ with initial and final energies $E_i=E_f=P^0$ will then be 
\beq\label{complete}
\bra{f,t_f}{P,t}\rangle\bra{P,t}i,t_i\rangle=\bra{f}P\rangle e^{-i(t_f-t_i)P^0}\bra{P}i\rangle
\eeq
The Fourier transform $t_f-t_i \to p^0$ will generate a pole at $p^0=P^0-i\eps$, with residue equal to a product of the final $\bra{f}P\rangle$ and initial $\bra{P}i\rangle$ wave functions, as indicated in \fig{pole}. This holds for any bound state, no matter how complicated. 

The rest frame ($\Pv=0$) energies of positronium $(e^+ e^-)$ atoms are known from Introductory Quantum Mechanics, 
\beq\label{bindenergy}
P^0=2m_e+E_b
\eeq
with binding energies $E_b = -\quart m_e\alpha^2/n^2 \simeq -6.8$ eV$/n^2$ (at lowest order in $\alpha$, for the principal quantum number $n=1,2,\ldots$). Hence the elastic $e^+ e^-$ amplitude $G(e^+ e^- \to e^+ e^-)$ has an infinite set of positronium poles just below threshold ($E_{CM}^{th}=2m_e$), and slightly below the real $s=E_{CM}^2\,$-axis due to the finite life-times. How are these poles generated by the Feynman diagrams describing $G$?

\begin{figure}[h]
\includegraphics[width=.8\columnwidth]{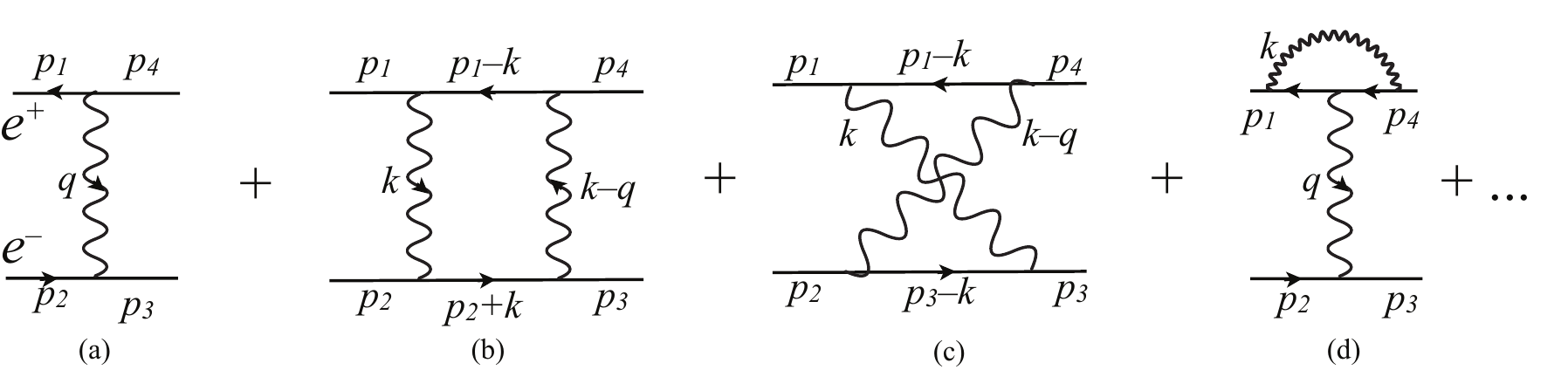}
\caption{Feynman diagrams contributing to elastic $e^+(p_1)e^-(p_2)$ scattering. The arrows indicate the fermion direction. The momentum of the upper line is in the antifermion ($e^+$) direction, thus $p_1^0 > 0$. 
{\label{feyndiags}}}
\end{figure}

We may regard the positions of the bound state poles in $s=(2m_e+E_b)^2$ as functions of $E_b$, \ie, of $\alpha$. A Feynman diagram of \order{\alpha^n} cannot have a pole in $\alpha$ at any finite order $n$. The only way to generate a bound state pole in $G$ is for the perturbative expansion to diverge\footnote{This divergence is distinct from that due to perturbative expansions being asymptotic series \cite{Dyson:1952tj}.}! This sounds surprising at first, since we are used to trusting QED perturbation theory. The poles exist for any $\alpha$, however small. Thus some nominally higher order diagrams, such as those in \fig{feyndiags}(b-d), must be effectively of the same order in $\alpha$ as the Born term (a). 
%
%

The breakdown of the perturbative expansion is actually familiar from classical physics, where phenomena involving many photons dominate. For example, the notion that opposite charges attract while like charges repel cannot be explained by just the Born term in \fig{feyndiags}. This diagram changes sign if the positron is replaced by an electron, so its absolute square is invariant. The product of diagrams (a) and (b), on the other hand, contributes with opposite signs to $\sigma(e^\pm e^- \to e^\pm e^-)$. Thus our everyday experience of attraction and repulsion originates from quantum interference effects.

Higher order diagrams have not only more vertices $\propto e$ but also more propagators, which are enhanced at low momenta. Typical momentum exchanges in atoms are of the order of the Bohr momentum\footnote{In calculations of higher order corrections to physical quantities other momentum scales must be considered as well.}, and electron energy differences then follow from non-relativistic dynamics:
\beq\label{bohrmom}
|\qv| \sim \alpha m_e  \hspace{2cm}     q^0 \sim \qv^2/2m_e \sim \halft\alpha^2 m_e
\eeq
The Born diagram of \fig{feyndiags}(a) scales with $\alpha$ as 
\beq\label{borndiag}
G[\ref{feyndiags}(a)] \sim \alpha/q^2 \sim \alpha/\qv^2 \sim 1/\alpha
\eeq
The box diagram \ref{feyndiags}(b) has four vertices, giving a factor $e^4\sim \alpha^2$. The two photon propagators contribute $1/\qv^2 \sim \alpha^{-2}$ each. The electron and positron propagators are off-shell on the order $q^0 \sim k^0$, each propagator being of \order{\alpha^{-2}}. The relevant region of loop momentum is $\int dk^0\,d^3\kv \sim \alpha^2\,(\alpha)^3 \sim \alpha^5$. Altogether,
\beq\label{lad2}
G[\ref{feyndiags}(b)] \sim \alpha^2\,(\alpha^{-2})^2\,(\alpha^{-2})^2\,\alpha^5 \sim 1/\alpha \sim G[\ref{feyndiags}(a)]
\eeq
A similar analysis shows that ``ladder'' diagrams with any number of photon exchanges are of \order{\alpha^{-1}} and thus of the same order in $\alpha$ as the Born diagram \eq{borndiag}. This allows the perturbative series to diverge for any $\alpha$. Note that the above counting requires the initial and final momenta $p_1,\ldots p_4$ of the scattering to themselves satisfy the scaling \eq{bohrmom}: As $\alpha \to 0$ the external momenta need to be correspondingly decreased. Conversely, the initial and final states do not couple to the bound states in a ``hard'' scattering process where the momentum exchange $|\qv| \gg \alpha m_e$. Then $\bra{P}i\rangle \sim \bra{f}P\rangle \simeq 0$ in \eq{complete} and bound state contributions can be ignored. In the following we shall see more such analogies to ``hard'' and ``soft'' processes in QCD. In QED we know how to deal with ``soft'' scattering, which might be helpful for understanding the properties of QCD.  

All except the ladder diagrams scale with a higher power of $\alpha$ than the Born term, and can thus be ignored in a lowest order calculation of non-relativistic bound states. We  shall not prove this, but just illustrate by the crossed ladder (c) and the vertex correction (d) in \fig{feyndiags}. Both have the same number of propagators and vertices as the straight ladder (b), and would give the same estimate as in \eq{lad2}. However, their leading contributions cancel in the loop integration. In \fig{feyndiags}(c), $-p_1^0 = -m_e+\morder{\alpha^2}$ whereas $p_3^0 = +m_e+\morder{\alpha^2}$. Hence the leading contribution comes from the negative energy pole in the ($-p_1+k$) propagator, and from the positive energy pole in the ($p_3-k$) propagator. The Feynman $\ieps$ prescription implies that both poles are in the Im$k^0>0$ hemisphere. Closing the $k^0$ contour in the Im$k^0<0$ plane these poles do not contribute. The situation is similar for the vertex diagram (d), whereas for the straight ladder (b) the integration contour is pinched by the two poles.

We have thus identified the ladders as leading Feynman diagrams in the rest frame, where \eq{bohrmom} applies. Since Feynman diagrams are Lorentz covariant the same diagrams will dominate in all frames. This will allow us to analyze positronium in motion in Section \ref{motion}.

\subsection{Evaluating ladder diagrams\label{ladsum}}

The standard Feynman rules give for the Born diagram \fig{feyndiags}(a),
\beq\label{L1}
L_1(p_1,p_2\to p_1-q,p_2+q) =\bar v(p_1)(-ie\gamma^\mu)v(p_4)D_{\mu\nu}(q)\bar u(p_3)(-ie\gamma^\nu)u(p_2)
\eeq
where the photon propagator is $D_{\mu\nu}(q)=-ig_{\mu\nu}/(q^2+\ieps)$ in Feynman gauge.

The double ladder \fig{feyndiags}(b) is similarly
\beqa\label{L2}
L_2(p_1,p_2\to p_1-q,p_2+q) &=& \int \frac{d^4k}{(2\pi)^4} \bar v(p_1)(-ie\gamma^\mu)i\frac{-\psl_1+\ksl+m}{(-p_1+k)^2-m^2+\ieps}(-ie\gamma^\rho)v(p_4)\nn\\[2mm]
&&\times\ D_{\mu\nu}(k)\,D_{\rho\sigma}(k-q) \nn\\
&&\times\ \bar u(p_3)(-ie\gamma^\sigma)i\frac{\psl_2+\ksl+m}{(p_2+k)^2-m^2+\ieps}(-ie\gamma^\nu)u(p_2)
\eeqa
The positive and negative energy poles of a fermion propagator may be separated using the identity
\beq
\frac{\psl+m}{p^2-m^2+\ieps}=\inv{2E_p}\sum_\lambda\Big[\frac{u(\pv,\lambda)\bar u(\pv,\lambda)}{p^0-E_p+\ieps}+\frac{v(-\pv,\lambda)\bar v(-\pv,\lambda)}{p^0+E_p-\ieps}\Big]
\eeq
where $E_p=\sqrt{\pv^2+m^2}$ and $\lambda=\pm\halft$ is the helicity. Note that on the rhs. $p^0$ appears only in the denominator.
In the region \eq{bohrmom} relevant for bound states at lowest order the fermion propagators are close to their mass-shell, so
\beqa\label{onapp}
(-p_1+k)^0 &\simeq& -E_{1k}\equiv -\sqrt{(\pv_1-\kv)^2+m^2} \hspace{1cm} \Longrightarrow \hspace{1cm} {\rm keep\ only\ the}\ v\bar v\ {\rm term}\nn\\
(p_2+k)^0 &\simeq& +E_{2k}\equiv\ \  \sqrt{(\pv_2+\kv)^2+m^2} \hspace{1cm} \Longrightarrow \hspace{1cm} {\rm keep\ only\ the}\ u\bar u\ {\rm term}
\eeqa
With this approximation we find
\beqa\label{L2app}
L_2(p_1,p_2\to p_1-q,p_2+q)&=&\sum_{\lambda_{int}}\int \frac{d^4k}{(2\pi)^4} \,
L_1(p_1,p_2\to p_1-k,p_2+k)\nn\\
&\times& S(p_1-k,p_2+k)L_1(p_1-k,p_2+k\to p_1-q,p_2+q)\nn\\
&\equiv& L_1\,S\,L_1
\eeqa
where the convolution is over the helicities $\lambda_{int}$ and momenta $k$ of the intermediate state which has propagator $S$,
\beq\label{ffbarprop}
S(p_1-k,p_2+k) = \frac{i}{-p_1^0+k^0+E_{1k}-\ieps}\, \frac{i}{p_2^0+k^0-E_{2k}+\ieps}\,  \inv{2E_{1k}\,2E_{2k}}
\eeq
and the $E_{ik}$ are defined in \eq{onapp}.

The same procedure will show that a ladder $L_n$ with $n>1$ rungs is obtained from the one with $n-1$ rungs as
\beq
L_n = L_{n-1}SL_1
\eeq
Summing over $n$ we find
\beq\label{DSeq}
L \equiv \sum_{n=1}^\infty L_n = L_1 + L\,S\,L_1
\eeq
with a convolution on the rhs. as in \eq{L2app}. This is a Dyson-Schwinger equation for $L$ with lowest-order propagator $S$ \eq{ffbarprop} and kernel $L_1$ \eq{L1}. Note that we did not need to specify the frame, the equation is valid for any $e^+e^-$ momentum $\Pv=\pv_1+\pv_2$.

\begin{wrapfigure}[7]{r}{0.5\textwidth}
  \vspace{-25pt}
  \begin{center}
    \includegraphics[width=0.48\textwidth]{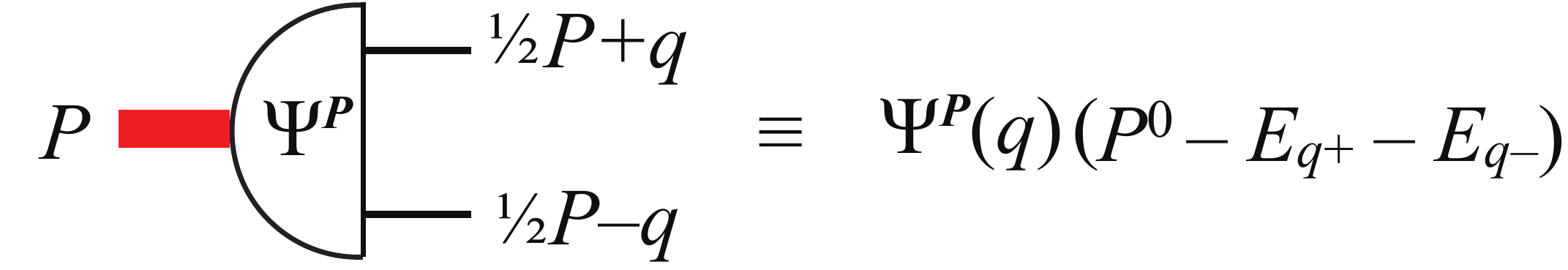}
  \end{center}
  \vspace{-10pt}
  \caption{A factor $P^0-E_{q+}-E_{q-}$ is included in the definition of the wave function, with $E_{q\pm}= \sqrt{(\halft\Pv\pm\qv)^2+m^2}$.}
\label{wfdef}
\end{wrapfigure}

If the ladder sum $L$ has a pole at $P^0=\sqrt{\Pv^2+M^2}$, 
with $M$ the rest mass of the bound state, the residue will factorize as shown in \fig{pole} and Eq. \eq{complete}, 
\beq\label{bspole}
L = \frac{\overline{\Psi}^{\Pv} \Psi^{\Pv}}{P^0-E_{P}}+\ldots
\eeq
Canceling common factors on the two sides of \eq{DSeq} and expressing the wave function as indicated in \fig{wfdef} we find the Bethe-Salpeter Equation (BSE)
\beq\label{wfconstraint}
\Psi^\Pv = \Psi^\Pv\,S\,L_1
\eeq
Denoting the propagator \eq{ffbarprop} $S(k)\equiv S(\halft P-k,\halft P+k)$, the kernel \eq{L1} $K(k,q) \equiv L_1(\halft P-k,\halft P+k \to \halft P-q,\halft P+q)$ and extracting (for later convenience) a factor $P^0-E_{q+}-E_{q-}$ from the wave function  $\Psi^\Pv(q)$ we have more explicitly,
\beq\label{BSeq}
\Psi^\Pv(q)(P^0-E_{q+}-E_{q-}) =
 \sum_{\lambda_{int}}\int \frac{d^4k}{(2\pi)^4} \,\Psi^\Pv(k)(P^0-E_{k+}-E_{k-})
S(k)  K(k, q)
\eeq

\subsection{Remarks on Positronium at higher orders\label{highord}}

In the previous subsection we derived the Bethe-Salpeter equation \eq{BSeq} at lowest order in $\alpha$. Much work has been devoted to obtaining more accurate predictions of QED bound states. These calculations are considerably more involved and will not be detailed in these lectures. However, I shall briefly describe the progress that has been made, and refer to the reviews \cite{Sapirstein} and \cite{Kinoshita} for a more complete account and references.

In 1951 Salpeter and Bethe \cite{Salpeter:1951sz} showed that \eq{BSeq} is formally exact provided one includes all corrections to the electron and positron propagators in $S$ and to the kernel $K$. 
The corresponding Bethe-Salpeter wave function of a positronium state $\ket{P}$ of 4-momentum $P=(\sqrt{M^2+\Pv^2},\Pv)$ can be defined to all orders in coordinate space through the time-ordered matrix element
\beq \label{BSwavef}
\bra{\Omega}\ T\left\{\bar{\psi}_\beta(x_2)\psi_{\alpha}(x_1)\right\}\ket{P} \equiv e^{-iP\cdot(x_1+x_2)/2}\, \Phi_{\alpha\beta}^\Pv(x_1-x_2)  
\eeq
where $\psi(x)$ is the electron field operator (in the Heisenberg picture) and $\ket{\Omega}$ is the vacuum state. The plane wave dependence on $x_1+x_2$ is specified by space-time translation invariance since the bound state has momentum $P$.

It turned out to be difficult in practice to calculate higher order corrections to bound state energies from the BSE \eq{BSeq}. The Lorentz covariant wave function \eq{BSwavef} cannot be expressed in closed form even when only the lowest order kernel (single photon exchange) is used\footnote{For a recent discussion of the solutions of the BSE see \cite{Carbonell:2013kwa}.}. However, because the equation involves {\em two} functions $S$ and $K$, there is a freedom in choosing either one, without affecting the validity of the equation \cite{Lepage:1977gd}. This is seen as follows.

Let $G_T$ be the Green function for a $2\to 2$ scattering process with the external propagators truncated. The perturbative expansion of $G_T$ in $\alpha$ may be calculated using the standard Feynman rules. We then {\em declare} a Dyson-Schwinger type equation by
\beq\label{dyson}
G_T = K + G_T\,S\,K
\eeq
where the products imply convolutions over four-momenta similar to that in \eq{BSeq}. This equation is valid provided the kernel satisfies
\beq
K=(1+G_T\,S)^{-1}G_T = G_T-G_T\,S\,G_T+...
\eeq 
Thus the ``propagator'' $S$ may in fact be chosen freely. The expansion of $K$ in $\alpha$ follows from the corresponding expansions of $S$ and $G_T$. As a consequence of unitarity the residues of the bound state poles of $G_T$ factorize into a product of wave functions similarly as in \eq{bspole}.
Since the finite order kernel $K$ in \eq{dyson} cannot have a bound state pole the Bethe-Salpeter wave function $\Phi^\Pv_T$ (with external propagators truncated) must satisfy
\beq \label{BSE}
 \Phi^\Pv_T(q) \equiv \int d^4x\, \Phi^\Pv_T(x)e^{iq\cdot x}= \int \frac{d^4k}{(2\pi)^4}\,\Phi^\Pv_T(k)\,S(k)\, K(k,q)\ 
\eeq
which is the all-orders equivalent\footnote{In \eq{BSeq} a factor $P^0-E_{q+}-E_{q-}$ was extracted from the wave function $\Psi(q)$.} of \eq{BSeq}.
With a suitable choice of the propagator $S$ analytic expressions for the wave functions are obtained when the lowest order kernel is used in the BSE. These solutions facilitate calculations of higher order corrections to the binding energies \cite{Sapirstein}.

The wide range of possibilities in the choice of propagator in the BSE motivated a search for an optimal approach based on physical arguments. The perturbative expansion relies on the non-relativistic nature of atoms, $v/c \simeq \alpha \ll 1$. This suggested the use of an effective QED Lagrangian (NRQED) \cite{Caswell:1985ui}, which is essentially an expansion of the standard Lagrangian in inverse powers of $m_e$. At the expense of introducing more interactions the NRQED Lagrangian allows to use non-relativistic dynamics, which is of great help in high order calculations \cite{Kinoshita}. The contribution of relativistic momenta ($p \sim m_e$) in positronium is only of \order{\alpha^5} $\sim 10^{-11}$, making NRQED very efficient.

The continuous development of theoretical and experimental techniques have allowed precision tests of QED using bound states.
Thus the hyperfine splitting in positronium, \ie, the energy difference $\Delta E$ between orthopositronium ($J^{PC}=1^{--}$) and parapositronium ($J^{PC}=0^{-+}$), expressed in terms of $\Delta\nu \equiv \Delta E/2\pi\hbar$, is calculated using NRQED methods to be \cite{Czarnecki:1999uk}
\vspace{-.2cm}\beqa
\Delta \nu_{QED}
&=& m_e\alpha^4 \left\{ \frac {7}{12} -
\frac{\alpha}{\pi} \left(\frac {8}{9}+\frac{\ln 2}{2}  \right)
\right.
\nonumber\\
&&
+\frac {\alpha^2}{\pi^2} \left[
- \frac{5}{24}\pi^2 \ln \alpha
+ \frac {1367}{648}-\frac{5197}{3456}\pi^2
 + \left( \frac{221}{144}\pi^2 +\frac {1}{2} \right) \ln 2
-\frac{53}{32}\zeta (3) \right]
\nonumber\\
&& \left.
-\frac{7\alpha^3}{8\pi}\ln^2\alpha
+\frac{\alpha^3}{\pi}\ln\alpha\left(\frac{17}{3}\ln2-\frac{217}{90}\right)
+\morder{\alpha^3} \right\}
= 203.39169(41) \mbox{ GHz}
\label{hfsfin}
\eeqa

\begin{wrapfigure}[16]{r}{0.4\textwidth}
  \vspace{-30pt}
  \begin{center}
    \includegraphics[width=0.4\textwidth]{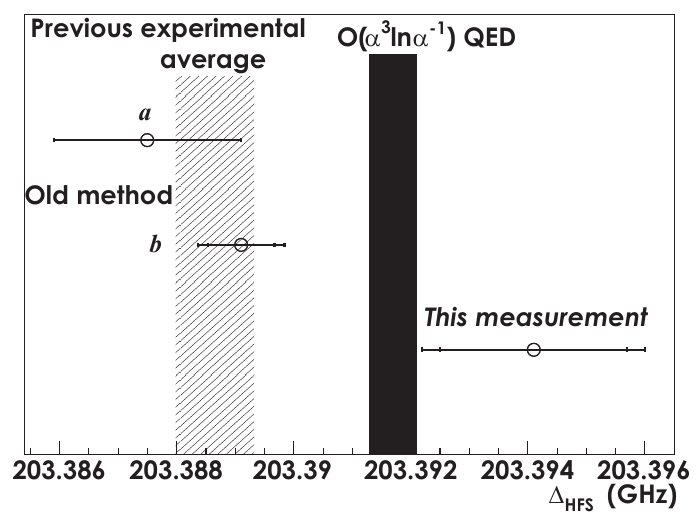}
  \end{center}
  \vspace{-15pt}
  \caption{Data on positronium hyperfine splitting compared to theory. Two previous results (a \cite{Mills:1983zzd}, b \cite{Ritter}) compared to a new measurement \cite{Ishida:2013waa} and QED \cite{Czarnecki:1999uk} (black band). Figure  from \cite{Ishida:2013waa}.}
\label{hypfig}
\end{wrapfigure}

The appearance of $\ln\alpha$ in \eq{hfsfin} demonstrates that bound state perturbation theory indeed differs from the usual expansions of scattering amplitudes. Such factors arise from apparent infrared divergences which are regulated by the neutrality of positronium at the scale of the Bohr radius $(\alpha m_e)^{-1}$.

The combined result of the two most precise measurements of the hyperfine splitting in positronium \cite{Mills:1983zzd,Ritter} is $\Delta\nu_{EXP} = 203.38865(67)$ GHz, which is more than $3\,\sigma$ from the QED value \eq{hfsfin}. Very recently a new measurement \cite{Ishida:2013waa} gave $\Delta\nu_{EXP} = 203.3941 \pm .0016 \pm .0011$ GHz, which is closer to the theoretical value. The present situation is illustrated in \fig{hypfig}.

Bound state poles in the photon propagator affect also standard perturbative calculations. The positronium contribution to the anomalous magnetic moment of the electron was recently evaluated \cite{Mishima:2013ama}. It was found to be of the same order as state-of-the-art five-loop calculations -- and several times bigger than the weak corrections.

The successes of QED have inspired the use of analogous methods for the other interactions. In particular, Bethe-Salpeter and Dyson-Schwinger equations have been extensively applied in QCD (see \cite{Roberts:2012sv} and references therein). Viewed as non-perturbative equations they give exact relations between Green functions but do not close -- an infinite set of functions are coupled to each other. Models based on judicious truncations have allowed studies of spontaneous chiral symmetry breaking and been successfully compared to hadron properties deduced from data and lattice calculations.

Effective theories analogous to NRQED have been formulated for heavy quarks with mass $m_Q \gg \lqcd$, and used to describe $Q\bar Q$ bound states \cite{Brambilla:2004jw}. These methods are particularly useful in the limit where the quarkonia have small enough radius for perturbative gluon exchange to dominate over the confining interaction.

\subsection{The Schr\"odinger equation\label{schreq}}

Let us now return to the lowest-order Bethe-Salpeter equation \eq{BSeq} and verify that it reduces to the Schr\"odinger equation in the rest frame, $\Pv=0$. Using \eq{bohrmom} the photon propagator of $L_1$ \eq{L1} to lowest order in $\alpha$ is
\beq
D_{\mu\nu}(q)=\frac{ig_{\mu\nu}}{\qv^2}
\eeq
Due to the non-relativistic kinematics the upper (lower) components of the $u$ ($v$) spinors dominate (in the Dirac representation). The main contribution to $L_1$ is then from the diagonal $\gamma$-matrix, \ie, $\mu=\nu=0$. The kernel of the BSE \eq{BSeq} is thus independent of $q^0$ (and helicity preserving),
\beq
 K(k,q) = -ie^2\frac{4m^2}{(\kv-\qv)^2}
\eeq
The other factors on the rhs. of the BSE also do not depend on $q^0$, consequently the wave function $\Psi^{\Pv=0}(q)$ is independent of $q^0$. This implies that the wave function is an {\em equal-time} wave function: Fourier transforming $q^0\to t_1-t_2$ we find, including the dependence on $t_1+t_2$ (\cf\ \eq{BSwavef}),
\beq
\Psi^{\Pv=0}(\qv;t_1,t_2) \equiv \int\frac{dq^0}{2\pi}e^{-i(\halfs M+q^0)t_1-i(\halfs M-q^0)t_2}\Psi^{\Pv=0}(q) \equiv \delta(t_1-t_2) e^{-iMt_1}\phi^{\Pv=0}_{e^+e^-}(\qv)
\eeq
This is a direct consequence of the fact that instantaneous Coulomb exchange dominates in the atomic rest frame. As we shall see, the situation is different for atoms with $\Pv\neq 0$.

In the integrand of the BSE \eq{BSeq} only the propagator $S$ depends on $k^0$,
\beqa
\phi^{\Pv=0}_{e^+e^-}(\qv)(M-2E_{q}) &=&\int \frac{d^4k}{(2\pi)^4} \,\phi^{\Pv=0}_{e^+e^-}(\kv)(M-2E_{k}) \frac{i}{-\halft M+k^0+E_{k}-\ieps}\, \frac{i}{\halft M+k^0-E_{k}+\ieps}\,  \frac{-ie^2}{(\kv-\qv)^2}\nn\\
&=& -e^2\int \frac{d^3\kv}{(2\pi)^3} \,\frac{\phi^{\Pv=0}_{e^+e^-}(\kv)}{(\kv-\qv)^2}
\eeqa
According to \eq{bohrmom}, $E_q \equiv \sqrt{\qv^2+m^2} \simeq m+ \qv^2/2m$ to leading order in $\alpha$. Defining the binding energy $E_b$ as in \eq{bindenergy} we get the Schr\"odinger equation in momentum space,
\beq\label{Schrmom}
\Big(E_b-\frac{\qv^2}{2m_R}\Big)\phi^{\Pv=0}_{e^+e^-}(\qv)=-4\pi\alpha \int \frac{d^3\kv}{(2\pi)^3} \,\frac{\phi^{\Pv=0}_{e^+e^-}(\kv)}{(\kv-\qv)^2}
\eeq
where $m_R=\halft m$ is the reduced mass. In coordinate space,
\beq
\Phi(\xv) \equiv \int \frac{d^3\qv}{(2\pi)^3}\phi^{\Pv=0}_{e^+e^-}(\qv)\, e^{i\qv\cdot\xv}
\eeq
the Schr\"odinger equation \eq{Schrmom} reads
\beq\label{Schrcoo}
\Big(-\frac{\nv^2}{2m_R}-\frac{\alpha}{|\xv|}\Big)\Phi(\xv) = E_b\Phi(\xv)
\eeq

\subsection{Positronium in motion\label{motion}}

The derivation of the bound state equation \eq{BSeq} in Section \ref{ladsum} was based on summing Feynman diagrams. The Lorentz covariance of these diagrams allows to consider the frame dependence of atomic wave functions. The following discussion is based on the work by Matti J\"arvinen \cite{Jarvinen:2004pi}, and is instructive for understanding how bound states transform under Lorentz boosts. It is frequently assumed that bound states Lorentz contract similarly to measuring sticks in classical relativity, and so high-momentum protons and nuclei are depicted as ovals. Only partial indications \cite{Brodsky:1968xc} were available before 2004 of how equal-time atomic wave functions actually transform. On the other hand, wave functions defined on the light front (at equal $t+z$) are boost invariant \cite{Brodsky:1997de}.

\subsubsection{Classical Lorentz contraction\label{classical}}

Let us start by recalling how Lorentz contraction arises in classical relativity, through a length measurement by two observers who are in relative motion. Each observer defines the length of a rod as the distance between its endpoints at an instant of time. The contraction arises because the concept of simultaneity is frame dependent. We may assume that Observer A is at rest with the rod  and that the frame of Observer B is reached by a boost $\zeta$ in the $x$-direction. If the endpoints of the rod are at $(0,0)$ and $(t,L_A)$ in the rest frame they transform under the boost as
\beqa
(0,0) &\to& (0,0)\nn\\
(t,L_A) &\to& (t\cosh\zeta+L_A\sinh\zeta,t\sinh\zeta+L_A\cosh\zeta)
\eeqa
Observer A measures the length of the rod at rest to be $L_A$, independently of the time $t$ of his measurement. Observer B makes his measurement at time zero on his clock, \ie, when
\beq
t\cosh\zeta+L_A\sinh\zeta=0
\eeq 
He thus finds the contracted length
\beq
L_B = t\sinh\zeta+L_A\cosh\zeta = \frac{L_A}{\cosh\zeta}
\eeq 

\subsubsection{Equal-time wave functions\label{equaltime}}

In atoms the ends of the rod correspond to the positions $\xv_1$ and $\xv_2$ of the electron and positron in the wave function \eq{BSwavef}. To study Lorentz contraction we need to consider {\em equal-time} wave functions, $x^0_1=x^0_2$ in all frames. Such wave functions have a non-trivial, dynamic frame dependence. In a Lorentz boost $x \to x'=\Lambda x$ the fermion field operator transforms as 
\beq
\psi(x) \to U(\Lambda)\psi(x)U^\dag(\Lambda) = S^{-1}(\Lambda)\psi(\Lambda x)
\hspace{2cm} \bar\psi(x) \to \bar\psi(\Lambda x)S(\Lambda)
\eeq
where $S(\Lambda)$ is the $4\times 4$ matrix which transforms the Dirac matrices as $S^{-1}(\Lambda)\gamma^\mu S(\Lambda) = \Lambda^\mu_{\ \nu}\gamma^\nu$. Using this in \eq{BSwavef} we find the Bethe-Salpeter wave function in a frame where the bound state momentum is $\Lambda P = ({P'}^0,\Pv')$,
\beq\label{bsrel}
\Phi^{\Pv'}(x_1'-x_2')=S(\Lambda) \Phi^{\Pv}(x_1-x_2) S^{-1}(\Lambda)
\eeq
When $\Lambda$ is a boost this relates wave functions defined at unequal times of the constituents ($x^0_1\neq x^0_2$ in at least one of the frames). Hence this transformation is not relevant for the issue of Lorentz contraction. 

In a Hamiltonian framework one usually quantizes the fields at equal time, with (anti-)commutation relations
\beq\label{acomrel}
\acom{\psi_\alpha^\dag(t,\xv)}{\psi_\beta(t,\yv)}=\delta_{\alpha\beta}\delta^3(\xv-\yv)
\eeq 
Correspondingly, the Fock expansion
\beq\label{Fockexp}
\ket{P}= \int d^3\xv_1\,d^3\xv_2\, \phi_{e^+e^-}^\Pv(\xv_1,\xv_2)\ket{e^+e^-,\Pv}+\int d(\cdots) \phi_{e^+e^-\gamma}^\Pv(\cdots)\ket{e^+e^-\gamma,\Pv}+\ldots
\eeq
defines a positronium state through its set of equal-time\footnote{An equal-time wave function describes the positions of the constituents at a common instant $t$ of ordinary time.} Fock state wave functions $\phi_{e^+e^-}^\Pv, \phi_{e^+e^-\gamma}^\Pv, \ldots$. The $\Pv$-dependence of the Fock wave functions is dynamic, since the notion of equal time depends on the frame (the Hamiltonian does not commute with the boost operators). For positronium at rest only the $\phi_{e^+e^-}^{\Pv=0}(\xv_1,\xv_2)$ wave function is non-vanishing at lowest order in $\alpha$, and satisfies the Schr\"odinger equation \eq{Schrcoo} with $\xv=\xv_1-\xv_2$. As we shall see, also $\phi_{e^+e^-\gamma}^\Pv$ contributes at lowest order when $\Pv \neq 0$.

\subsubsection{Contribution from transversely polarized photon exchange\label{transverse}}

Let us return to the lowest order bound state equation \eq{BSeq}. In the rest frame ($\Pv = 0$) analysis we made use of two simplifications:
\begin{enumerate}
\item The electrons moved non-relativistically, hence the upper (lower) components of the $u$ ($v$) spinors were dominant. This allowed us to keep only Coulomb photon exchange ($\mu=\nu=0$) in the kernel $L_1$ of \eq{L1} and \fig{feyndiags}(a).
\item According to \eq{bohrmom} the exchanged energy $q^0$ could be neglected compared to the three-momentum $\qv$.
\end{enumerate}
Neither of these assumptions is valid for a general bound state momentum $\Pv$. The vertex factors $\bar v \gamma^\mu v$ and $\bar u \gamma^\mu u$ in \eq{L1} transform as 4-vectors and reduce to $2m\,g^{\mu 0} \simeq P^\mu$ in the rest frame (for helicity non-flip). Hence in any frame,
\beq\label{vertices}
\bar u(\halft P+q,\lambda) \gamma^\mu u(\halft P+k,\lambda') \simeq \bar v(\halft P-k,\lambda) \gamma^\mu v(\halft P-q,\lambda') \simeq P^\mu\,\delta_{\lambda,\lambda'}
\eeq
In Coulomb gauge the photon propagator is,
\beq\label{Cgauge}
D^{00}(q)=\frac{i}{\qv^2}\hspace{1cm} D^{0j}(q)=D^{j0}(q)=0 \hspace{1cm} D^{jk}(q)= \frac{i}{q^2}\Big(\delta^{jk}-\frac{q^jq^k}{\qv^2}\Big)
\eeq 
The transverse part $D^{jk}(q^0,\qv)$ depends on $q^0$, and hence (after a Fourier transform) $D^{jk}(t,\qv)$ depends on $t$: Transverse photons propagate at finite speed. When the transverse photon is in flight the Fock state is $\ket{e^+e^-\gamma}$, and described by the wave function $\phi_{e^+e^-\gamma}^\Pv$ in \eq{Fockexp}.

\begin{wrapfigure}[12]{R}{0.25\textwidth}
  \vspace{-20pt}
  \begin{center}
    \includegraphics[width=0.2\textwidth]{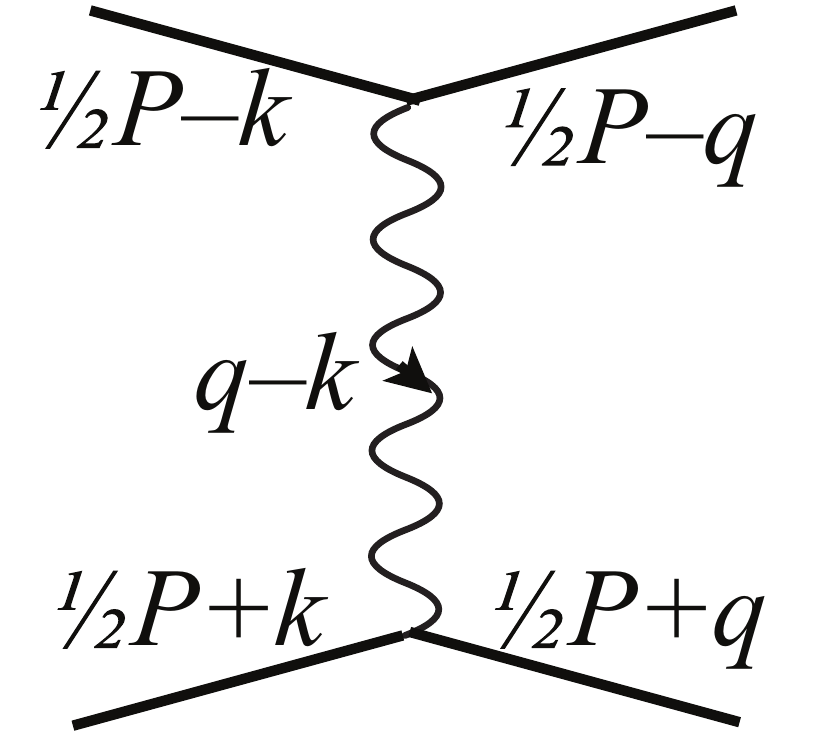}
  \end{center}
  \vspace{-15pt}
  \caption{Single photon exchange amplitude $A$. The charged particles are taken to be scalars.}
\label{fig2to2}
\end{wrapfigure}

It is perhaps worthwhile to convince ourselves with the help of a simple example that the transverse photon contribution cannot be neglected. Let us compare the rest frame expression for the $2 \to 2$ amplitude (\fig{fig2to2}) with that in a general frame. For simplicity we may assume the charged particles to be mass $m$ scalars, and assume $90^\circ$ scattering in the CM:
\beq\label{restmom}
\halft P=(m\sqrt{1+\alpha^2},\bs0) \hspace{1cm} k=(0,0,0,\alpha m)  \hspace{1cm} q=(0,\alpha m,0,0)
\eeq
Using Feynman gauge the exact Lorentz invariant amplitude is easily found to be
\beq\label{fullsamp}
A = 4\pi\,\frac{2+3\alpha^2}{\alpha}
\eeq
After a boost $\zeta$ in the $z$-direction the momenta \eq{restmom} are
\beq\label{genmom}
\halft P=m\sqrt{1+\alpha^2}(\cosh\zeta,0,0,\sinh\zeta) \hspace{1cm} k=\alpha m(\sinh\zeta,0,0,\cosh\zeta)  \hspace{1cm} q=(0,\alpha m,0,0)
\eeq
The propagator \eq{Cgauge} contributes a Coulomb ($C$) and transverse ($T$) part to the scattering amplitude,
\beq\label{CTamp}
A_C=\frac{4\pi}{\alpha}\,\frac{(4+3\alpha^2)\cosh^2\zeta+\alpha^2}{\cosh^2\zeta+1}  \hspace{1cm} 
A_T=-\frac{8\pi}{\alpha}\,\frac{\sinh^2\zeta-\alpha^2}{\cosh^2\zeta+1}
\eeq 
which together form the complete amplitude of \eq{fullsamp}, $A=A_C+A_T$. In the CM ($\zeta=0$) the leading contribution to $A$ is from $A_C$ for small $\alpha$, but in a general frame $A_C$ and $A_T$ are comparable.

The $q^0$-dependence of the transverse propagator in the kernel $K=L_1$ of the bound state equation \eq{BSeq} implies that $\Psi^\Pv(q)$ depends on $q^0$, so in the integrand $\Psi^\Pv(k)$ depends on $k^0$. Hence the integral equation cannot be easily reduced to a time-independent equation, as was the case in the rest frame. This reflects the fact that there are intermediate states with propagating, transverse photons. We must time-order the interactions to find the equal-time Fock state wave functions of positronium in motion.

\subsubsection{Time ordering\label{timeordering}}

In a time-ordered description the ``life-time'' $\Delta t$ of each intermediate state is inversely proportional to its difference in energy from the initial state, $\Delta t \sim 1/\Delta E$. The energies of $\ket{e^+e^-}$ Fock states differ from the positronium energy by approximately the binding energy, thus $\Delta t_{e^+e^-} \sim 1/\alpha^2$. The energy of a transverse photon with Bohr momentum $\qv\sim\alpha m$ is\footnote{Recall that in a time ordered picture $E=\pm\sqrt{\pv^2+m^2}$  for all particles, \cf\ the propagator \eq{ft}. The life-times are Lorentz dilated in boosts, but this does not affect their order of $\alpha$.} $E_q=|\qv|$, so $\Delta t_{e^+e^-\gamma} \sim 1/\alpha$. At small $\alpha$ the positronium atom consequently propagates most of the time as an $\ket{e^+e^-}$ Fock state, with only an \order{\alpha} probability to find a transverse photon in flight. While the scattering amplitude \eq{CTamp} showed that this contribution nevertheless cannot be neglected, the probability that {\em two} transverse photons are in flight {\it simultaneously} is suppressed by a further power of $\alpha$. Similarly the contribution where an instantaneous Coulomb photon is exchanged during the flight of a transverse photon can be neglected at lowest order. Hence only the $\ket{e^+e^-,\Pv}$ and $\ket{e^+e^-\gamma,\Pv}$ Fock states contribute.

\begin{wrapfigure}[15]{r}{0.35\textwidth}
  \vspace{-20pt}
  \begin{center}
    \includegraphics[width=0.3\textwidth]{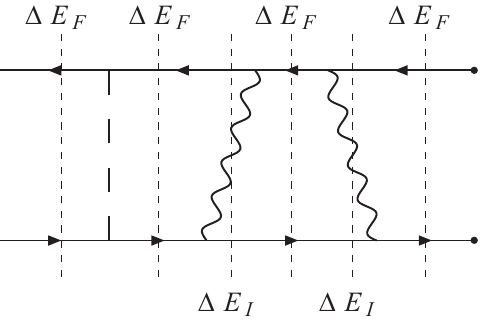}
  \end{center}
  \vspace{-20pt}
  \caption{Positronium propagates mostly as an $e^+e^-$ state (time slices are indicated by the short dashed lines). Transverse photons are exchanged only \order{\alpha} of the time, since $\Delta E_F \propto \alpha^2$ while $\Delta E_I \propto \alpha$. Contributions with overlapping photon exchanges may be neglected at lowest order. Figure from \cite{Jarvinen}.}
\label{thesis2}
\end{wrapfigure}

Multiple, overlapping photon exchanges do contribute at higher orders. This is one of the aspects that complicate bound state perturbation theory. For example, the vertex correction in \fig{feyndiags}(d) contributes to the Lamb shift at \order{\alpha^5}. At this order any number of Coulomb exchanges may be exchanged while the transverse photon is in flight. In section \ref{feyndirac} we shall see that to find the Dirac equation by summing Feynman diagrams we must likewise include diagrams with any number of overlapping photon exchanges.

\begin{figure}
\includegraphics[width=\columnwidth]{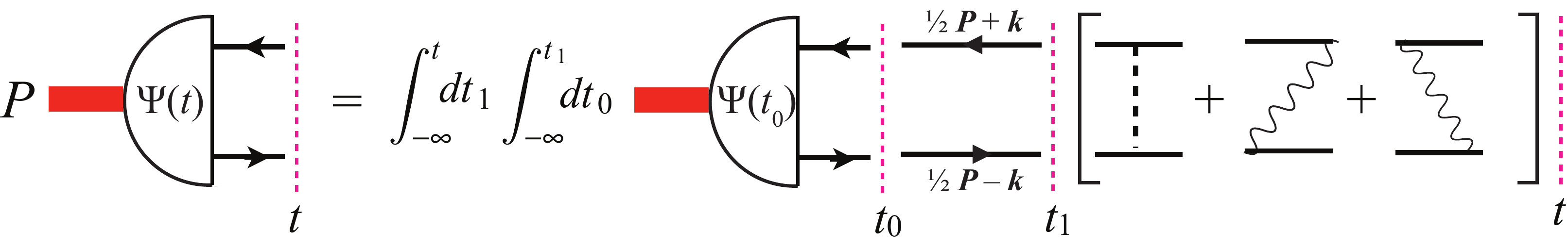}
\caption{Time-ordered version of the bound state equation \eq{BSeq}: $\Psi^\Pv=\Psi^\Pv\,S\,L_1$. 
{\label{timeord}}}
\end{figure}

We now time-order the bound state equation \eq{BSeq} as shown in \fig{timeord}, taking advantage of the non-overlapping photon exchanges. The $e^+e^-$ Fock state wave function $\phi_{e^+e^-}^\Pv$ of \eq{Fockexp} is given by the wave function of \fig{wfdef} at equal time of the constituents,
\beqa\label{equaltwf}
\Psi^\Pv(t,\qv)&=&\int\frac{dq^0}{2\pi}\Psi^P(q)\,e^{-it(\halfs P^0+q^0)-it(\halfs P^0-q^0)}\nn\\
&=&e^{-itP^0}\int\frac{dq^0}{2\pi}\Psi^\Pv(q) \equiv e^{-itP^0}\phi_{e^+e^-}^\Pv(\qv)
\eeqa
The time ordering of the $e^-$ and $e^+$ propagators in $S$ \eq{ffbarprop} is, for $t>0$ and with $E_{k\pm}=\sqrt{(\halft\Pv\pm \kv)^2+m^2}$,
\beqa
S_f(t,\halft \Pv+\kv) &=& \int\frac{dk^0}{2\pi}\,\frac{ie^{-it(\halfs P^0+k^0)}}{\halft P^0+k^0-E_{k+} +\ieps}=e^{-itE_{k+}}\nn\\[2mm]
S_{\bar f}(t,\halft \Pv-\kv) &=& \int\frac{dk^0}{2\pi}\,\frac{ie^{-it(\halfs P^0-k^0)}}{-\halft P^0+k^0+E_{k-} -\ieps}=-e^{-itE_{k-}}
\eeqa
The time-ordered bound state equation then takes the form indicated in \fig{timeord},
\beq
\Psi^\Pv(t,\qv)(P^0-E_{q+}-E_{q-})=\int_{-\infty}^t dt_1 \int_{-\infty}^{t_1} dt_0\,\int\frac{d^3\kv}{(2\pi)^3}\,\Psi^\Pv(t_0,\kv)(P^0-E_{k+}-E_{k-})S(t_1-t_0)L_1(t-t_1)
\eeq
The time-ordered kernel $L_1(t-t_1)$ has contributions from instantaneous Coulomb exchange $\propto \delta(t-t_1)$ and from the transverse photon propagator in \eq{Cgauge}.
The Fourier transform of the factor $1/q^2$ in the transverse photon propagator 
has, as in \eq{ft}, two contributions, depending on whether the photon propagates forward or backward in time. Using \eq{vertices} for the vertex factors and \eq{equaltwf} for the wave functions the bound state equation becomes
\beqa
&& \hspace{-.8cm} e^{-itP^0}\phi_{e^+e^-}^\Pv(\qv)(P^0-E_{q+}-E_{q-}) =  \int\frac{d^3\kv}{(2\pi)^3}\int_{-\infty}^t dt_1 \int_{-\infty}^{t_1} dt_0\,e^{-it_0P^0}\phi_{e^+e^-}^\Pv(\kv)(P^0-E_{k+}-E_{k-}) \nn\\[2mm]
&&\times\ \frac{-1}{2E_{k+}2E_{k-}}e^{-i(t_1-t_0)(E_{k+}+E_{k-})}\\[2mm]
&&\times (-ie^2)\left\{i\frac{(P^0)^2}{(\qv-\kv)^2}\delta(t-t_1) + \left[\Pv^2-\frac{\big(\Pv\cdot(\qv-\kv)\big)^2}{(\qv-\kv)^2} \right]\frac{e^{-i(t-t_1)(E_{k-}+E_{q+}+|\qv-\kv|)}+e^{-i(t-t_1)(E_{k+}+E_{q-}+|\qv-\kv|)}}{2|\qv-\kv|}\right\}\nn
\eeqa
When the time integrals are done we have a bound state equation for the equal-time wave function of the $\ket{e^+e^-}$ Fock state,
\beqa\label{BSt}
&&\hspace{-.8cm}\phi_{e^+e^-}^\Pv(\qv)(P^0-E_{q+}-E_{q-}) = -e^2 \int\frac{d^3\kv}{(2\pi)^3}\phi_{e^+e^-}^\Pv(\kv)\,\frac{1}{2E_{k+}2E_{k-}}\\[2mm]
&&\times\left\{\frac{(P^0)^2}{(\qv-\kv)^2}+\inv{2|\qv-\kv|}\left[\inv{P^0-E_{k-}-E_{q+}-|\qv-\kv|}+\inv{P^0-E_{k+}-E_{q-}-|\qv-\kv|}\right]\left[\Pv^2-\frac{\big(\Pv\cdot(\qv-\kv)\big)^2}{(\qv-\kv)^2} \right] \right\}\nn
\eeqa
Since this is a time-ordered equation it is not explicitly Lorentz covariant. Thus it is not obvious that the energy eigenvalue $P^0$ has the $\Pv$-dependence required by Poincar\'e invariance, nor that the wave function Lorentz contracts. We shall now verify these properties in the range of validity of the equation, \ie, at lowest order in $\alpha$.

%
%

\subsubsection{Reduction to the Schr\"odinger equation\label{redschr}}

Let us first identify the leading power of $\alpha$ on both sides of the equation. On the lhs. $P^0-E_{q+}-E_{q-}$ is of the order of the binding energy, hence of \order{\alpha^2}. On the rhs. the (boosted) Bohr momenta  $|\kv|,|\qv|\propto \alpha$. Hence in the numerator $e^2\int d^3\kv \propto \alpha^4$ while in the denominator $(\qv-\kv)^2 \propto \alpha^2$. The leading powers of $\alpha$ agree, and subleading powers may be ignored.

We denote the electron energy at zeroth order in $\alpha$ by $E$ and the corresponding Lorentz factor by $\gamma$, 
\beq
E \equiv \sqrt{(\halft \Pv)^2+m^2}  \hspace{2cm} \gamma\equiv\frac{E}{m}
\eeq
The binding energy $E_b$ is defined in accordance with \eq{bindenergy},
\beq\label{Pbinden}
P^0=\sqrt{\Pv^2+(2m+E_b)^2} = 2E+\frac{m}{E}\,E_b+\morder{\alpha^4} = 2E+\inv{\gamma}\,E_b + \morder{\alpha^4}
\eeq
The fermion energies are
\beq
E_{q\pm}=\sqrt{(\halft\Pv\pm\qv)^2+m^2}= E\pm\inv{2E}\,\qv\cdot(\Pv\pm\qv)-\inv{8E^3}\,(\qv\cdot\Pv)^2+\morder{\alpha^3}
\eeq
The factor on the lhs. of \eq{BSt} is then
\beq
(P^0-E_{q+}-E_{q-})=\inv{E}\left(mE_b-q_\perp^2-\inv{\gamma^2}\,q_\parallel^2\right)+\morder{\alpha^3}
\hspace{2cm} (\Pv\cdot\qv \equiv |\Pv|\,q_\parallel)
\eeq
where we defined the $\parallel$ and $\perp$ directions wrt. $\Pv$. The energy denominators in \eq{BSt} are
\beqa
P^0-E_{k-}-E_{q+}-|\qv-\kv|&=&\inv{P^0}\,\Pv\cdot(\kv-\qv)-|\qv-\kv|+\morder{\alpha^2} \nn\\
P^0-E_{k+}-E_{q-}-|\qv-\kv|&=&-\inv{P^0}\,\Pv\cdot(\kv-\qv)-|\qv-\kv|+\morder{\alpha^2}
\eeqa
so that
\beq
\inv{P^0-E_{k-}-E_{q+}-|\qv-\kv|}+\inv{P^0-E_{k+}-E_{q-}-|\qv-\kv|}=\frac{(P^0)^2}{(\qv-\kv)^2}
\frac{-2|\qv-\kv|}{\Pv^2-\frac{\big(\Pv\cdot(\qv-\kv)\big)^2}{(\qv-\kv)^2}+4m^2}\Big[1+\morder{\alpha^2}\Big]
\eeq
Substituting this in \eq{BSt} and noting that $2E_{k+}2E_{k-}\simeq (P^0)^2$ the bound state equation becomes
\beq\label{schrmov}
\phi_{e^+e^-}^\Pv(\qv)\left(mE_b-q_\perp^2-\inv{\gamma^2}\,q_\parallel^2\right)=-\frac{e^2m}{\gamma}\int\frac{d^3\kv}{(2\pi)^3}\frac{\phi_{e^+e^-}^\Pv(\kv)}{(\qv-\kv)_\perp^2+\inv{\gamma^2}(\qv-\kv)_\parallel^2}
\eeq
This is the same as the rest frame equation \eq{Schrmom} when the longitudinal components of $\qv$ and $\kv$ are scaled by $\gamma$. We conclude that the binding energy $E_b$ is independent of $\Pv$, so that the energy \eq{Pbinden} of the bound state has the correct frame dependence. The wave function Lorentz contracts classically in coordinate space since the longitudinal components of the relative momenta scale with the Lorentz factor $\gamma$.

\begin{wrapfigure}[17]{r}{0.5\textwidth}
  \vspace{-20pt}
  \begin{center}
    \includegraphics[width=0.45\textwidth]{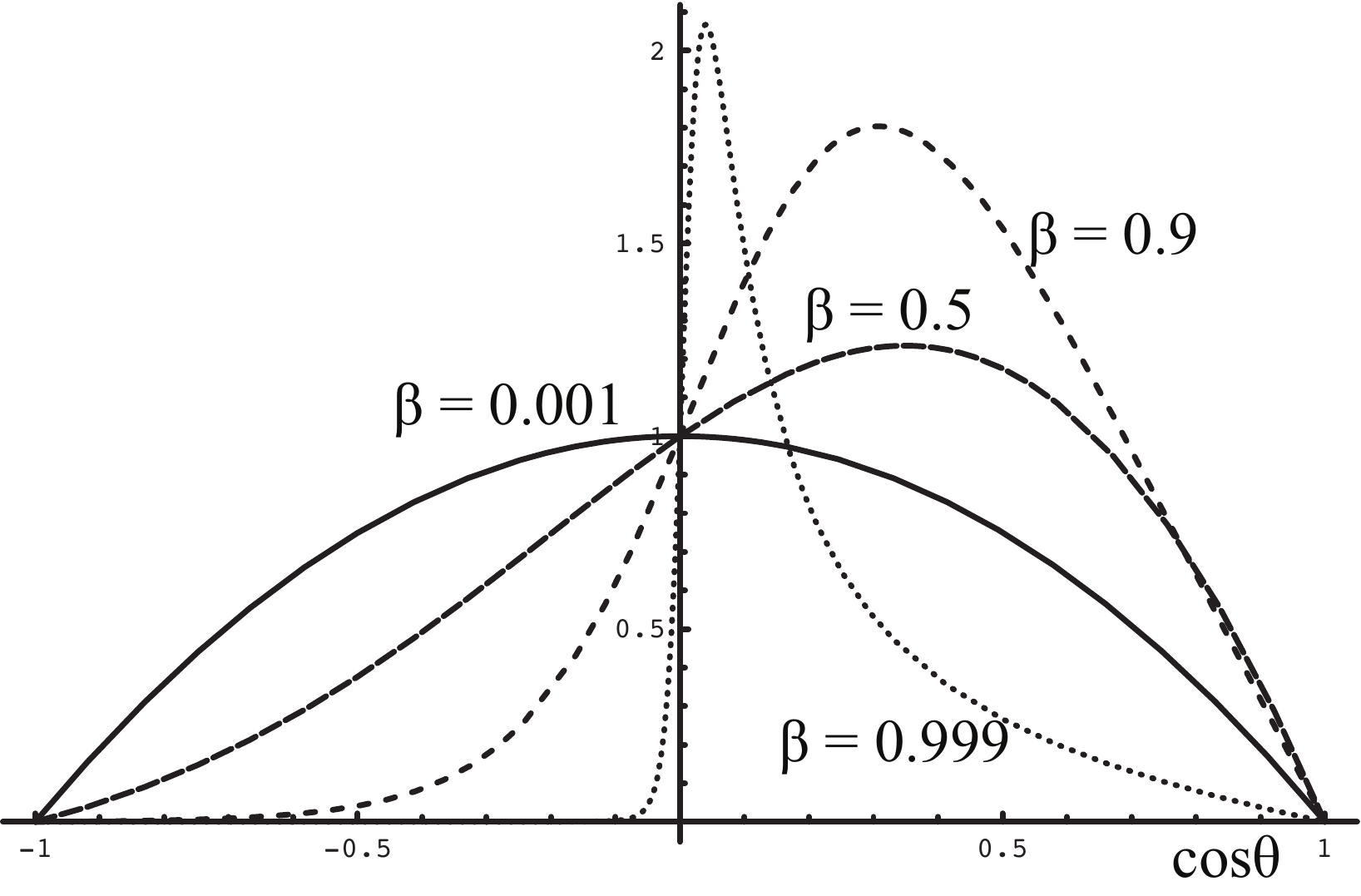}
  \end{center}
  \vspace{-15pt}
  \caption{Angular distribution of the photon wrt. to $\Pv$ in the $\ket{e^+e^-\gamma,\Pv}$ Fock state, for various values of the bound state velocity $\beta$ \cite{Jarvinen:2004pi}. The contraction effect on $\cos\theta$ is removed by dividing the longitudinal momentum of the photon by $\gamma=E/m$.}
\label{fig8}
\end{wrapfigure}

The wave function $\phi_{e^+e^-\gamma}$ of the $\ket{e^+e^-\gamma}$ Fock component is given by the sum of the amplitudes for the radiation of the photon from the electron and the positron \cite{Jarvinen:2004pi}. With increasing bound state momentum $\Pv$ the photon is emitted preferentially in the forward direction, as shown in \fig{fig8}. In the infinite momentum frame the result agrees with the wave function of Light-Front quantization, where the photon is never emitted in the backward direction. 

As we have seen, the $\ket{e^+e^-\gamma,\Pv}$ equal-time Fock state occurs with a small probability of \order{\alpha} in a positronium atom, but contributes at leading order to the binding energy when $\Pv \neq 0$. In strongly bound states with $\alpha$ of \order{1} transverse photon exchange will be more prominent. Although perturbative methods are then insufficient we may expect that the frame dependence will be less similar to classical contraction \cite{Rocha:2009xq}.

An equal-time formulation allows to study bound states both in the rest frame (with rotational symmetry) and in the infinite momentum frame, which corresponds \cite{Weinberg:1966jm,Brodsky:1997de} to quantization on the light front (equal $t+z$). When the Coulomb field of the rest frame is boosted it generates a transverse field component. The underlying Poincar\'e invariance of QED ensures that all physical quantities have the correct frame dependence, even though equal-time wave functions transform non-trivially under boosts.

\subsection{Hamiltonian formulation\label{Hform}}

\subsubsection{Ladder diagrams sum to a classical EM field\label{irprobs}}

We have seen that the bound state poles of QED $S$-matrix elements arise from the divergence of an infinite series of Feynman diagrams. This is rather remarkable, as the calculation is based on a {\em perturbative} expansion where higher order terms should be small corrections. In section \ref{divsum} we found that ladder diagrams are not suppressed in the region where the soft momenta scale with $\alpha$ as in \eq{bohrmom}. This is superficially similar to QCD, where bound state (hadron) physics becomes manifest in soft scattering processes. Let us analyze the reason for the phenomenon in QED, which is readily understood.

\begin{wrapfigure}[11]{r}{0.5\textwidth}
  \vspace{-20pt}
  \begin{center}
    \includegraphics[width=0.45\textwidth]{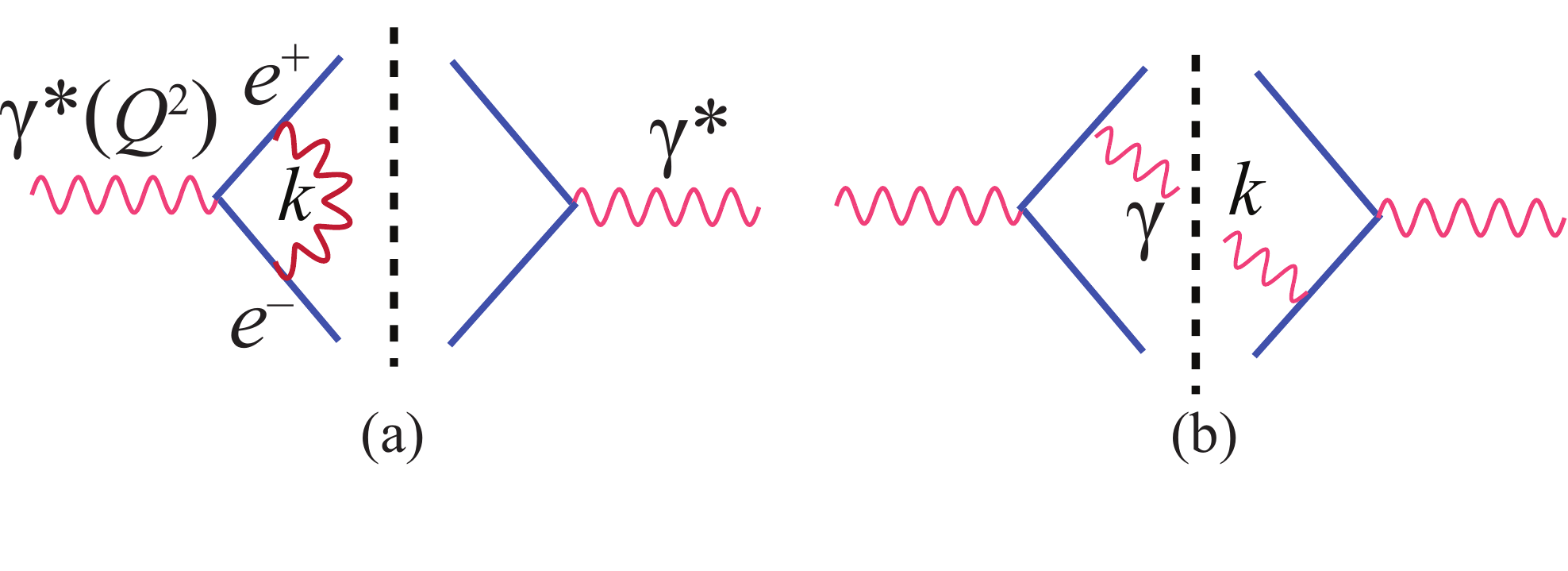}
  \end{center}
  \vspace{-30pt}
  \caption{(a) An \order{\alpha} correction to the squared electron form factor. (b) An $e^+e^-\gamma$ final state contribution needed to cancel $k\to 0$ singularities. Only representative diagrams are shown.}
\label{optical}
\end{wrapfigure}

In order to get simple expressions for $S$-matrix elements standard perturbation theory expands around non-interacting $\ket{in}$ and $\ket{out}$ states. Thus an incoming electron of momentum $\pv$ is described by the state $b_{in}^\dag(\pv)\ket{0}$ with no comoving photon field. The photons are in principle generated by the interactions during the (infinite) time interval from the initial $t_{in}= -\infty$ to the time of a scattering process. However, at the lowest order of the perturbative expansion the photon field remains absent -- one expands around an unphysical state.

The use of non-interacting asymptotic states in perturbation theory leads to infrared problems. These can, however, be ``fixed'' by considering only processes that are inclusive of soft photons.  The loop correction to the squared electron form factor ($\gamma^* \to e^+e^-$) shown in \fig{optical}(a) is a well-known example. The integral over the loop momentum $k$ is logarithmically divergent at $k = 0$. Adding contributions like \fig{optical}(b) with a photon of momentum $k$ in the final state ($\gamma^* \to e^+e^-\gamma$) cancels the singularity provided one integrates over all $k<\mu$ of the additional photon, where $\mu$ is a finite parameter\footnote{The cancellation occurs because the contributions shown in \fig{optical} build the imaginary part of an electron loop correction to the photon propagator. The initial and final (transverse) photons are neutral, physical states, so the photon propagator is IR finite.}. For small $\mu$ and large virtuality $Q^2$ of the $\gamma^*$ one needs to sum the leading logarithms of any number of loops and of soft, final state photons. This results in the Sudakov form factor of the electron \cite{Collins:1989bt},
\beq\label{sudakov}
F(Q^2/\mu^2) = \exp\left[-\frac{\alpha}{4\pi}\ln^2(Q^2/\mu^2)\right]
\eeq 
which vanishes as $\mu \to 0$, \ie, as one considers cross sections where the momentum of any photon in the final state approaches zero. Physically, this tells us that charged particles are always accompanied by a cloud of soft photons. In practice, all measurements are inclusive of (undetectable) soft photons. Note also that the Born level form factor has no IR singularities and approximates a sufficiently {\em inclusive} measurement. The situation is similar in QCD.

Atoms are bound by the very same soft photons that are neglected in the asymptotic states of the perturbative expansion. Hence Feynman diagrams do not have bound state poles -- individual diagrams do not even give a first approximation to bound state physics. Bound states are characterized by being stationary in time, which requires the presence of a soft photon cloud around the electrons. Fortunately, perturbation theory allows us to identify the leading contributions to bound states, namely the ladder diagrams like \fig{feyndiags}(a) and 2(b). Their sum tells us something we might have expected: The soft photon cloud is equivalent to the classical electromagnetic field generated by the electric charges. Thus the $-\alpha/r$ potential in the Schr\"odinger equation \eq{Schrcoo} is more simply obtained using Gauss' law for the Coulomb potential $A^0$.

The field theoretical Schr\"odinger equation
\beq\label{ftsch}
\hat H\ket{E}=E\ket{E}
\eeq
is the exact expression of the stationarity in time of the bound state $\ket{E}$. The Hamiltonian operator $\hat H$ is determined by the QED action as the generator of time translations. Its interaction term $\hat H_{int}$ creates and annihilates electrons, positrons and photons. Correspondingly, a positronium eigenstate $\ket{E}$ is an infinite superposition of Fock states with any number of $e^+e^-$ pairs and photons. We cannot solve \eq{ftsch} exactly, but the above arguments show that in the $\alpha\to 0$ limit the leading, ``Born level'' bound state involves (in the rest frame) only one non-relativistic $e^+e^-$ pair and a photon field which is given by the classical Coulomb field $A^0$ of the electron and positron. We shall next demonstrate, using operator methods, how the exact equation \eq{ftsch} reduces to the standard $c$-numbered Schr\"odinger equation postulated in introductory courses on quantum mechanics.

Due to its instantaneity, the Coulomb ($A^0$) interaction brings no propagating (transverse) photon constituents into $\ket{E}$. This simplifies the analysis of processes where $A^0$ dominates over $\Av^\perp$. In Section \ref{diracbound} we discuss relativistic Dirac states bound by an $A^0$ potential, and in Section \ref{ffstatesec} we consider positronium in $D=1+1$ dimensions, where $\Av^\perp=0$. In both cases the bound states contain any number of $e^+e^-$ pairs, but can nevertheless be described by explicit wave functions. In Section \ref{3dpotential} we study whether a classical $A^0$ field can describe confinement in $D=3+1$ dimensions. Allowing 
an additional homogeneous solution of Gauss' law results in a linear potential. The requirements of Poincar\'e and gauge invariance are fulfilled in a novel way.

\subsubsection{The QED Hamiltonian in $\nv\cdot\Av=0$ gauge\label{hameigen}}

We begin by briefly recalling \cite{406190} some basic relations in Coulomb gauge ($\nv\cdot\Av=0$). QED is defined by its action
\beq\label{qedaction}
{\mathcal S}_{QED} = \int d^4y\,\mL(y) =  \int d^4y \left[\bar\psi(i\slashed{\partial}-e\hat{\slashed{A}}-m)\psi-\quart F_{\mu\nu} F^{\mu\nu}\right]
\eeq 
We use a hat on the electromagnetic field operator $\hat A^\mu$ to distinguish it from the $c$-number (classical) field $A^\mu$. The equation of motion for the Coulomb field gives Gauss' law,
\beq\label{gausslaw}
\frac{\delta{\mathcal S}_{QED}}{\delta \hat A^0(t,\xv)}=0 \hspace{1cm} \Rightarrow \hspace{1cm}
-\nv^2 \hat A^0(t,\xv)=e\psi^\dag(t,\xv)\psi(t,\xv)
\eeq
This allows to express the Coulomb field in terms of the electron field,
\beq\label{a0sol}
\hat A^0(t,\xv)= \int d^3\yv\,\frac{e}{4\pi|\xv-\yv|}\psi^\dag\psi(t,\yv)
\eeq 

To obtain the QED Hamiltonian from the action \eq{qedaction} we first identify the conjugate fields. The conjugate of the electron field is
\beq\label{elconj}
\frac{\delta{\mathcal S}_{QED}}{\delta \partial_0\psi(x)}=i\psi^\dag(x)
\eeq
The conjugate of the vector potential $\hat A^j$ is
\beq\label{ajconj}
\pi_j=\frac{\delta{\mathcal S}_{QED}}{\delta \partial_0 \hat A^j}=F_{j0}=\partial_j \hat A^0+\partial_0\hat A^j=-E^j
\eeq
where $\bs{E}$ is the electric field operator. Since the action is independent of $\partial_0\hat A^0$ the field conjugate to $\hat A^0$ vanishes, $\pi^0=0$. The Hamiltonian density is then obtained from the Lagrangian density in the standard way,
\beq\label{ham1}
\mH=i\psi^\dag\partial_0\psi+\pi_j\partial_0 A^j-\mL = \bar\psi(-i\nv\cdot\gv+e\hat{\slashed{A}}+m)\psi+\halft(\bs{E}^2+\bs{B}^2)+\bs{E}\cdot\nv A^0
\eeq
where $\bs{B}=\nv\times\hat\Av$ is the magnetic field. If we express the electric field in terms of $A^0$ and $\Av$ as in \eq{ajconj} and use Gauss' law \eq{gausslaw} we obtain, after partial integrations (neglecting contributions from spatial infinity),
\beq\label{ham2}
H = \int d^3\xv\,\mH=\int d^3\xv\big[\bar\psi(-i\nv\cdot\gv+\halft e\gz\hat A^0 -e\gv\cdot\hat\Av+m)\psi +\halft(\partial_0\hat\Av)^2+\halft\bs{B}^2\big]
\eeq
We may interpret the factor $\halft$ in front of the Coulomb interaction term as due to a partial cancellation between the fermion Coulomb interaction and the energy of the Coulomb field.

\vspace{-.3cm}

\subsubsection{Positronium as an eigenstate of the QED Hamiltonian\label{hameigen2}}

\vspace{-.1cm}

A positronium state may be expressed as a superposition of $e^+e^-$ pairs specified by an equal-time wave function $\Phi_{\alpha\beta}$ with $4\times 4$ $c$-numbered Dirac components,
\beq\label{posstate}
\ket{e^+e^-,t} = \int d^3\xv_1\,d^3\xv_2\,\bar\psi_\alpha(t,\xv_1)\Phi_{\alpha\beta}(\xv_1-\xv_2)\psi_{\beta}(t,\xv_2)\ket{0}
\eeq
This is a state at rest, $\hat\Pv\ket{e^+e^-,t}=0$, since the effect of a translation $e^{i\hat\Pv\cdot\bs{a}}\psi(t,\xv_i)e^{-i\hat\Pv\cdot\bs{a}}=\psi(t,\xv_i+\bs{a})$ of the fields can be eliminated by a coordinate transformation, $\xv_i \to \xv_i+\bs{a}$.

Using \eq{a0sol} we see that a state with a single electron at $\zv$ is an eigenstate of $\hat A^0$, 
\beq\label{fermionpot}
\hat A^0(t,\xv)\psi_{\alpha}^\dag(t,\zv)\ket{0} = \int d^3\yv\,\frac{e}{4\pi|\xv-\yv|}\psi_\beta^\dag(t,\yv)\acom{\psi_\beta(t,\yv)}{\psi^\dag_\alpha(t,\zv)}\ket{0} = 
\frac{e}{4\pi|\xv-\zv|}\psi_{\alpha}^\dag(t,\zv)\ket{0}
\eeq
where the neglect of $e^+e^-$ pair production is justified for non-relativistic dynamics.

\begin{wrapfigure}[11]{r}{0.2\textwidth}
  \vspace{-35pt}
  \begin{center}
    \includegraphics[width=0.2\textwidth]{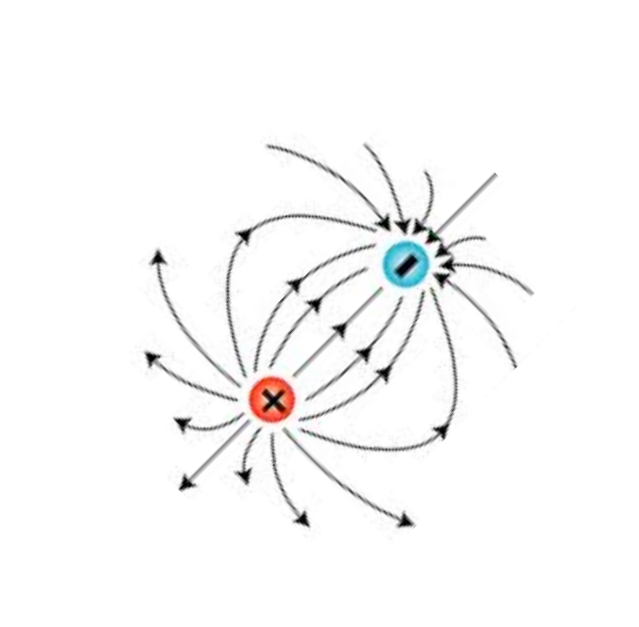}
  \end{center}
  \vspace{-20pt}
  \caption{The dipolar $A^0$ field \eq{a0expr} for specific positions of the charges.}
\label{Dipole}
\end{wrapfigure}

Similarly to \eq{fermionpot}, the components of the positronium state \eq{posstate} are eigenstates of $\hat A^0$,  
\beq\label{a0eigen}
\hat A^0(t,\xv)\,\bar\psi(t,\xv_1)\psi(t,\xv_2)\ket{0} =A^0(\xv;\xv_1,\xv_2)\bar\psi(t,\xv_1)\psi(t,\xv_2)\ket{0}
\eeq
Since the positron contributes with the opposite sign due to the anticommutation relation we have (\fig{Dipole})
\beq\label{a0expr}
A^0(\xv;\xv_1,\xv_2) = \frac{e}{4\pi}\left(\inv{|\xv-\xv_1|}-\inv{|\xv-\xv_2|}\right)
\eeq
The dipolar Coulomb field $A^0(\xv;\xv_1,\xv_2)$ is seen to depend on the positions $\xv_1,\xv_2$ of the charges and, due to the charge screening, falls faster than $1/|\xv|$ at large $|\xv|$.
This approach differs from the standard discussion of the Hydrogen atom, where one reduces the two-body problem to that of a single charge in a fixed $-\alpha/r$ potential.

Bound states are by definition stationary in time, and hence eigenstates of the Hamiltonian as in \eq{ftsch}. In section \ref{possection} we saw that the sum of ladder diagrams generates (in the rest frame) a classical $A^0$ potential, which satisfies Gauss' law \eq{gausslaw}. At leading order we may thus neglect the vector field $\Av$ in the Hamiltonian \eq{ham2}. Using \eq{a0sol} the interaction term becomes 
\beq\label{intham}
H_{int}(t)= \int d^3\xv d^3\yv\,\big[\psi^\dag(t,\xv)\psi(t,\xv)\big]\big[\psi^\dag(t,\yv)\psi(t,\yv)\big]\frac{\alpha}{2|\xv-\yv|}
\eeq

In evaluating the action of $H$ on the state \eq{posstate} we may discard pair 
production, due to the non-relativistic dynamics. Hence two pairs of fermion fields need to annihilate, setting $\xv=\xv_1$ and $\yv=\xv_2$ in \eq{intham} or {\it vice versa} -- this gives a factor 2. Altogether,
\beqa
H(t)\ket{e^+e^-,t}&=&\int d^3\xv_1\,d^3\xv_2\,\bar\psi(t,\xv_1)\Big[(-i\lnab_1\cdot\gamma^0\gv+m\gamma^0)\Phi(\xv_1-\xv2)-\Phi(\xv_1-\xv2)(-i\rnab_2\cdot\gamma^0\gv+m\gamma^0)\nn\\ &-& \frac{\alpha}{|\xv_1-\xv_2|}\Big]\psi(t,\xv_2)
\eeqa
After partial integrations the state has the same form as the positronium state \eq{posstate}. The eigenstate condition \eq{ftsch} becomes
\beq\label{bseq1}
i\nv\cdot\acom{\gamma^0\gv}{\Phi(\xv)}+m\com{\gamma^0}{\Phi(\xv)} = \big[E-V(\xv)\big]\Phi(\xv)
\eeq
with the standard potential
\beq\label{coulpot}
V(\xv)=-\frac{\alpha}{|\xv|}
\eeq

Although the bound state equation \eq{bseq1} has a relativistic, ``double Dirac'' appearance, it was derived assuming the dynamics of non-relativistic positronium at rest ($\bs{A}=0$, no pair production). It should reduce to the Schr\"odinger equation similarly as the standard Dirac equation. As in \eq{bindenergy} we express the total energy as $E=2m+E_b$ and identify the relative magnitudes at small $\alpha$:
\beq
m=\morder{\alpha^0} \hspace{2cm} \nv=\morder{\alpha} \hspace{2cm}  E_b,\ V=\morder{\alpha^2} 
\eeq
Writing the $4\times 4$ wave function $\Phi$ in $2\times 2$ block form the bound state condition \eq{bseq1} becomes, using the Dirac representation of the $\gamma$ matrices and $\sv=(\sigma_x,\sigma_y,\sigma_z)$,
\beq 
i\nv\cdot\left[\begin{array}{cc} 
\sv\Phi_{21}+\Phi_{12}\sv& \sv\Phi_{22}+\Phi_{11}\sv \\ \sv\Phi_{11}+\Phi_{22}\sv & \sv\Phi_{12}+\Phi_{21}\sv
\end{array} \right] + 
2m\left[\begin{array}{cc} 
0 & \Phi_{12} \\ -\Phi_{21} & 0
\end{array} \right]
=(2m+E_b-V)\left[\begin{array}{cc} 
\Phi_{11}& \Phi_{12} \\ \Phi_{21} & \Phi_{22}
\end{array} \right]  \hspace{1cm}
\eeq
The terms with the \order{\alpha^0} coefficient $m$ require that $\Phi_{11},\,\Phi_{22}$ and $\Phi_{21}$ be suppressed by at least a factor $\alpha$ compared to $\Phi_{12}$. Taking $\Phi_{12}$ to be of \order{\alpha^0} the conditions
\beqa
i\nv\cdot\sv\Phi_{21} + i\nv\cdot\Phi_{12}\sv &=& 2m\Phi_{11} \nn\\
i\nv\cdot\sv\Phi_{12} + i\nv\cdot\Phi_{21}\sv &=& 2m\Phi_{22}
\eeqa
imply that $\Phi_{11}$ and $\Phi_{22}$ are \order{\alpha} and that $2i\,m\nv\cdot\Phi_{11}\sv=2i\,m\nv\cdot\sv\Phi_{22}=-\nv^2\Phi_{12}$. Then the \order{\alpha^2} condition
\beq
i\nv\cdot\sv\Phi_{22}+i\nv\cdot\Phi_{11}\sv = (E_b-V)\Phi_{12}
\eeq
gives the Schr\"odinger equation \eq{Schrcoo} for $\Phi_{12}$. As expected, the relative magnitudes of the components of $\Phi$ are consistent with those of the product of two non-relativistic spinors,
\beq
\Phi \sim u\,v^\dag \sim \left[\begin{array}{c}1 \\ \morder{\alpha}\end{array}\right]
\left[\begin{array}{cc}\morder{\alpha} & 1 \end{array}\right] =
\left[\begin{array}{cc}\morder{\alpha} & 1 \\ \morder{\alpha^2} & \morder{\alpha} \end{array}\right]
\eeq
The Schr\"odinger equation is the same for all four component of the $2\times 2$ matrix $\Phi_{12}$, reflecting the spin independence of non-relativistic dynamics.

Compared to the standard approach of Introductory Quantum Mechanics the field theory derivation of atomic bound states has the advantage of being based on the QED action, not requiring to postulate the Schr\"odinger equation. It is applicable also for relativistic systems, as we shall discuss next in terms of Dirac bound states.

\section{Dirac bound states\label{diracbound}}

The Dirac equation describes the dynamics of an electron in an external field, which I shall assume to arise from a static charge $-eZ$ at the origin. The charge generates the Coulomb field 
\beq\label{diracsource}
A_Z^0(\xv)= -\frac{eZ}{4\pi|\xv|}
\eeq
whose interaction with the electron is described by the potential
\beq\label{coulpotZ}
V_Z(\xv) = -\frac{\alpha Z}{|\xv|}
\eeq
The Dirac equation for a bound state of energy $E$, described by the c-numbered, 4-component wave function $\Psi(\xv)$ is
\beq\label{direq}
(-i\nv\cdot\gamma^0\gv+m\gamma^0)\Psi(\xv)=(E-V_Z)\Psi(\xv)
\eeq

It is important to distinguish the c-numbered Dirac equation \eq{direq} from the {\em operator-valued} equation of motion which is obtained by varying the QED action \eq{qedaction} wrt. $\bar\psi$,
\beq\label{opdirac}
(i\slashed{\partial}-e{\hat{\slashed{A}}}-m)\psi=0
\eeq
This relation is exact for all matrix elements of physical states, whereas loop effects are neglected in \eq{direq}.

We first recall how the Dirac equation may be obtained by summing Feynman diagrams, generalizing the corresponding derivation of the Schr\"odinger equation presented in section \ref{possection}. We then show how the equation may be derived using the field theoretic Hamiltonian method, which gives further insight into the structure of a Dirac state. Finally we discuss the explicit solutions of the Dirac equation in $D=1+1$ dimensions, where the Coulomb potential is linear.

\subsection{The Dirac equation from Feynman diagrams\label{feyndirac}}

For relativistic effects to be relevant the electron binding energy must be of the order of its mass $m$, which implies $\alpha Z\sim$ \order{1} in \eq{coulpotZ}. At the same time we need $\alpha\ll 1$ to justify the neglect of higher order (loop) corrections in $\alpha$. The Dirac equation is then obtained by summing all straight {\em and} crossed ladder diagrams in the limit where the source particle (of charge $-eZ$) is very massive, $M\to\infty$  \cite{diracref}.

\begin{wrapfigure}[12]{r}{0.5\textwidth}
  \vspace{-30pt}
  \begin{center}
    \includegraphics[width=0.45\textwidth]{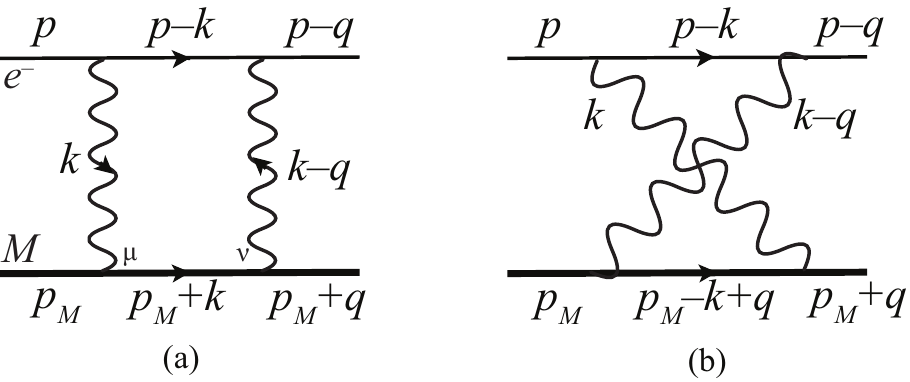}
  \end{center}
  \vspace{-20pt}
  \caption{Uncrossed (a) and crossed (b) ladder diagrams describing electron (upper line) scattering in the Coulomb field of a heavy particle (lower line).}
\label{diracfig}
\end{wrapfigure}

With the notations of \fig{diracfig} we take $p_M=(M,\bs{0})$ to be the initial momentum of the heavy particle. After the exchange of momentum $q$ its final momentum is $p_M+q=(M+q^0,\qv)$. The on-shell condition $(p_M+q)^2=M^2$ gives $q^0\simeq \qv^2/2M \to 0$ as $M \to \infty$. For single photon exchange the lower vertex $\bar u_M(\qv)\gamma^\mu u_M(\bs{0}) \simeq \delta^{\mu,0}2M$ since the spinor $u_M$ of the heavy particle is non-relativistic. The Born diagram thus becomes
\beq\label{t1amp}
T_1= 2M\bar u(\pv-\qv)(-ie\gamma^0)(ieZ) \frac{i}{\qv^2}u(\pv)
\eeq

As in section \ref{divsum} we note that the crossed ladder diagram in \fig{diracfig}(b) does not contribute to non-relativistic bound states since the positive energy poles in $k^0$ of the $p-k$ and $p_M-k+q$ propagators are on the same side of the $k^0$ integration contour. This argument does not apply to Dirac states, where also the negative energy pole of the relativistic electron propagator contributes.

The Dirac algebra of the heavy particle in the ladder of \fig{diracfig}(a) gives
\beqa\label{largenum}
\bar u_M(\qv)\gamma^\nu(\slashed{p}_M+\slashed{k}+M)\gamma^\mu u_M(\bs{0})\hspace{2.5cm}&&\nn\\
\simeq M \bar u_M(\bs{0})\gamma^\nu(1+\gamma^0)\gamma^\mu u_M(\bs{0})
=(2M)^2\delta^{\mu,0}\delta^{\nu,0}\hspace{.3cm}&&
\eeqa
where the loop momentum $k$ could be ignored compared to $M$ since the loop integral converges. The same result is obtained for the crossed ladder diagram in \fig{diracfig}(b). The denominators $(p_M+k)^2-M^2+\ieps$ and $(p_M-k+q)^2-M^2+\ieps$ contribute, respectively,
\beqa
\inv{\big(M+k^0-\sqrt{M^2+\kv^2}+\ieps\big)\big(M+k^0+\sqrt{M^2+\kv^2}-\ieps\big)} &\simeq& \inv{2M}\,\inv{k^0+\ieps}\nn\\
\inv{\big(M-k^0+q^0-\sqrt{M^2+(\kv-\qv)^2}+\ieps\big)\big(M-k^0+q^0+\sqrt{M^2+(\kv-\qv)^2}-\ieps\big)} &\simeq& \inv{2M}\,\inv{-k^0+\ieps}
\eeqa 
The other factors of the two diagrams in \fig{diracfig} are identical, so we may add these terms, giving $-2\pi i\delta(k^0)/2M$. The sum of the diagrams is thus
\beq\label{t2amp}
T_2 = 2M\,\bar u(\pv-\qv) \int\frac{d^3\kv}{(2\pi)^3} \Big[(-ie\gamma^0)(ieZ)\frac{i}{\kv^2}\,i\frac{\slashed{p}-\slashed{k}+m}{(p-k)^2-m^2+\ieps}(-ie\gamma^0)(ieZ)\frac{i}{(\kv-\qv)^2}\Big]u(\pv)
\eeq

The expression \eq{t1amp} for single photon exchange and that of \eq{t2amp} for two-photon exchange describe scattering from a time-independent external charge $-eZ$. The analysis can be generalized to any number of photon exchanges, provided {\em all} crossed photon diagrams are included: $n!$ diagrams must be added for $n$-photon exchange. The result (with the factor $2M$ and the spinors $\bar u(\pv-\qv)$ and $u(\pv)$ removed) is of the form
\beq
\Vsl_Z+\Vsl_Z\,S\,\Vsl_Z + \ldots = \Vsl_Z\,\inv{1-S\,\Vsl_Z} = \Vsl_Z\,S^{-1}\, \inv{S^{-1}-\Vsl_Z}
\eeq
where the products involve 3-momentum convolutions, $S$ is the free Dirac propagator and $\Vsl_Z= \gz V_Z$ is given by the external potential \eq{coulpotZ} (in momentum space). Bound state poles can occur when
\beq
S^{-1}-\gz V_Z =0
\eeq
which implies the Dirac equation \eq{direq} for states that are stationary in time.

Just as for positronium, bound state poles in the scattering amplitude arise not from any single Feynman diagram but from the divergence of their sum. With each additional photon exchange there are more photons which cross each other. A standard Bethe-Salpeter approach (\cf\ \eq{BSeq}) is based on iterating a kernel $K$. In a kernel of \order{\alpha^n} one photon can cross at most $n-1$ others. This means that the Dirac equation, which requires any number of crossings, cannot be obtained from the usual Bethe-Salpeter equation with a kernel of finite order.

\begin{wrapfigure}[15]{r}{0.5\textwidth}
  \vspace{-20pt}
  \begin{center}
    \includegraphics[width=0.45\textwidth]{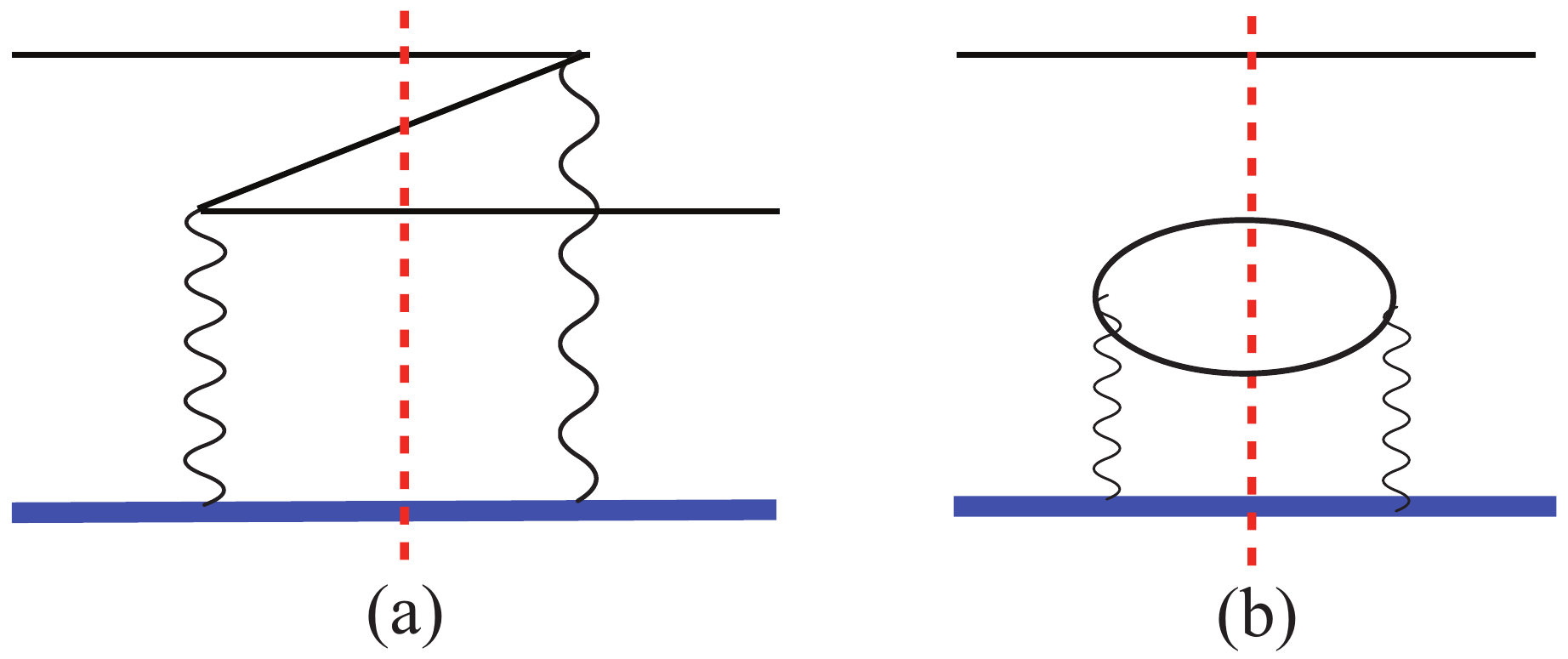}
  \end{center}
  \vspace{-20pt}
  \caption{(a) Time-ordered version of \fig{diracfig}(b) (time is running from left to right). The dashed line indicates an intermediate time with an additional $e^+e^-$ pair. If the dashed line is viewed as a unitarity cut the diagram represents the product of two scatterings with pair creation/annihilation. (b) Squaring the pair production amplitude in (a) gives a loop diagram.}
\label{pairfig}
\end{wrapfigure}

Coulomb photon exchanges are instantaneous in time. When a crossed photon diagram like \fig{diracfig}(b) is time-ordered it turns into the diagram of \fig{pairfig}(a). At the intermediate time indicated by the dashed line there is an extra $e^+e^-$ pair. Higher order diagrams contribute several pairs, so a relativistic bound state must have Fock components with any number of pairs. Thus the Dirac wave function should not be thought of as a single particle wave function, as known already from the Klein paradox \cite{Hansen:1980nc}. Even though $\Psi(\xv)$ has the degrees of freedom of a single particle it describes the spectrum of a relativistic state with many constituents. This is similar to hadrons, whose quantum numbers are found to be given by their valence quarks, even though hadrons have a sea of $q\bar q$ pairs.

Ladder diagrams like those in \fig{diracfig} which build the Dirac states are distinguished by being of leading order in $\alpha Z$. Loop corrections on the electron and photon propagators are \order{\alpha} and neglected. However, a loop correction on the target line (\fig{pairfig}(b)) is of leading order in $\alpha Z$. It factorizes from the electron scattering dynamics since a photon exchange between the loop and the electron would be of \order{\alpha}. Such target corrections nevertheless affect the Dirac wave function via interference effects. If the amplitude on the left side of the dashed line in \fig{pairfig}(a) is squared it gives both diagram (a) and the loop diagram (b): Once an $e^+e^-$ pair is created the state has two electrons which are indistinguishable and interfere. 

As shown by Weinberg \cite{406190}, regardless of its interpretation the Dirac wave function should be normalized to unity when the normalization integral converges. In section \ref{dirsol} we shall see that the normalization integral does not converge in $D=1+1$ dimensions, where the QED$_2$ potential is linear. The norm of the wave function tends to a constant at large distances from the source, reflecting abundant pair fluctuations in the strong potential.

\subsection{Dirac states as eigenstates of the Hamiltonian\label{hamdirac}}

Instead of summing Feynman diagrams we may, in analogy to positronium \eq{posstate}, express a Dirac bound state as
\beq\label{dirstate}
\ket{E,t}= \int d^3\xv\, \psi^\dag(t,\xv)\Psi(\xv)\ket{0}
\eeq
where $\Psi(\xv)$ is the 4-component, $c$-numbered Dirac wave function of \eq{direq}. The fixed external Coulomb field \eq{diracsource} takes the place of $\hat A^0$ in the QED Hamiltonian \eq{ham2}. The Dirac equation \eq{direq} for $\Psi(\xv)$ follows from
\beq\label{hamdirstate}
H\ket{E,t}= \int d^3\xv\com{H}{\psi^\dag(t,\xv)}\Psi(\xv)\ket{0} = E\ket{E,t}
\eeq
where we needed
\beq\label{zvac}
H\ket{0}=0
\eeq
The negative energy components of $\Psi(\xv)$ in \eq{dirstate} are connected to the positron annihilation term $v^\dag d$ in the electron field $\psi^\dag(t,\xv)$. Thus by keeping the $d\ket{0}$ contribution we implicitly include the pair production effects shown in the time-ordered diagram \fig{pairfig}(a). These unusual rules imply that we are using retarded boundary conditions, which may be justified as follows \cite{Hoyer:2009ep}.

The retarded electron propagator is obtained by changing the $\ieps$ prescription at the negative energy pole,
\beq\label{sr}
S_{R}(p^0,\pv) = i\frac{\slashed{p}+m_{e}}{(p^0-E_{p}+\ieps)(p^0+E_{p}+\ieps)}
\eeq
where $E_{p}=\sqrt{\pv^2+m_{e}^2}$. This means that backward propagation is inhibited, both positive and negative energy electrons move forward in time,
\beq\label{srt}
S_{R}(t,\pv) = \frac{\theta(t)}{2E_{p}}\left[(E_{p}\gamma^0-\pv\cdot\gv+m_{e})e^{-iE_{p}t} +(E_{p}\gamma^0+\pv\cdot\gv-m_{e})e^{iE_{p}t}\right]
\eeq
Consequently the $Z$-diagram of \fig{pairfig}(a) is absent: only a single (positive or negative energy) electron is present at any intermediate time, and it is described by the Dirac wave function.

\begin{wrapfigure}[8]{r}{0.3\textwidth}
  \vspace{-30pt}
  \begin{center}
    \includegraphics[width=0.3\textwidth]{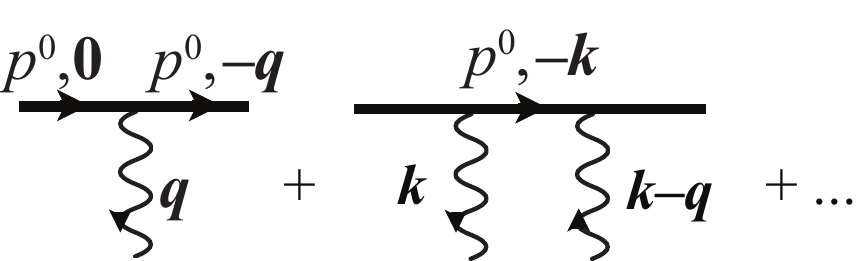}
  \end{center}
  \vspace{-15pt}
  \caption{The $p^0$ component of the electron momentum is unchanged in scattering from a static source.}
\label{retfig}
\end{wrapfigure}

Scattering from a static source does not change the energy component of the electron momentum. Hence the initial $p^0$ of the electron remains unchanged throughout the scattering, as indicated in \fig{retfig}. When $p^0 > 0$ the negative energy pole of the electron propagator \eq{sr} is not reached ($p^0+E_p > 0$), so its $\ieps$ prescription is irrelevant\footnote{The external potential \eq{coulpotZ} shifts the poles of the electron propagator. The lowest energy eigenvalue $E_0$ of the Dirac equation \cite{406190}
\beq\label{direigen}
E_0=m\,\frac{(1-Z^2\alpha^2)^{1/4}}{(Z^2\alpha^2+\sqrt{1-Z^2\alpha^2}\,)^{1/2}}
\eeq
reaches $E_0=0$ for $Z\alpha= 1$, and is complex at larger couplings. The gap between the positive and negative energy poles of the electron propagator vanishes when $E_0=0$, making the $\ieps$ prescription relevant. With Feynman prescription the pinch between the two poles gives the propagator an imaginary part. This is seen only in the full, resummed propagator, not in the single diagrams of \fig{retfig}.}. Consequently each diagram and their sum $G(p^0,\qv)$ are the same for Feynman and retarded propagators. In particular, the positions and residues of the bound state poles are prescription independent,
\beq\label{dirpole}
G_{\alpha\beta}(p^0,\qv) = \frac{\Psi_\alpha^\dag(\qv)\Psi_\beta(\qv)}{p^0-E} + \ldots
\eeq
Retarded and Feynman propagators do not give the same result for loop corrections on the electron or photon lines, \eg, for diagrams like \fig{feyndiags}(d), since the loop integral probes both positive and negative energy poles. In fact, electron loops vanish with the retarded propagator \eq{sr}, since the electron cannot move backward in time to its starting point. The Dirac bound states are not affected by this since they do not include loop effects.

In an operator formulation the retarded propagator 
\beq\label{sr2}
S_{R}(x-y)= {_{R}\bra{0}}\,T[\psi(x)\bar\psi(y)]\,\ket{0}_{R}
\eeq
is obtained with a ``retarded vacuum'' $\ket{0}_{R}$. The requirement that $S_R(x^0<0)=0$ implies
\beq\label{retcond}
\psi(x)\ket{0}_{R}=0
\eeq
which is formally realized by defining
\beq\label{retvac}
\ket{0}_{R} \propto\prod_{\pv,\lambda} d_{\pv,\lambda}^\dag \ket{0}
\eeq

The Dirac state \eq{dirstate} may be understood to be built on the retarded vacuum, \ie,
\beq\label{dirstateR}
\ket{E,t} = \int d^3\xv\, \psi^\dag(t,\xv)\Psi(\xv)\ket{0}_{R}
\eeq
The condition \eq{retcond} then implies $H\ket{0}_R=0$, justifying \eq{hamdirstate}.

Positive and negative energy states are created by the electron creation ($b^\dag$) and the positron annihilation ($d$) operators, respectively, in the retarded vacuum \eq{retvac}. Hence 
\beq
\int d^3\xv\,\psi^\dag\psi = \int \frac{d^3\pv}{(2\pi)^3}\,(b^\dag_\pv b_\pv+d_\pv d^\dag_\pv)
\eeq
is the number (rather than charge) operator. The expectation value of $\psi^\dag\psi$ in the Dirac state \eq{dirstateR} is
\beq\label{dirdensity}
\bra{E,t}\psi_\alpha^\dag(t,\xv)\psi_\alpha(t,\xv)\ket{E,t} = \Psi^\dag(\xv)\Psi(\xv)\ {_{R}\bra{0}}0\rangle_{R}
\eeq
Thus $|\Psi(\xv)|^2$ may be interpreted as the density of positive and negative energy electrons.

\subsection{Properties of the Dirac wave functions in $D=1+1$\label{dirsol}}

It is instructive to study the properties of the Dirac wave functions in $D=1+1$ dimensions \cite{Dietrich:2012un}. In $A^1=0$ gauge the QED$_2$ potential is linear:
\beq
-\partial_x^2 A^0(x) = -eZ\delta(x)\ \ \Longrightarrow\ \ V(x)\equiv eA^0(x) = \halft e^2Z|x|
\eeq
The charge $e$ has the dimension of mass in $D=1+1$, so the relevant dimensionless parameter is $m/e$, with $m$ the electron mass\footnote{The dimension of $e$ is readily deduced from the requirement that the QED$_2$ action $S=\int d^2x \mathcal{L}$ be dimensionless.}. It is convenient to use units where $e^2Z=1$ so that
\beq\label{qedtwopot}
V(x)=\halft |x|
\eeq
In $D=3+1$ the potential \eq{coulpotZ} was negative, which for $\alpha Z>1$ led to complex energy eigenvalues, as seen in \eq{direigen}. The positive potential \eq{qedtwopot} ensures that the energy eigenvalues are positive and real for all values of $m/e$.

The Dirac matrices may be represented in terms of the $2\times 2$ Pauli matrices,
\beq\label{dirrep}
\gz=\sigma_3 \hspace{1cm} \gamma^1=i\sigma_2 \hspace{1cm} \gz\gamma^1=\sigma_1
\eeq
The Dirac equation \eq{direq} is then, with the energy eigenvalue denoted $M$,
\beq\label{direq2}
(-i\sigma_1\partial_x+m\sigma_3)
\left[\begin{array}{c}\vphi(x) \\ \chi(x)\end{array}\right] = (M-V)
\left[\begin{array}{c}\vphi(x) \\ \chi(x)\end{array}\right]
\eeq
The analytic solutions are known since long \cite{sauter}. Since $V(x)=V(-x)$ we may consider solutions with definite parity,
\beq\label{paritycond}
\vphi(x) = \eta\vphi(-x) \hspace{1cm} \chi(x) = -\eta\chi(-x) \hspace{1cm} (\eta=\pm 1)
\eeq
It is then sufficient to consider solutions for $x\ge 0$ only, with a continuity requirement at $x=0$,
\beqa\label{contcond}
\partial_x\vphi(0)&=&\chi(0)=0 \hspace{1cm} (\eta=+1)\nn\\
\partial_x\chi(0)&=&\vphi(0)=0 \hspace{1cm} (\eta=-1)
\eeqa

Eliminating $\chi(x)$ in \eq{direq2} gives
\beq\label{phieq}
\partial_x^2\vphi(x)+\frac{1}{2(M-V+m)}\partial_x\vphi(x)+\big[(M-V)^2-m^2\big]\vphi(x)=0,
\eeq
For large $x$ this takes the asymptotic form
\beq\label{asphieq}
\partial_x^2\vphi+\quart x^2\vphi=0\hspace{1cm} (x\to \infty)
\eeq
which implies an oscillating (rather than exponentially suppressed) behavior,
\beq\label{asphi}
\vphi(x\to\infty) \sim \exp(\pm i x^2/4).
\eeq 
The component $\chi(x)$ has a similar behavior, as may be seen from \eq{direq2}. The fact that $|\vphi(x\to\infty)|$ is indeed a constant is verified in the exact solution below. Consequently the normalization integral $\int dx \big(|\vphi(x)|^2+|\chi(x)|^2\big)$  diverges linearly at large $x$. According to the interpretation \eq{dirdensity} of $\Psi^\dag\Psi(x)$ this implies a constant density of virtual $e^+e^-$ pairs at large $x$, which is consistent with the linearly rising potential energy.

The wave function is potentially singular at $M-V(x)+m=0$, where the coefficient of $\partial_x\vphi(x)$ in \eq{phieq} diverges. Assuming $\vphi(x) \sim (M-V+m)^\beta$ gives $\beta=0$ or $2$. Thus $\vphi(x)$ is regular at this point, and in fact for all finite $x$. 

We may choose the phases such that $\vphi(x)$ is real and $\chi(x)$ is imaginary. The solution of the second-order differential equation for $\vphi$ then has two real parameters. {\it E.g.}, for $\eta=+1$ in \eq{paritycond} one parameter would determine the overall normalization through the value of $\vphi(0)$ and the other be adjusted to ensure $\partial_x\vphi(0)=0$. In the case of the non-relativistic Schr\"odinger equation the integral of the norm of the wave function provides a third condition. In the absence of this condition, due to the divergence of the normalization integral, the Dirac mass spectrum $M$ is {\it continuous}.

It was actually realized already in the 1930's \cite{plesset} that the Dirac wave function cannot be normalized and that the mass spectrum is continuous for {\it any polynomial potential}. The sole exception is the $V = -\alpha/r$ potential in $D=3+1$. Textbooks often discuss this solution, but rarely mention the general case.

The analytic solution of the Dirac equation is conveniently expressed by replacing $x$ with the variable
\beq\label{tdirac}
\dsi = (M-V)^2, \hspace{2cm} \partial_x = \frac{d\dsi}{dx}\partial_\dsi= -(M-V)\partial_\dsi \hspace{2cm} (x>0)
\eeq
For $M-V(x)>0$ the Dirac equation \eq{direq2} becomes
\beqa\label{diract}
i\partial_\dsi \chi(\dsi) &=& \Big(1-\frac{m}{\sqrt{\dsi}}\Big)\vphi(\dsi)\nn\\[2mm]
i \partial_\dsi \vphi(\dsi) &=& \Big(1+\frac{m}{\sqrt{\dsi}}\Big)\chi(\dsi)
\eeqa
Combining the real $\vphi(x)$ and imaginary $\chi(x)$ components of the wave function into the single complex function
\beq\label{phidef}
\psi(\dsi) \equiv \vphi(\dsi)+\chi(\dsi)  \hspace{.5cm} {\rm and\ conversely
} \hspace{.5cm} \vphi(\dsi) = {\rm Re}\big[\psi(\dsi)\big], \hspace{1cm} \chi(\dsi) = i\,{\rm Im}\big[\psi(\dsi)\big].
\eeq
the general solution of the Dirac equation \eq{diract} is \cite{Dietrich:2012un,sauter}
\beq\label{phisol}
\psi(\dsi) \equiv \vphi(\dsi)+\chi(\dsi) = \left[(a+ib){_1}F_1\Big(-\frac{im^2}{2},\inv{2},2i\dsi\Big)+(b+ia)2m\,(M-V)\,{_1}F_1\Big(\frac{1-im^2}{2},\frac{3}{2},2i\dsi\Big)\right]\exp(-i\dsi)
\eeq
where ${_1}F_1$ is the confluent hypergeometric function and the real constants $a,\,b$ are determined by the continuity conditions \eq{contcond} and the value of $\psi$ at $x=0$. The solution \eq{phisol} is valid for all values of $M$.

The behavior of the wave function at large $x$ is consistent with \eq{asphi},
\beqa\label{psilimit}
\lim_{x\to\infty}\psi(\dsi) &=& \sqrt{\pi}e^{\pi m^2/4}\left[\frac{a+ib}{\Gamma[\halft(1+im^2)]}+
\frac{a-ib}{\Gamma(1+\halft im^2)}\frac{me^{-i\pi/4}}{\sqrt{2}}\right](2\dsi)^{im^2/2}e^{-i\dsi} \nn\\[2mm]
&+& \frac{\sqrt{\pi}}{\sqrt{2\dsi}}e^{\pi (m^2-i)/4}\left[\frac{a+ib}{\Gamma(-\halft im^2)}-\frac{a-ib}{\Gamma[\halft(1-im^2)]}\frac{me^{i\pi/4}}{\sqrt{2}}\right](2\dsi)^{-im^2/2}e^{i\dsi}+\morder{\inv{\dsi}},
\eeqa

In the non-relativistic limit $m \gg e$ the coordinate $x$ and binding energy $E_b=M-m$ scale with increasing mass as
\beq\label{nrscaling}
 x \sim E_b \sim m^{-1/3}
\eeq
As shown in Appendix \ref{NRlimit} the two independent solutions of the wave function \eq{phisol} then reduce (if $a+b \neq 0$) to the same, normalizable Airy function\footnote{Due to a typo the corresponding expression (2.21) in \cite{Dietrich:2012un} has an incorrect factor $\exp(im^2)$.},
\beq\label{dirnrlimit}
\psi(\dsi) = (1+i)(a+b)\sqrt{\pi}m^{1/3}e^{\pi m^2/2-i\pi/4}\mathrm{Ai}\big[m^{1/3}(|x|-2E_b)\big]\Big[1+\morder{m^{-2/3}}\Big] 
\eeq
This non-relativistic limit of $\psi(\dsi)$ agrees with the solution $\rho(x)$ of the Schr\"odinger equation,
\beqa\label{seq}
 -\inv{2\moe}\partial_x^2 \rho(x)&+&\frac{1}{2} |x|\, \rho(x) = E_b\,\rho(x) \nn\\[2mm]
\rho(x)  &=& N\,\mathrm{Ai}[\moe^{1/3}(x-2E_b)] \qquad (x>0) 
\eeqa

The single parameter $a+b$ in \eq{dirnrlimit} is determined by the normalization condition $\int dx |\psi|^2 = 1$. The continuity condition at $x=0$ then allows only discrete values of $M$.
The compatibility of the continuous energy spectrum of the Dirac equation with the discrete spectrum of the Schr\"odinger equation is resolved in an interesting fashion: In the non-relativistic limit solutions with a continuous range of $M$ are found only for parameter values $a = -b$. The approach to the Schr\"odinger solution \eq{dirnrlimit} is quite fast. {\it E.g.,} for $m/e = 2.5$ solutions with a continuous range of $M$ are found for $a/b = -1.041 \pm 10^{-6}$.

%
\begin{figure}[t]
\includegraphics[width=10cm]{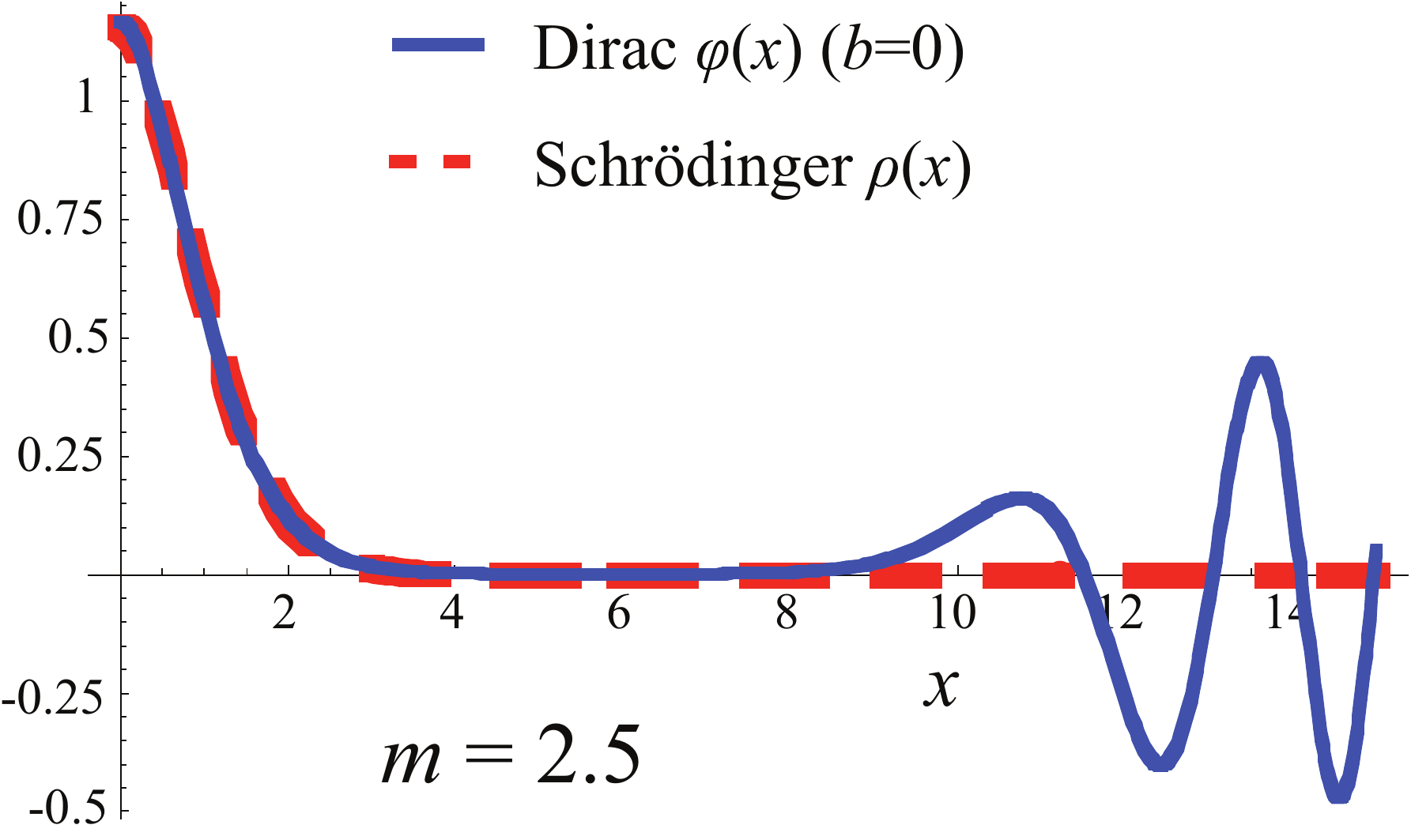}
\caption{\label{Diracfig}The upper component $\vphi(x)$ of the Dirac wave function in \eq{direq2} (continuous blue line) compared to the solution $\rho(x)$ of the Schr\"odinger equation \eq{seq} (dashed red line). The potential is \eq{qedtwopot} and the fermion mass is $m=2.5\,e$. The parameter $b=0$ in the expression \eq{phisol} of the Dirac wave function. Both solutions are normalized to unity in the range $0 \leq x\,e \leq 6$. The potential reaches $V(x)=2m$ at $x\,e=10$.}\label{Diracwf}
\end{figure}

In \fig{Diracwf} the upper component $\vphi(x)$ of the Dirac wave function (evaluated with $b=0$ in \eq{phisol}) is compared to the Schr\"odinger wave function \eq{seq} for $m/e = 2.5$. The two solutions are seen to be very similar in the non-relativistic range $V(x) \lsim m$. Starting from $x \simeq 10/e$ (where $V(x) = 2m$) the $e^+e^-$ pair fluctuations in the field of the external charge (\cf\ \fig{pairfig}) manifest themselves as a resurgence and oscillations of the Dirac wave function.

The Dirac wave functions corresponding to different eigenvalues $M$ are orthogonal \cite{Dietrich:2012un}. As for plane waves, wave functions with the full (continuous) range of $M$ form a complete set of functions \cite{titchmarsh}. 

Since the Dirac equation with a confining potential has a continuous mass spectrum it is not obviously useful for hadron phenomenology. The external potential furthermore breaks translation invariance, so the bound states do not have well-defined momenta. In the next section we shall see how both features are improved when two particles are bound by their respective Coulomb fields.

\section{Summary (Part 1)\label{disc1}}

QED bound states (atoms) are perturbative in the fine structure constant $\alpha \simeq 1/137$. Precision calculations of binding energies have been successfully compared with data, as illustrated in \eq{hfsfin} and \fig{hypfig}. Bound states can be identified as poles in scattering amplitudes. The poles are not present in any single Feynman diagram, but are created by the divergence of the perturbative expansion. The expansion can diverge, however small is the coupling, because bound state momenta scale with $\alpha$, \eg, the Bohr momentum $p_B = \alpha m$. Higher powers of $\alpha$ from the vertices are thus compensated by propagator denominators. The leading diagrams may be identified as iterations of single photon exchange (ladder) diagrams, which are all of \order{\alpha^{-1}}. Since Feynman diagrams are Lorentz invariant the ladder diagrams give the leading contribution in any frame.

QED perturbation theory expands around non-interacting $in$ and $out$ states, in which electrons are unaccompanied by any electromagnetic field. Such states violate the QED equations of motion and are in this sense unphysical. Consequently the perturbative expansion does not converge for physical processes which are sensitive to soft photons, such as bound states. The sum of ladder diagrams restores the classical electromagnetic fields of the charges. This brings bound state poles to scattering amplitudes, with residues that satisfy the Bethe-Salpeter equation (BSE) \eq{BSeq} with a single photon exchange kernel. In the rest frame the BSE reduces to the Schr\"odinger equation \eq{Schrcoo}.

In a frame where the atom is moving with relativistic velocity the wave function $\Psi^\Pv(q)$ of the BSE \eq{BSeq} depends on $q^0$, or equivalently on the relative time between the constituents. This is because transverse photon exchange contributes to the kernel, as illustrated in \eq{CTamp} and \fig{fig2to2}. While Coulomb ($A^0$) photons are instantaneous (in Coulomb gauge), transverse photons propagate with the (finite) speed of light. Positronium in flight therefore has both $\ket{e^+e^-}$ and $\ket{e^+e^-\gamma}$ Fock components. The BSE may be reduced to a relation involving only the equal-time $\ket{e^+e^-}$ component through time-ordering. The resulting equation reduces to a Schr\"odinger equation \eq{schrmov} where the longitudinal distances are Lorentz contracted. The wave function of the $\ket{e^+e^-\gamma}$ component, on the other hand, does not transform simply under boosts (\fig{fig8}).

The field theory condition for time stationarity of a bound state $\ket{\Psi}$ is $\hat H\ket{\Psi}= E\ket{\Psi}$. The QED (operator) Hamiltonian $\hat H$ is determined by the Lagrangian, and involves the (operator) gauge field ${\hat A}^\mu$. According to the perturbative analysis of Feynman diagrams the leading order interaction is mediated by single photon exchange, which is equivalent to the classical field. Thus we may replace ${\hat A}^\mu$ by the classical potential ${A}^\mu$ found from Maxwell's equations. In the rest frame only the $\ket{e^+e^-}$ component of Positronium contributes at leading order, so the state may be parametrised in terms of a $4\times 4$ wave function $\Phi_{\alpha\beta}$ as in \eq{posstate}. In Section \ref{hameigen2} we verified that the leading components of $\Phi$ satisfy the Schr\"odinger equation in the non-relativistic limit ($\alpha\to 0$). 

\vspace{4mm}

The Dirac equation describes the bound states of an electron in an external potential. The external field breaks translation invariance, so Dirac states are not Poincar\'e covariant (unless the external field is transformed as well). The Coulomb Dirac equation may be obtained by summing Feynman diagrams of leading power in the charge $Ze$ of the particle which acts as the source of the external field, in the limit where the mass of that particle tends to infinity. All photon exchanges between the (relativistic) electron and the heavy (source) particle must be taken into account, including all possible crossed exchanges. In a time-ordered framework crossed (instantaneous) Coulomb photon exchanges imply intermediate states with $e^+e^-$ pairs. Thus one explicitly sees that the Dirac wave function describes a multi-particle state, as is well-known from the Klein paradox.

The Dirac spinor wave function has both positive energy ($u$) and negative energy ($v$) components. The negative energy components are related to Fock states with $e^+e^-$ pairs -- the positron may be viewed as a negative energy electron. Whereas each Fock state wave function depends on the positions of all its constituents (\cf\ \eq{Fockexp}), the Dirac wave function appears to describe only a single electron. The absence of the degrees of freedom corresponding to the pair constituents makes the Dirac spectrum an interesting analog of the hadron spectrum. The experimentally determined spectra of hadrons reflect only their valence quark d.o.f's ($q\bar q$ and $qqq$), in spite of their sea quark (and gluon) constituents.

The Born-level Feynman diagrams describing electron scattering in a static potential are independent of the $\ieps$ prescription at the negative energy pole of the electron propagator. Thus the Dirac spectrum is equally obtained using retarded propagators, with both positive and negative energy electrons moving only forward in time. With the retarded prescription there is only a single electron at any intermediate time, and the Dirac wave function describes the distribution of that electron. Whereas in the standard vacuum the operator $d^\dag$ creates a positive energy positron, $d$ creates a negative energy electron in the retarded prescription. The expectation value of $\psi^\dag(t,\xv)\psi(t,\xv)$ in a Dirac state gives the density of positive and negative energy electrons at $\xv$. 

In Section \ref{hamdirac} we saw how to define a Dirac eigenstate \eq{dirstate} of the QED Hamiltonian, with ${\hat A}^\mu$ replaced by the external potential. The Dirac equation is obtained for the wave function $\Psi(\xv)$ provided the state is built on a vacuum which itself is an eigenstate of the Hamiltonian. This is the case for the vacuum \eq{retvac} which gives retarded electron propagators.

Surprisingly, the energy spectrum of the Dirac equation is {\it continuous} for potentials $V(r)$ which are polynomial in $r$ or $1/r$. The norm of the wave function tends to a constant at large $r$, consequently the integral of the norm diverges. This was established already in the early 1930's \cite{plesset}. Textbooks usually mention only the sole (albeit important) exception $V(r) \propto 1/r$, appropriate for a point source in $D=3+1$ dimensions. In Section \ref{dirsol} we studied in some detail the solutions of the Dirac equation in $D=1+1$, with the linear potential $V(x) \propto |x|$ of QED$_2$. It is apparent (see \fig{Diracwf}) that the constant norm reflects the density of $e^+e^-$ pairs created by the strong potential at large $|x|$, as indicated by the expectation value of $\psi^\dag\psi$ in \eq{dirdensity}. An analogous conclusion was reached in \cite{Giachetti:2007vq}, for the normalizable solutions having complex eigenvalues $E$: Im$\,E$ agrees with the rate of pair production in the potential.

\vspace{1cm}

\centerline{\Large PART 2: Research level}


\section{Relativistic $f\bar f$ bound state in QED$_2$\label{ffstatesec}}

The features of the non-relativistic positronium and relativistic (Dirac) electron that we discussed in the previous sections are mostly well-known. We saw how the standard results can be obtained using a Hamiltonian method, by defining the states as in \eq{posstate} and \eq{dirstate}. For non-relativistic atoms pair production in the vacuum is suppressed. Also in the Dirac case the vacuum had to be an eigenstate of the Hamiltonian.

We now apply these methods to a relativistic $f\bar f$ system, bound by the electromagnetic field of the fermions themselves. This takes us beyond textbook topics. In this section we study QED in $D=1+1$ dimensions, also known as the ``massive Schwinger model'' \cite{Coleman:1976uz}. We consider states at ``Born'' level, bound by a classical (linear) Coulomb potential without loop corrections. The bound states, defined at equal time of the constituents, turn out to have a hidden, exact Poincar\'e invariance. This allows studies of the frame dependence of the wave functions and of scattering dynamics. In section \ref{3dpotential} we apply this method to QED and QCD in $D=3+1$ dimensions. The approach summarized here is described in more detail in \cite{Hoyer:2009ep,Dietrich:2012un,Dietrich:2012iy}.

\subsection{Bound state equation for QED$_2$\label{qed2bse}}

We consider $f\bar f$ states in $D=1+1$, defined at equal time of the constituents in all frames. For simplicity we take the fermions to have equal mass\footnote{See \cite{Dietrich:2012un} for the more general case of unequal masses.} $m$. A bound state of energy $E$ and CM-momentum $P$ is defined (at $t=0$) in analogy to the positronium state of \eq{posstate} as
\beq \label{statedef}
 \ket{E,P;t=0}\equiv \int dx_1 dx_2\, \exp\big[\halft iP(x_1+x_2)\big]\bar\psi(t=0,x_1)\Phi(x_1-x_2)\psi(t=0,x_2)\ket{0}
\eeq
Now $\hat P \ket{E,P}=P \ket{E,P}$ since the state picks up a phase $\exp(iPa)$ in the coordinate transformation $x_i \to x_i+a$.

The QED Hamiltonian \eq{ham2} in $D=1+1$ and $A^1=0$ gauge is
\beq\label{qed2ham}
H(t) = \int dy\,\psi^\dag(t,y)\Big[-i\gf\rder_y+\halft eA^0(y)+m\gamma^0\Big]\psi(t,y)
\eeq
where $\rder_y$ indicates $\partial/\partial y$ operating to the right. The Hamiltonian generates time translations,
\beqa\label{hcom}
\partial_t\bar\psi(t,x_1) &=& i\com{H}{\bar\psi(t,x_1)} = i\bar\psi(t,x_1)\Big[-i\gf\lder_1+\halft eA^0(x_1)+m\gamma^0\Big]\nn\\
&& \hspace{10cm} (\partial_j \equiv \partial/\partial x_j)\\
\partial_t\psi(t,x_2) &=& i\com{H}{\psi(t,x_2)} = i\Big[i\gf\rder_2-\halft eA^0(x_2)-m\gamma^0\Big]\psi(t,x_2)\nn
\eeqa

As in the Dirac case \eq{zvac} we do not consider pair production in the vacuum: $H\ket{0}=0$. The states thus obtained may be regarded as asymptotic states at $t=\pm\infty$, analogous to the usual $in$ and $out$ states (\cf\ the form factor expression \eq{formfac}). Corrections due to string breaking and higher orders in the coupling are generated in the time development from the asymptotic to finite times (\cf\ Section \ref{disc2}). 

The bound state condition 
\beq\label{eigen1}
H\ket{E,P}=\int dx_1 dx_2\,e^{i (x_1+x_2)P/2}\Big\{\com{H}{\bar\psi(0,x_1)}\Phi(x)\psi(0,x_2)+\bar\psi(0,x_1)\Phi(x)\com{H}{\psi(0,x_2)}\Big\}\ket{0} =E\ket{E,P}
\eeq
requires that the wave function $\Phi(x_1-x_2)$ in \eq{statedef} satisfy
\beq\label{eom1}
i\partial_x\acom{\gf}{\Phi(x)}-\halft P\com{\gf}{\Phi(x)}+m\com{\gamma^0}{\Phi(x)}=(E-V)\Phi(x)
\eeq
where $V(x_1-x_2) = \halft [eA^0(x_1)-eA^0(x_2)]$. The component $\bar\psi(t,x_1)\psi(t,x_2)\ket{0}$ is an eigenstate of the field operator $\hat A^0(y)$ determined by Gauss' law $-\partial_y^2 \hat A^0(y)=e\psi^\dag\psi(y)$. The eigenvalue equals the classical field (\cf\ \eq{a0expr}),
\beq\label{aexpr}
A^0(y) = -\frac{e}{2}|y-x_1|+\frac{e}{2}|y-x_2|
\eeq
so that
\beq\label{qed2pot}
V(x_1-x_2) = \frac{e^2}{2}|x_1-x_2| \equiv \frac{1}{2}|x_1-x_2| 
\eeq
Since the charge $e$ has the dimension of mass in $D=1+1$ we may measure all energies and masses in units of $e$, in effect setting\footnote{The linear potential \eq{qed2pot} should, strictly speaking, be regarded as analogous to \eq{linpot}, arising from a non-vanishing boundary condition (of scale $\Lambda$) on Gauss' law. Then small $m/\Lambda$ does not imply a strong coupling $e$, and perturbative corrections remain under control.} $e=1$.

We may expand the $2\times 2$ wave function $\Phi$ in the basis formed by the Pauli matrices \eq{dirrep},
\beq\label{Phiexpr}
\Phi=\phi_0+\gf\phi_1+\gz\phi_2+\go\phi_3
\eeq
Since $\acom{\gf}{\gz}=\acom{\gf}{\go}=0$ only $\phi_0$ and $\phi_1$ are differentiated in \eq{eom1}. Expressing $\phi_2$ and $\phi_3$ in terms of $\phi_0$ and $\phi_1$ gives
\beq\label{wf1}
\Phi=\phi_0+\gf\phi_1-\frac{2m}{\sigma}\phi_1\gf\slashed{\Pi}^\dag
\eeq
where $\Pi$ is the ``kinetic'' 2-momentum and $\sigma$ is its square,
\beq\label{sigdef}
\Pisl(x)=(E-V)\gamma^0-P\gamma^1 \hspace{2cm}
\sigma(x) = \Pi^2 = (E-V)^2-P^2
\eeq
Note that the function $\sigma=\sigma(x)$ is frame dependent and in the rest frame ($P=0$) reduces to the variable \eq{tdirac} that we used in the solution of the Dirac equation.

Inserting the expression \eq{wf1} into the BSE \eq{eom1} gives the condition on $\phi_0$ and $\phi_1$,
\beq\label{eom2}
2i\partial_x(\phi_0\gf+\phi_1)=(E-V)\Big[\phi_0+\Big(1-\frac{4m^2}{\sigma}\Big)\phi_1\gf\Big]
\eeq
For the linear potential \eq{qed2pot} we have (when $x>0$) $\partial_x = -(E-V)\partial_\sigma$. Canceling the common factor $E-V$ all dependence on $E$ and $P$ in \eq{eom2} appears only through the variable $\sigma$,
\beq\label{eom3}
-2i\partial_\sigma\phi_1=\phi_0 \hspace{2cm}-2i\partial_\sigma\phi_0=\Big(1-\frac{4m^2}{\sigma}\Big)\phi_1
\eeq
The solution of these coupled equations specifies the wave function in an arbitrary frame, in the sense that $\phi_0$ and $\phi_1$ are the same functions of $\sigma$ in all frames. The frame dependence in terms of $x$ is seen from
\beq
dx = -\frac{d\sigma}{E-V(x)}\hspace{1cm} (x>0)
\eeq
Since the wave function is invariant in terms of $\sigma$ it will Lorentz contract in $x$. However, the contraction factor $1/(E-V)$ becomes the classical $1/E$ only when $V \ll E$. Moreover, when $E < V$ the Lorentz contraction turns into an expansion\footnote{In this region the string-breaking corrections to the wave function will be important, however.}!

The $2 \times 2$ wave function $\Phi$ has also an explicit frame dependence through the factor $\Pisl^\dag$ in \eq{wf1}. In terms of the ($x$-dependent) boost parameter $\zeta$ defined by
\beq\label{zetadef}
\cosh\zeta=\frac{E-V}{\sqrt{\sigma}} \hspace{2cm} \sinh\zeta=\frac{P}{\sqrt{\sigma}}
\eeq
the kinetic momentum $\Pisl$ in \eq{sigdef} is 
\beq\label{piframe}
\Pisl = \exp(\halft\gf\zeta) \sqrt{\sigma}\,\gz \exp(-\halft\gf\zeta)
\eeq
Since $\phi_0$ and $\phi_1$ are frame independent functions of $\sigma$ the wave function $\Phi^{(P)}$ of \eq{wf1}, in the frame with CM momentum $P$, may be expressed in terms of the rest frame wave function $\Phi^{(P=0)}$ as
\beq\label{wfframe}
\Phi^{(P)}(\sigma) = \exp(-\halft\gf\zeta) \Phi^{(P=0)}(\sigma) \exp(\halft\gf\zeta) 
\eeq
The transformation \eq{wfframe} is similar to the standard one for a boost along the ``1''-axis, except that $E-V$ appears instead of $E$ in the expression \eq{zetadef} of the boost parameter.

The covariant frame dependence of the wave function \eq{wfframe} is hidden in the original form \eq{eom1} of the bound state equation. However, it allows us to write an equivalent covariant equation for $\Phi^{(P)}$ \cite{Hoyer:1986ei},
\beq\label{eom4}
-i\partial_\sigma \big\{\gf,\Phi^{(P)}(\sigma)\big\}+\frac{m}{\sigma}\big[\Pisl^\dag,\Phi^{(P)}(\sigma)\big]=\Phi^{(P)}(\sigma)
\eeq
The equivalence of this equation with \eq{eom1} is readily seen for $P=0$ (where $\zeta=0$) and the solution of \eq{eom4} for any $P$ is given by \eq{wfframe} as required. 

Poincar\'e covariance is a necessary requirement for bound state dynamics, and is non-trivial for extended states in quantum field theory. The exact frame dependence \eq{wfframe} is the only case known for relativistic, equal-time wave functions. In Appendix \ref{sqedapp} we show that $\sigma$ appears as an ``invariant distance'' also for the bound states of scalar QED$_2$.

The above analysis is compatible with, but did not require the standard relation between energy and momentum,
\beq\label{eprel}
E^2=P^2+M^2
\eeq
where $M$ is the rest mass. We shall next verify that the frame dependence \eq{wfframe} of the wave function agrees with that obtained by boosting the bound state, which establishes \eq{eprel}.

\subsection{Bound state boost\label{boost}}

A boost with an infinitesimal parameter $\xi$ transforms the fermion coordinates in the state \eq{statedef} as
\beq\label{boostdef}
(t_1=0,x_1) \to (\xi x_1,x_1) \hspace{2cm} (t_2=0,x_2) \to (\xi x_2,x_2)
\eeq 
The corresponding boost operator transforms the fermion fields as\footnote{In \cite{Dietrich:2012iy} we included a gauge transformation with the boost generator, to ensure that $A^1=0$ after the boost. This was necessary for the Poincar\'e Lie algebra to close. The gauge parameter satisfied $\partial_x\theta(x)=eA^0(x)$, which with $A^0$ given by \eq{aexpr} implied $\theta(x)=-\quart e^2\big[(x-x_1)|x-x_1|-(x-x_2)|x-x_2|\big]$. Since $\theta(x_1)-\theta(x_2)=0$ this gauge transformation does not affect the state \eq{statedef}.}
\beqa\label{fieldboost}
U(\xi)\bar\psi(0,x_1)U^\dag(\xi)&=& \bar\psi(\xi x_1,x_1)S(\xi)= \bar\psi(0,x_1)(1+\halft\xi \gf)+i\xi x_1\com{H}{\bar\psi(0,x_1)} \nn\\[2mm]
U(\xi)\psi(0,x_2)U^\dag(\xi)&=&(1-\halft \xi\gf)\psi(0,x_2)+i\xi x_2\com{H}{\psi(0,x_2)}
\eeqa
With $U(\xi)\ket{0}=\ket{0}$ the boost operator acts on the state \eq{statedef} as
\beqa\label{boost1}
U(\xi)\ket{E,P}&=&\int dx_1 dx_2\,e^{i (x_1+x_2)P/2}\Big\{
\bar\psi(0,x_1)\Big(\Phi^{(P)}(x)+\halft \xi\big[\gf,\Phi^{(P)}(x)\big]\Big)\psi(0,x_2) \nn\\[2mm]
&+& i\xi x_1\com{H}{\bar\psi(0,x_1)}\Phi^{(P)}(x)\psi(0,x_2)
+   i\xi x_2\bar\psi(0,x_1)\Phi^{(P)}(x)\com{H}{\psi(0,x_2)}\Big\}\ket{0}
\eeqa
where $x=x_1-x_2$.
Noting that $x_{1,2}=\halft(x_1+x_2)\pm\halft x$ the coefficient of $i\xi\halft(x_1+x_2)$ has the same form as for the Hamiltonian operator in \eq{eigen1} and thus gives a factor $E$ (after partial integrations). This term shifts the momentum of the plane wave to $\Pxi$,
\beq\label{momshift2}
e^{i (x_1+x_2)P/2}\big[1+i\halft\xi(x_1+x_2)E\big]=e^{i (x_1+x_2)(\Pxi)/2}
\eeq
The coefficient of $i\xi\halft x$ in \eq{boost1} involves the difference between the commutators of the Hamiltonian with $\bar\psi$ and $\psi$. In this combination the potential energy cancels, $eA^0(x_1)+eA^0(x_2)=0$ according to \eq{aexpr}. In the partial integration which shifts the derivatives from the fields onto the wave function the differentiation of the coefficients $\xi x_1$ and $\xi x_2$ changes the sign of the term $\halft \xi\com{\gf}{\Phi(x)}$ in \eq{boost1}. The remaining terms appear with commutators and anticommutators interchanged as compared to the BSE in \eq{eom1},
\beqa\label{boost2}
U(\xi)\ket{E,P}&=&\int dx_1 dx_2\,\bar\psi(0,x_1) e^{i (x_1+x_2)(\Pxi)/2}
\Big\{\Phi^{(P)}(x)-\halft \xi\comb{\gf}{\Phi^{(P)}(x)} \nn\\
&& +\halft i\xi x\Big(i\partial_x\comb{\gf}{\Phi^{(P)}(x)}-\halft P\acomb{\gf}{\Phi^{(P)}(x)}
+m\acomb{\gamma^0}{\Phi^{(P)}(x)}\Big)\Big\}\psi(0,x_2)\ket{0} \nn\\[2mm]
&=& \int dx_1 dx_2\,\bar\psi(0,x_1) e^{i (x_1+x_2)(\Pxi)/2}
\Big\{\Phi^{(P)}(x)-\halft i\xi xP\gf\Big(1-\frac{2m}{\sigma}\Pisl^\dag\Big)\phi_0(\sigma) \nn\\
&& -\xi xP\Big[\halft i \Big(1-\frac{4m^2}{\sigma}\Big)+\frac{2m}{\sigma^2}\gf\Pisl^\dag\Big]\phi_1(\sigma)-2m\frac{\phi_1(\sigma)}{\sigma}\gf\delta\Pisl^\dag \Big\}\psi(0,x_2)\ket{0}
\eeqa
The second equality follows using \eq{wf1} for $\Phi^{(P)}$, \eq{eom3} for the derivatives of $\phi_{0,1}$ and $\partial_x V = \halft$ ($x>0$). 
For convenience we separated the term involving $\delta\Pi$, the frame dependence of $\Pi(x)$:
\beq\label{delpi}
\delta\Pisl^\dag \equiv \Pisl_{\Pxi}^\dag(x)-\Pisl_{P}^\dag(x) = (E+\xi P-V)\gamma^0+(P+\xi E)\gamma^1 -\Pisl_P^\dag(x)=-\xi\gf\Pisl_P^\dag(x)+\xi V\gamma^1
\eeq

We need to verify that $U(\xi)\ket{E,P}$ in \eq{boost2} equals the state \eq{statedef} with boosted energy and momentum,
\beq\label{newframestate}
\ket{E+\xi P,\Pxi}\equiv \int dx_1 dx_2\, e^{i (x_1+x_2)(\Pxi)/2}\bar\psi(0,x_1)\Phi^{(\Pxi)}(x_1-x_2)\psi(0,x_2)\ket{0}
\eeq
The $\xi$-dependence of the wave function $\Phi^{(\Pxi)}(x)$ at fixed $x$ arises from the frame dependence \eq{delpi} of $\Pi(x)$ and that of $\sigma(x)$,
\beq\label{delsig}
\delta\sigma \equiv \sigma_{\Pxi}(x)-\sigma_{P}(x) = (E+\xi P-V)^2-(P+\xi E)^2 - \sigma_{P}(x)=-2\xi PV(x) = -\xi x P
\eeq

Hence from \eq{wf1},
\beq\label{delPhi}
\delta\Phi(x) \equiv \Phi^{(\Pxi)}(x)-\Phi^{(P)}(x) = \delta\sigma\Big[\partial_\sigma(\phi_0+\gf\phi_1)-2m\Big(\partial_\sigma\frac{\phi_1}{\sigma}\Big)\gf\Pisl_P^\dag\Big]-2m\frac{\phi_1}{\sigma}\gf\delta\Pisl^\dag
\eeq  
The last term is explicit in \eq{boost2}. Substituting $\partial_\sigma\phi_{0,1}$ from \eq{eom3} in \eq{delPhi} the expression within curly brackets in \eq{boost2} is found to be
\beq
\big\{\eq{boost2}\big\} = \Phi^{(P)}(x)+\delta\Phi(x)=\Phi^{(\Pxi)}(x)
\eeq
which establishes the (infinitesimal) boost covariance
\beq\label{boostcov}
U(\xi)\ket{E,P} =\ket{E+\xi P,\Pxi}
\eeq

In the above demonstration the linearity of the potential $V(x)=\halft |x|$ was repeatedly used. This requirement is not unexpected, since Gauss' law implies a linear potential in $D=1+1$. The appearance of the kinetic 2-momentum $\Pi$ and its square $\sigma=\Pi^2$ in \eq{eom4} suggests the possibility of an explicitly covariant framework.

\subsection{Bound state properties\label{property}}

\subsubsection{Analytic solution\label{ansol}} 

The coupled differential equations \eq{eom3} for $\phi_0(\sigma)$ and $\phi_1(\sigma)$ can be solved in terms of confluent Hypergeometric functions of the first (${_1}F_1$) and second ($U$) kind. Eliminating $\phi_0$ gives the second order equation
\beq\label{bse4}
4\partial_\dsi^2\phi_1+\left(1-\frac{4m^2}{\dsi}\right)\phi_1=0
\eeq
The general solution is 
\beq\label{Phi1sol}
\phi_1(\dsi)= \dsi\, e^{-i\dsi/2}\big[a\,\kum(1-im^2,2,i\dsi)+b\, U(1-im^2,2,i\dsi)\big]
\eeq
where $a$ and $b$ are constants. The full wave function $\Phi$ in \eq{wf1} is regular at $\sigma=0$ only provided $b=0$. With this constraint the eigenvalues $E$ are discrete.
This is a crucial difference compared to the solutions \eq{phisol} of the Dirac equation, which are regular for all values of the two parameters, implying a continuous spectrum.

According to the differential equations \eq{eom3} we may choose $\phi_1$ to be real and $\phi_0$ to be imaginary. The constant $a$ in \eq{Phi1sol} is then real, as seen from the integral representation of the $\kum$-function,
\begin{equation}\begin{split}\label{Phiexp}
\phi_1(\dsi)&= 
a\, \dsi\, 
e^{-i\dsi/2}\kum(1-im^2,2,i\dsi) = \frac{a\sinh(\pi m^2)}{\pi m^2}\,
\dsi  
\,e^{-i\dsi/2} \int_0^1 du\, e^{i\dsi u}u^{-i\moe^2}(1-u)^{i\moe^2} = \phi_1^*(\dsi)
\\
\phi_0(\dsi)&= 
-\phi_1(\dsi)-2ia  
e^{-i\dsi/2}\kum (1-i\moe^2,1,i\dsi) = -\phi_0^*(\dsi)
\end{split}\end{equation}

%
\begin{wrapfigure}[20]{r}{0.5\textwidth}
  \vspace{-20pt}
  \begin{center}
    \includegraphics[width=0.5\textwidth]{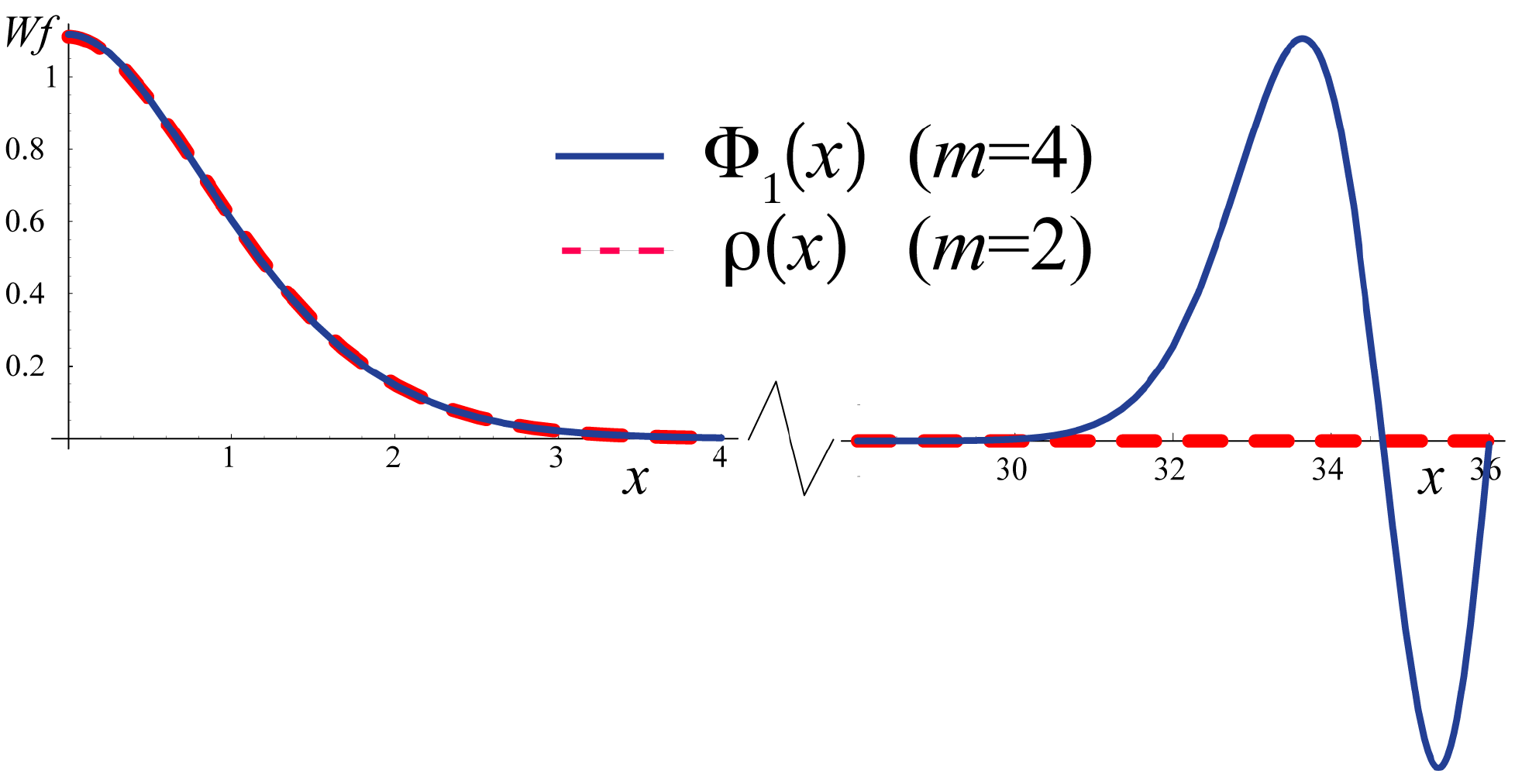}
  \end{center}
  \vspace{-15pt}
  \caption{The ground state $f\bar f$ wave function $\phi_1(x)$ in \eq{Phiexp} for $P=0$ and $m=4$ (solid blue line) compared to the nonrelativistic Schr\"odinger wave function $\rho(x)$ in \eq{seq} with the reduced mass $m=2$ (dashed red line). Both wave functions are normalized to unity in the region $0\leq x \leq 5$. Note that the $x$-axis scale is different in the two domains.}
\label{ffwf}
\end{wrapfigure}

According to \eq{eom2} solutions of definite parity $\eta=\pm 1$ may be defined by,
\beq\label{parity}
\phi_1(-x) = \eta~\phi_1(x), \hspace{1cm} \phi_0(-x) = -\eta~\phi_0(x)
\eeq
If \eq{Phiexp} is taken to be a solution for $x>0$ of parity $\eta$ we must impose the continuity constraints
\beq\label{cont}
 \partial_x\phi_1(x=0)=0\ \ (\eta=+1),  \hspace{.5cm} \phi_1(x=0)=0\ \ (\eta=-1)
\eeq
The discrete eigenvalues $E$ which are compatible with continuity at $x=0$ are then determined by the values $\sigma=\sigma_0$ for which $\phi_1$ or its derivative vanishes. For example, if $\partial_\sigma\phi_1(\sigma_0)=0$ we may consider $\sigma_0$ to correspond to $x=0$ for a solution of positive parity. Since $V(0)=0$ the bound state mass $M$ is frame-independent,
\beq\label{masscond}
\sigma_0(x=0) = E^2-P^2=M^2
\eeq

\subsubsection{Non-relativistic limit} 

In the non-relativistic limit ($m \gg e$) the coordinate $x$ and the binding energy $E_b=M-2m$ scale with $m$ as in \eq{nrscaling}. As shown in Appendix \ref{NRlimit} the confluent hypergeometric functions in \eq{Phi1sol} turn into solutions of the non-relativistic Schr\"odinger equation \eq{seq},
\beqa
\dsi e^{-i\dsi/2} \kum(1-im^2,2,i\dsi) &=& 
\left(\frac{2}{m}\right)^{2/3}e^{\pi m^2}
\, \mathrm{Ai}\left[\left(\halft m\right)^{1/3}
(|x|-2E_b)\right]\left[1+\morder{m^{-4/3}}\right]
\\
\dsi e^{-i\dsi/2}\, U(1-im^2,2,i\dsi) &=&
-(2 m^2)^{2/3} 
\frac{\pi\, e^{-\pi m^2}}{\Gamma(1-im^2)}
\left\{
\mathrm{Ai}\left[(\halft m)^{1/3}(|x|-2E_b)\right] + i\,\mathrm{Bi}\left[(\halft m)^{1/3}(|x|-2E_b)\right]\right\} \nn\\&&\times\left[1+\morder{m^{-4/3}}\right]
\eeqa
The result for the $U$ function involves the non-normalizable Bi Airy function. A comparison between the exact solution \eq{Phiexp} and the Ai Airy function solution of the Schr\"odinger equation is shown in \fig{ffwf} for $m=4e$. At this value of $m/e$ the corresponding binding energies differ by ca. 1.4\%. The wave functions are very similar in the region where $V(x) \ll 2m$, but the relativistic solution oscillates at large $|x|$.

\subsubsection{No parity doublets for $m\to 0$} 

The bound state spectrum can be solved analytically in the limit of small fermion mass, $m \ll e$,
\begin{equation}\label{spectrum}
 M^2_n = e^2\pi n + 2 m^2\big[\log(\pi n)-\mathrm{Ci}(\pi n)+ \gamma_E\big] + {\cal O}\left(m^4\right) \ ;  \qquad  \eta=(-1)^{n+1}
\end{equation}
where $n = 0,1,2,\ldots$, Ci is the cosine integral function and $\gamma_E$ is Euler's constant. The states lie on nearly linear trajectories and have alternating parity. It is interesting to note that there is no parity degeneracy as $m\to 0$, even though chiral symmetry implies parity doubling at $m=0$. The bound state equation \eq{eom1} is manifestly chirally symmetric for $m=0$: If $\Phi$ is a solution then so is $\gf\Phi$. According to \eq{wf1} these two solutions differ by $\phi_0 \leftrightarrow \phi_1$, which have opposite parity according to \eq{parity}.

The reason that the spectrum breaks chiral symmetry for any $m \neq 0$ may be traced to the form of the bound state equation \eq{bse4}. The singularity at $\sigma=0$ which requires a discrete spectrum is absent when $m=0$. Thus the spectrum is continuous (and in particular parity doubled) only when $m=0$ exactly. 

\subsubsection{Duality\label{dualsec}} 

In the rest frame the variable $\sigma=(M-V)^2$ is mirror symmetric around $V(x)=M$ for $0 \leq V(x) \leq 2M$, which explains the symmetry of the wave function in \fig{ffwf}. For $V(x) > 2M$ the wave function begins to oscillate, similarly to the Dirac wave function in \eq{psilimit}. For large $|\sigma|$ the solutions \eq{Phiexp} behave as
\begin{equation}\begin{split}\label{aschi}
\phi_1(\dsi \to \pm\infty) &\simeq
\sqrt{\frac{2}{\pi}}\,\frac{a}{m}
\sqrt{e^{2\pi m^2}-1}\, 
e^{-\pi\moe^2\theta(-\dsi)} \sin\Big[{\frac{\dsi}{2}}-\moe^2\log(|\dsi|)+\arg\Gamma(1+i\moe^2)\Big][1+\morder{\dsi^{-1}}], 
\\[1mm]
\phi_0(\dsi \to \pm\infty) &\simeq
-i\sqrt{\frac{2}{\pi}}\,\frac{a}{m}
 \sqrt{e^{2\pi m^2}-1}\,
 e^{-\pi\moe^2\theta(-\dsi)}\cos\Big[{\frac{\dsi}{2}}-\moe^2\log(|\dsi|)+\arg\Gamma(1+i\moe^2)\Big][1+\morder{\dsi^{-1}}],
\end{split}\end{equation} 
where $\theta(\sigma)$ is the step function: $\theta(\sigma>0)= 1$ and $\theta(\sigma<0)= 0$. The norm $|\Phi(x)|$ is constant at large $|x|$ which, as for the Dirac wave function, suggests its interpretation as the inclusive density of the virtual pairs created by the linear potential. Wave functions corresponding to different eigenvalues are orthogonal \cite{Dietrich:2012un}.

%
\begin{wrapfigure}[10]{r}{0.5\textwidth}
  \vspace{-20pt}
  \begin{center}
    \includegraphics[width=0.5\textwidth]{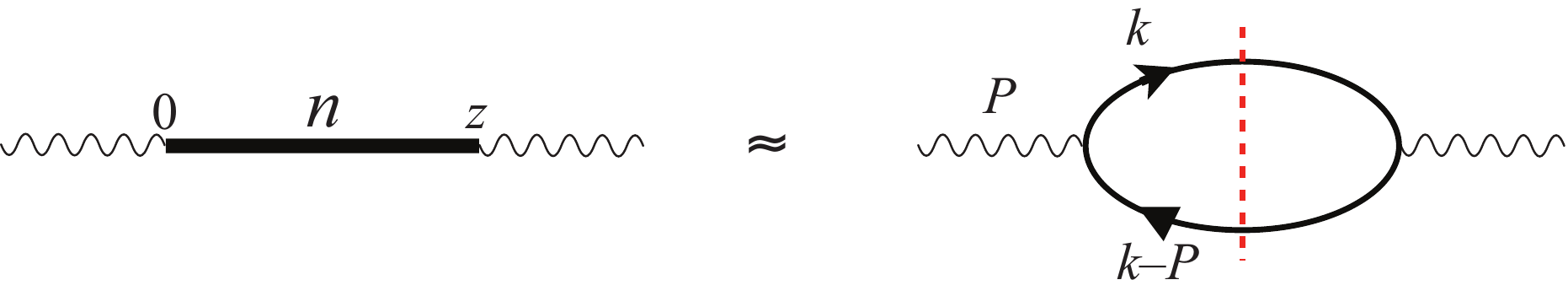}
  \end{center}
  \vspace{-15pt}
  \caption{Duality between resonance and fermion-loop contributions to the imaginary part of a current propagator. The relation should hold in a semi-local sense, and become more accurate
at high excitations.}
\label{dual}
\end{wrapfigure}

For highly excited states the normalization of the wave function at $x=0$ may be determined by duality. As indicated in \fig{dual} we expect the bound state contributions to current propagators to be (in an average sense) equal to the imaginary part of the free fermion loop. Scalar, pseudoscalar, vector and pseudovector currents give the same result. For states of parity $\eta$,
\beq\label{wfnorm}
|\phi_0^{\eta=-}(x=0)|^2 = |\phi_1^{\eta=+}(x=0)|^2 = \frac{\pi}{2}
\eeq
Recall that $\phi_0^{\eta=+}(x=0)=\phi_1^{\eta=-}(x=0)=0$ according to \eq{cont}.

Parton-hadron duality turns out to hold also at finite $x$, provided $V(x) \ll M$. For large $M$ we may use the asymptotic expressions \eq{aschi} which are plane waves in $x$, given that $\sigma \simeq M^2-2MV(x)$. The bound state turns out to consist of only positive energy $f\bar f$ pairs,
\beq\label{reststate4}
\ket{M,0}= \frac{\sqrt{2\pi}}{2M}\Big(\eta\, b^\dag_{M/2}\,d^\dag_{-M/2} + b^\dag_{-M/2}\,d^\dag_{M/2}\Big)\ket{0} 
\eeq
where $b^\dag_{M/2}\ (d^\dag_{M/2})$ creates a positive energy (anti-)fermion of momentum $M/2$. The $b^\dag b,\ d d^\dag$ and $db$ components of the general bound state \eq{statedef} do not contribute in \eq{reststate4}, allowing the parton interpretation. It is the oscillating (non-normalizable) behavior \eq{aschi} of the wave function which gives plane waves and thus partons of definite momenta. This duality is valid at any bound state momentum $P$.

\subsubsection{Frame dependence} 

The components $\phi_0(\sigma)$ and $\phi_1(\sigma)$ of the bound state wave function $\Phi$ in \eq{wf1} are frame independent functions of $\sigma=(E-V)^2-P^2$, where $V(x)=\halft |x|$ and $E=\sqrt{P^2+M^2}$. The $P$-dependence of $\sigma(x)$ introduces a frame dependence when the wave function is expressed in terms of the separation $x$ between the fermions. Corresponding to each $\sigma$ there are two values of $x$,
\beq\label{xfromsigma}
x= 2\Big(E \pm \sqrt{P^2+\sigma}\,\Big)
\eeq
The wave functions are defined for $x\geq 0$ by the bound state equation \eq{eom1} and for $x\leq 0$ by their parity \eq{parity}. Continuity at $x=0$ is imposed through \eq{cont}, which by \eq{masscond} determines the bound state mass $M$ through the zeros of $\phi_1$ or its derivative at $\sigma=\sigma_0 = M^2$.

In the rest frame \eq{xfromsigma} reads 
\beq
x= 2(M\pm \sqrt{\sigma}) \hspace{1cm} (P=0)
\eeq
so $\sigma=\sigma_0$ corresponds to $x=0$ and $x=4M$. As $\sigma$ decreases ($\sigma<\sigma_0$) the two solutions approach each other and meet for $\sigma=0$ at $x=2M$. This accounts for the mirror symmetry of the wave function in \fig{ffwf} for $0 \leq x \leq 4M$. For $\sigma > \sigma_0$ only the upper sign in \eq{xfromsigma} gives $x>0$. This solution corresponds to the large $x$ region with oscillations in \fig{ffwf}.

%
\begin{wrapfigure}[14]{r}{0.5\textwidth}
  \vspace{-20pt}
  \begin{center}
    \includegraphics[width=0.5\textwidth]{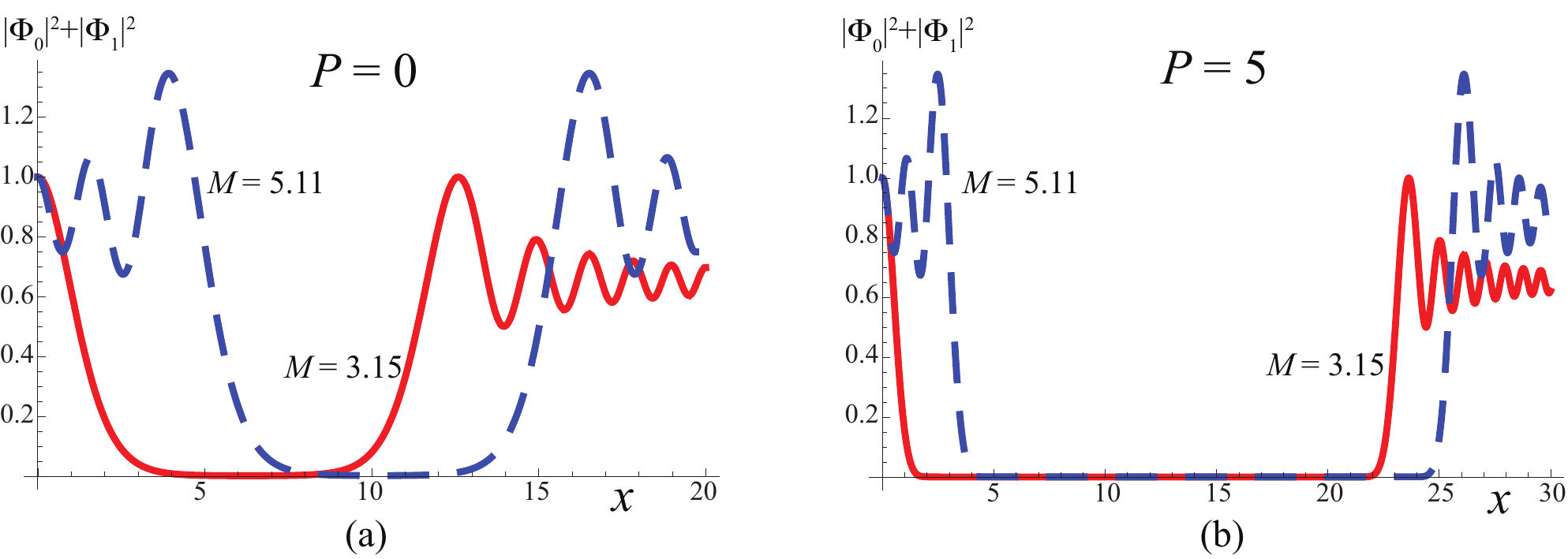}
  \end{center}
  \vspace{-15pt}
  \caption{(a) The density $|\phi_0|^2+|\phi_1|^2$ as a function of the distance $x$ between the constituents for the ground state ($M=3.15$, solid red line) and for an excited state ($M=5.11$, dashed blue line). The constituent masses are $m_1=1.0$ and $m_2=1.5$. (b) The densities in (a) plotted in the case of nonvanishing center-of-mass momentum, $P=5.0$. The densities are symmetric under $x \to -x$ and normalized to unity at $x=0$.}\label{m1m2MP}
\end{wrapfigure}

At large $P$, in the Infinite Momentum Frame (IMF), the relation \eq{xfromsigma} becomes (for $|\sigma|\ll P$)
\beq\label{largeP}
x \simeq 2(E\pm P) \pm \frac{\sigma}{P} \simeq  \left\{
\begin{array}{c} 4P + \sigma/P \\[2mm] (M^2-\sigma)/P \end{array} \right.
 \hspace{1cm} (P \to \infty)
\eeq
As $\sigma$ decreases from $\sigma_0$ the lower sign gives a Lorentz-contracted wave function, as expected for an equal-time state. The upper sign gives an asymptotically large $x \simeq 4P$. The separation of these two parts of the wave function with increasing $P$ is illustrated in \fig{m1m2MP}.

Figs. \ref{ffwf} and \ref{m1m2MP} indicate that the oscillations at large $x$ reflect pair production, which in time-ordered perturbation theory occurs via $Z$-diagrams such as in \fig{pairfig}a. With increasing CM momentum $P$ the energy required to create the pair increases due to the boost of its momentum. This qualitatively explains why $V(x) \propto P$ in the region of pair fluctuations. The large separations $x$ are allowed by the uncertainty principle due to the time dilation of the virtual pair life-time, and are required for Lorentz covariance.

In the $P \to \infty$ limit the term $\propto \Pisl^\dag=(E-V)\gz+P\go$ in \eq{wf1} gives the leading contribution to $\Phi$ when $\sigma$ is fixed. Retaining only the Lorentz contracting part of the wave function ($x \propto 1/P$, the lower solution in \eq{largeP}) the IMF wave function is
\beq\label{imfwf}
\Phi_{IMF}(\sigma)=2am\, P\gamma^+ e^{-i\dsi/2}\kum(1-im^2,2,i\dsi) \hspace{2cm} (\gamma^+=\gz+\go)
\eeq
where $\sigma \simeq M^2-P|x|$. In the $\sigma \to \infty$ limit $\Phi_{IMF}$ is suppressed by $1/\sigma$ compared to the limit \eq{aschi} of the complete solution. Hence the oscillations at large $x$ are suppressed and the normalization integral $\int dx\,|\Phi_{IMF}|^2$ is finite. The $P\to\infty$ (IMF) and $|x| \to \infty$ limits do not commute.

\vspace{-.3cm}

\subsubsection{Gauge covariance} 

The state \eq{statedef} involves fermion fields at points separated in space ($x_1$ and $x_2$) which are not connected by a gauge field exponential (Wilson line). In order for the state to be invariant under gauge transformations we need to transform the wave function $\Phi(x_1-x_2)$ accordingly. Here we only consider time-independent gauge transformations, to preserve our formulation of bound states defined at equal time.

In a space dependent gauge transformation
\beq\label{gaugetransform}
\psi(t=0,x) \to U(x)\psi(t=0,x) \hspace{2cm} \bar\psi(t=0,x) \to \bar\psi(t=0,x)U^\dag(x)
\eeq
where $U(x)$ is a phase in a $U(1)$ gauge theory and a $3\times 3$ color matrix in QCD. In the new gauge the state \eq{statedef} is described by the wave function
\beq\label{wfgauge}
\Phi_U(x_1,x_2) \equiv U^\dag(x_1)\Phi(x_1-x_2)U(x_2)
\eeq
Standard atomic wave functions in QED have the same gauge dependence.

\subsection{Form factor and parton distribution\label{ffpdf}}

\subsubsection{Electromagnetic form factor} 

The Poincar\'e covariance of the bound states \eq{statedef} allows to include them as $in$ and $out$ states of scattering processes. Let us consider the electromagnetic form factor\footnote{In the following $P=(E,P^1)$ denotes the 2-momentum.} 
\beq\label{formfac}
F^\mu_{AB}(z) = \bra{B(P_B);t=+\infty}j^\mu(z)\ket{A(P_A);t=-\infty}=e^{i(P_B-P_A)\cdot z}\bra{B(P_B);t=0}j^\mu(0)\ket{A(P_A);t=0}
\eeq
where the electromagnetic current
\beq\label{currentdef}
j^\mu(z)= \bar\psi(z)\gamma^\mu\psi(z) = e^{i\hat P\cdot z} j^\mu(0) e^{-i\hat P\cdot z}
\eeq
was shifted to the origin using translation invariance. We also translated the states $\ket{A}$ and $\bra{B}$ to the common time $t=0$, ignoring an irrelevant overall phase.

Using the equal-time anticommutation relations between the fields gives, with $j^\mu\ket{0}=0$,
\beqa
F^\mu_{AB}(z) &=& 
e^{i(P_B-P_A)\cdot z}\int dx_1 dx_2 dy_1 dy_2 e^{i(x_1+x_2) P_A^1/2-i(y_1+y_2) P_B^1/2}\nn\\
&\times& 
\bra{0}\psi^\dag(0,y_2)\Phi_B^\dag(y_1-y_2)\gamma^0 \psi(0,y_1)\big[\bar\psi(0,0)\gamma^\mu\psi(0,0)\big] \bar\psi(0,x_1)\Phi_A(x_1-x_2)\psi(0,x_2)\ket{0} 
\label{ff1a}\\[2mm]
&=& e^{i(P_B-P_A)\cdot z}\int_{-\infty}^{\infty} dx\, e^{i(P_B^1-P_A^1)x/2}\,
\Big\{\tr\big[\Phi_B^\dag(x)\gamma^\mu\gamma^0\Phi_A(x)\big]-\eta_A\eta_B\tr\big[\Phi_B(x)\gamma^0\gamma^\mu\Phi_A^\dag(x)\big]\Big\} \label{ff1}
\eeqa
In the second term of \eq{ff1} we used the parity relation $\Phi(-x)=\eta\Phi^*(x)$ which follows from \eq{parity}. 

The invariance of $F^\mu_{AB}(z)$ under gauge transformations follows by using the property \eq{wfgauge} of the wave functions in \eq{ff1}. Consequently we must have 
\beq\label{gaugecond}
G_{AB}(z) \equiv \partial_\mu F_{AB}^{\mu}(z) =0
\eeq
This implies that the form factor in $D=1+1$ can be expressed as
\beq\label{invff} 
F^\mu_{AB}(q)\equiv\int d^2z F^\mu_{AB}(z)e^{-iq\cdot z}= (2\pi)^2 \delta^2(P_B-P_A-q) \varepsilon^{\mu\nu}q_\nu F_{AB}(Q^2)
\eeq
where $Q^2 = -q^2$ and $\varepsilon^{\mu\nu}$ is the anti-symmetric tensor with $\varepsilon^{01}=1$.
Solving this for $F_{AB}(Q^2)$ with $\mu=0$, using Eq.~\eqref{ff1} for the left-hand side and the expression \eq{wf1} for $\Phi$ we obtain
\beq
F_{AB}(Q^2) = 
-4i\frac{1-\eta_A\eta_B}{q^1}\int_0^\infty dx\, \sin\Big(\frac{q^1x}{2}\Big)\Big[\Phi_{0B}^*(x)\Phi_{0A}(x)+\Phi_{1B}^*(x)\Phi_{1A}(x)\Big(1+\frac{4m^2}
{\dsi_{A}\dsi_{B}} 
\,\tilde \dpi_{A}\cdot \dpi_{B}\Big)\Big] \label{ffsym1}
\eeq
where $\tilde \dpi=(E-V,-P^1)$.
According to the asymptotic behavior \eq{aschi} of the wave functions the leading term for $x\to\infty$ in the square bracket of \eq{ffsym1} is $\propto \cos\big[\halft(\sigma_B-\sigma_A)\big] = \cos\big[\halft(M_B^2-M_A^2)-\halft x(E_B-E_A)\big]$. The integral may thus be regulated similarly to plane waves, and $F_{AB}(Q^2)$ is well defined.

\subsubsection{Gauge invariance of the form factor} 

It is instructive to verify the consequence \eq{gaugecond} of gauge invariance explicitly. The contribution of the first trace in \eq{ff1} to $G$ is
\beq\label{gexpr}
G_{AB}^{(1)}(0) = i\int dx\, e^{i(P_B^1-P_A^1)x/2}\,
\tr\big[\Phi_B^\dag(x)(\Psl_B-\Psl_A)\gamma^0\Phi_A(x)\big].
\eeq
The bound state equations \eq{eom1} for $\Phi_A$ and $\Phi_B^\dag$ are
\beqa\label{eomAB}
i\partial_x\acom{\gf}{\Phi_A}-\halft P_A^1\com{\gf}{\Phi_A}+m\com{\gamma^0}{\Phi_A} &=&(E_A-V)\Phi_A \nn\\[2mm]
-i\partial_x\acom{\gf}{\Phi_B^\dag}+\halft P_B^1\com{\gf}{\Phi_B^\dag}-m\com{\gamma^0}{\Phi_B^\dag} &=&(E_B-V)\Phi_B^\dag
\eeqa
Multiplying the first equation by $-\Phi_B^\dag$ from the left, the second by $\Phi_A$ from the right, and taking the trace of their sum gives
\beq
-i\partial_x\tr\left(\gf\acom{\Phi_B^\dag}{\Phi_A}\right)+\halft(P_B^1-P_A^1)\tr\left(\gf\com{\Phi_B^\dag}{\Phi_A}\right)=(E_B-E_A)\tr\left(\Phi_B^\dag\Phi_A\right).
\eeq
Using $\big[\Phi_B^\dag,\Phi_A\big]=\big\{\Phi_B^\dag,\Phi_A\big\}-2\Phi_A\Phi_B^\dag$ and multiplying both sides by $\exp[i(P_B^1-P_A^1)x/2]$ we find
\beq\label{totder}
-i\partial_x\Big[e^{i(P_B^1-P_A^1)x/2}\,\tr\left(\gf\acom{\Phi_B^\dag}{\Phi_A}\right)\Big]
= e^{i(P_B^1-P_A^1)x/2}\,\tr\big[\Phi_B^\dag(\Psl_B-\Psl_A)\gamma^0\Phi_A\big]
\eeq
Integrating both sides over $x$ the right-hand side becomes $-iG_{AB}^{(1)}(0)$ and the left-hand side vanishes (assuming that the integral over the oscillating wave functions is regularized as $|\xv| \to \infty$, similarly as for plane waves). This proves the gauge condition \eq{gaugecond} for the first trace in \eq{ff1}. Similarly we may show that the second trace satisfies the gauge condition. This demonstration of gauge invariance is easily generalized to $D=3+1$ dimensions \cite{Dietrich:2012un}.

\subsubsection{Boost covariance of the form factor\label{formboost}} 

The Lorentz invariance of the right hand side of \eq{ffsym1} is not explicit, but may be verified numerically. Let us study analytically how the form factor \eq{ff1} transforms under boosts. We consider the infinitesimal transformation 
\beq\label{momshift}
(E,P) \to (E',P') = (E+\xi P,P+\xi E)
\eeq
under which the fermion fields transform according to \eq{fieldboost} and the states satisfy \eq{boostcov}. The expression
\beq\label{formfac2}
F^0_{AB}(0) = \bra{B(P_B)}U^\dag(\xi)\big[U(\xi)j^0(0)U^\dag(\xi)\big]U(\xi)\ket{A(P_A)} = \bra{B(P'_B)}[j^0(0)-\xi j^1(0)]\ket{A(P'_A)}
\eeq
of the $\mu=0$ form factor in terms of the boosted states must agree with the definition \eq{formfac}.

The $P$-dependence \eq{delPhi} of the wave function involves factors  of \order{m^n} multiplying the components $\phi_0$ and $\phi_1$, with $n=0,1,2$. Since $m$ is arbitrary and there is no simple relation between $\phi_0$ and $\phi_1$ the equivalence of the expressions \eq{formfac2} and \eq{formfac} must hold for each power of $m$ separately. Here we consider only the terms with factors of \order{m^0}. Then it suffices to write \eq{delPhi} as
\beq\label{delPhi2}
\delta\Phi(x) = \big[\halft i\xi xP\go\gz + \morder{m}\big]\Phi(x)
\eeq

The contribution of \order{\xi} to \eq{formfac2} arises from the  $j^1(0)$ term, from the shift $\xi(E_B-E_A)$ of $P_B^1-P_A^1$ in the exponent of \eq{ff1} and from the change \eq{delPhi2} of $\Phi_A$ and $\Phi_B$. The first trace in \eq{ff1} contributes (up to terms with coefficients of \order{m}),
\beqa
\delta F^{0\;(1)}_{AB}(0) &=& \xi\int dx\,e^{i(P_B^1-P_A^1)x/2}\Big\{-\tr\big[\Phi_B^\dag\gamma^1\gamma^0\Phi_A\big]+\halft ix(E_B-E_A)\tr\big[\Phi_B^\dag\Phi_A\big]-\halft ix(P_B^1-P_A^1)\tr\big[\Phi_B^\dag\gamma^1\gamma^0\Phi_A\big]\Big\} \nn\\
&=& \xi\int dx \Big\{e^{i(P_B^1-P_A^1)x/2}\,\tr\big[\Phi_B^\dag\gamma^0\gamma^1\Phi_A\big]+ \halft x\partial_x\Big[e^{i(P_B^1-P_A^1)x/2}\,\tr\big(\gz\go
\{\Phi_B^\dag,\Phi_A\big\}\big)\Big]\Big\} = 0
\eeqa
In the second equality we used the identity \eq{totder}. The expression vanishes after a partial integration since $\com{\gamma^0\gamma^1}{\Phi}$ contributes only to coefficients of \order{m}. A similar analysis should demonstrate the Lorentz covariance of all contributions to $F^\mu_{AB}$.  

\subsubsection{Parton distribution} 

We describe deep inelastic scattering by $e(k_1)+A(P_A) \to e(k_2)+B(P_B)$, given by the form factor \eq{ffsym1}. The Bjorken limit is as usual defined by $Q^2=-q^2 \to \infty$ (where $q=k_1-k_2$) with $\xbj=Q^2/(2P_A\cdot q)$ held fixed. The discrete bound state $B$ of large mass $M_B^2 = Q^2\big(1/\xbj-1\big)$ describes the inclusive final state according to Bloom-Gilman duality \cite{Bloom:1970xb}. In section \ref{dualsec} we noted the simple description \eq{reststate4} of highly excited states in terms of nearly free partons. This allowed us to determine the normalization of the wave functions using the duality relation shown in \fig{dual}.

Transcribing the usual relation between the DIS cross section and the parton distribution  $f(x_{Bj})$ to $D=1+1$ dimensions we find \cite{Dietrich:2012un}
\beq \label{partondistr}
 f(\xbj) = \frac{1}{8\pi m^2} \inv{\xbj}|Q^2 F_{AB}(Q^2)|^2
\eeq
In the Breit frame, defined by $q=(0,-Q)$, the bound state momenta are
\beq\label{diskin}
P_A = \frac{Q}{2\xbj}(1,1), \hspace{1.0cm} P_B=\frac{Q}{2\xbj}(1,1-2\xbj)
\eeq
The dominant contribution to the form factor $F_{AB}(Q^2)$ in \eq{ffsym1} is found to come from fermion separations $x \propto 1/Q$. In terms of the scaling variable $v=xQ/2$,
\beq \label{taappr}
 \dsi_A \simeq M_A^2-\frac{|v|}{\xbj} \equiv \tau_A \hspace{2cm} 
 \dsi_B \simeq Q^2\Big(\inv{\xbj}-1\Big)-\frac{|v|}{\xbj} \propto Q^2
\eeq  
Making use of the asymptotic expressions \eq{aschi} for $\Phi_B(\sigma_B)$ in the $Q^2 \to \infty$ limit the form factor \eq{ffsym1} becomes
\beq \label{ffinv3}
Q^2 F_{AB}(\eta_B=-) \simeq  -4i\sqrt{2\pi}(1+\eta_A)\int_{0}^\infty dv\, \sin v\Big[\cos\Big(\frac{v}{2\xbj}\Big) i\Phi_{0A}(\tau_A) - \sin\Big(\frac{v}{2\xbj}\Big)\Phi_{1A}(\tau_A)\Big(1+\frac{2\moe^2}{\xbj\tau_A}\Big)\Big]
\eeq
%
\begin{figure}[h]
\includegraphics[width=14cm]{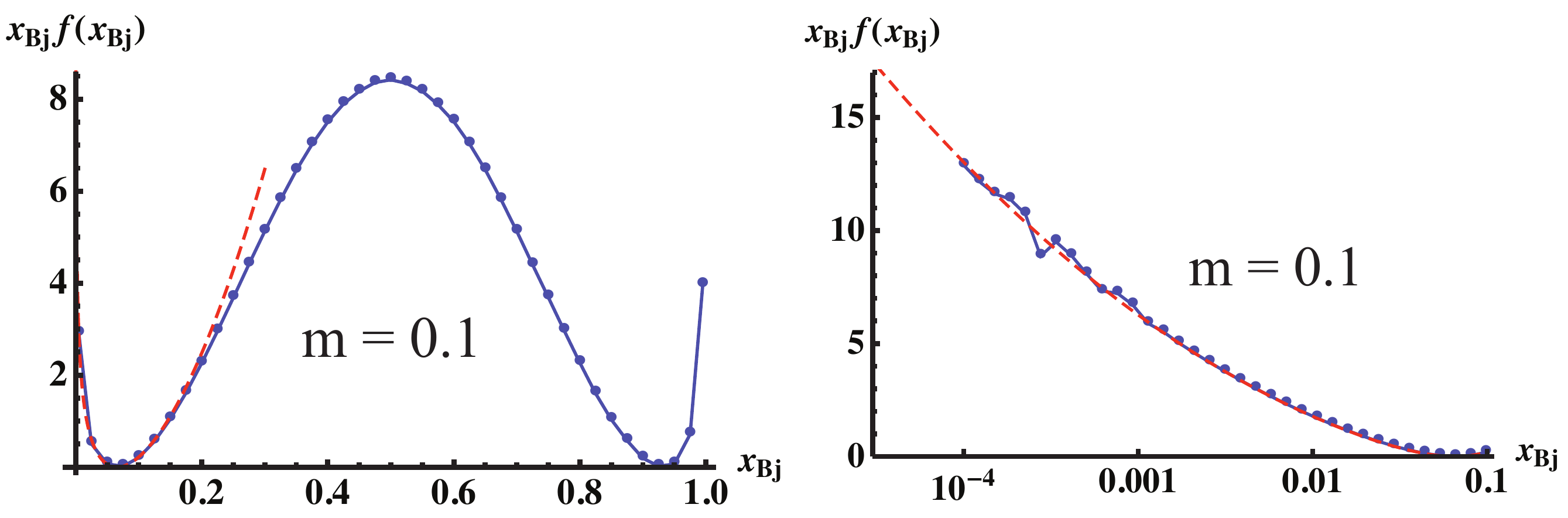}
\caption{The parton distribution of the positive parity ground state $A$, numerically evaluated for fermion mass $m=0.1\, e$ using \eq{partondistr} and \eq{ffinv3}. The distribution is shown on a linear scale in $x_{Bj}$ on the left and on a logarithmic scale on the right. The dashed red curves show an analytic calculation of $f(\xbj)$ valid for small $\xbj$, which neglects terms of \order{x_{Bj}^2}.}\label{partondist}
\end{figure}

The parton distribution of the ground state $A$ shown in \fig{partondist} has a ``sea''-like enhancement for $\xbj\to 0$. The enhancement is only present at small fermion masses $m$ and indicative of scattering from fermion pairs in the bound state. When $\xbj$ is small the argument $\tau_A$ \eq{taappr} of $\Phi_A$ is large and negative, allowing the use of the asymptotic expressions \eq{aschi}. For small $m$ the $\xbj\to 0$ form factor becomes
\beq
Q^2 F_{AB}(\eta_B=-) \propto \int_0^\infty dv\,\sin v \cos\big[\halft M_A^2+m^2\log\xbj\big] = 
\cos\big[\halft M_A^2+m^2\log\xbj\big]
\eeq
According to \eq{spectrum} $M_A^2 \simeq (2n+1)\pi$ for $\eta_A=+$, thus
\beq\label{seadist}
f(\xbj) \propto \frac{\log^2(\xbj)}{\xbj} \hspace{1cm} (\xbj \to 0)
\eeq
The analytic approximation shown as a red dashed curve in \fig{partondist} includes also terms suppressed by \order{\xbj} wrt. the leading behavior \eq{seadist}.

In the Breit frame the virtual photon probes distance scales $x \propto 1/Q$, as expected since $q^1 \simeq -Q$. Since the target momentum $P_A \propto Q$ is large in the Bj limit one would expect that the parton distribution would be determined by the IMF wave function $\Phi_{IMF}^A$ given in \eq{imfwf}. However, the mass of the final state $M_B \propto Q$. Hence the variable $\sigma_B \propto Q^2$, making the IMF limit \eq{imfwf} (taken at fixed $\sigma$) inapplicable for $B$. In fact, the first two terms in \eq{wf1} give the leading contribution to $\Phi_B(\sigma_B)$, but they are orthogonal to $\Phi_{IMF}^A$ in the trace $\tr\big[\Phi_B^\dag(x)\Phi_A(x)\big]$ of the form factor \eq{ff1}. Thus the scaling contribution to the parton distribution arises from the leading term in $\Phi_B$ combined with the next-to-leading term in $\Phi_A$, and {\it vice versa}. In particular, the enhancement for $\xbj \to 0$ in \fig{partondist} does not arise from $\Phi_{IMF}^A$.

\section{Relativistic bound states in $D=3+1$\label{3dpotential}}

So far we considered three examples of ``Born level'' bound states in abelian gauge theory. In this approximation the gauge field is determined by the classical field equations and explicit pair production is ignored ($H\ket{0}=0$).
\begin{enumerate}
\item {\it QED atoms.} For small $\alpha$ the ladder diagrams (Figs. \ref{feyndiags}a, \ref{feyndiags}b, \ldots) dominate near the bound state poles of the elastic $e^+e^-$ amplitude. Their sum generates the classical $-\alpha/r$ potential. The Schr\"odinger equation follows from $H_{QED}\ket{E}=E\ket{E}$, when the state $\ket{E}$ is defined as in \eq{posstate} and the classical $A^0$ field is used in $H_{QED}$.

\item {\it The Dirac equation}. A static point charge generates the confining potential $V(x)=\halft |x|$ in $D=1+1$ dimensions. The state \eq{dirstate} is an eigenstate of $H_{QED}$ if the wave function $\Psi(x)$ satisfies the Dirac equation. Virtual $e^+e^-$ pairs (\fig{pairfig}) appear for $V(x) \gsim 2m$ (\fig{Diracfig}), giving a constant particle density $|\Psi(x)|^2$ at large $|x|$.

\item {\it $f\bar f$ states in $D=1+1$}. The state $\ket{E,P}$ of \eq{statedef} is bound when its equal-time wave function $\Phi$ satisfies \eq{eom1}, with $V(x)$ determined by Gauss' law. A hidden boost covariance ensures that electromagnetic form factors are Poincar\'e as well as gauge invariant. There is no parity doubling as $m\to 0$.

\end{enumerate}
In this Section we consider how this approach may be extended to QCD hadrons in $D=3+1$ dimensions. 

Gribov \cite{Gribov:1998kb,Dokshitzer:2003bt} found a critical coupling in gauge theories,
\beq\label{critcoup}
\alpha^{crit}({\rm QED})=\pi\left(1-\sqrt{\frac{2}{3}}\right) \simeq 0.58 \hspace{2cm}
\alpha_s^{crit}({\rm QCD})=\frac{\pi}{C_F}\left(1-\sqrt{\frac{2}{3}}\right) \simeq 0.43
\eeq
at which the Coulomb interaction between light fermions becomes strong enough to cause a rearrangement of the perturbative vacuum. In QED $\alpha\simeq 1/137$ is well below the critical value, whereas $\as(m_\tau)\simeq 0.33$ in QCD \cite{Beringer:1900zz}. $\as(Q)$ approaches the critical value \eq{critcoup} for $Q < m_\tau$. 

Dokshitzer \cite{Dokshitzer:ws2013} has argued that confinement may be described by a classical field. In order to preserve Poincar\'e invariance such a field must satisfy the QCD equations of motion. Gauss' law fixes $A^0$ up to a boundary condition. We shall consider solutions with a constant, universal field strength $|\nv A^0|$ at large distances. A mass scale ($\Lambda_{QCD}$) can only arise from a boundary condition when loop effects are neglected.

\subsection{The abelian case\label{3dabelian}}

We begin by illustrating the procedure for $U(1)$ gauge theory (even though it is not relevant for QED). In our discussion of Positronium we recalled that stationarity of the action imposes Gauss' law \eq{gausslaw} on the Coulomb field. The operator solution \eq{a0sol} for $\hat A^0$ assumes that the field vanishes at spatial infinity. Let us consider the possibility of including a homogeneous solution\footnote{The formulation below differs technically from that in \cite{Hoyer:2009ep}, but the principles and results are the same.},
\beq\label{a0sol2}
\hat A^0(t,\xv)= \int d^3\yv\left(\kappa\, \xv\cdot\yv+ \frac{e}{4\pi|\xv-\yv|}\right)\psi^\dag\psi(t,\yv)
\eeq 
Here $\kappa$ is an $\xv$-independent parameter which will be determined to give a universal field strength for $|\xv| \to \infty$. The \order{\kappa} contribution will be considered leading compared to the perturbative \order{e} term.

As in \eq{a0eigen} we find (taking $\hat A^0\ket{0}=0$) that $\bar\psi(t,\xv_1)\psi(t,\xv_2)\ket{0}$ is an eigenstate of $\hat A^0(t,\xv)$ with eigenvalue
\beq\label{a0eigen2}
A^0(\xv) =\left[\kappa\,\xv\cdot(\xv_1-\xv_2)+\frac{e}{4\pi}\left(\inv{|\xv-\xv_1|}-\inv{|\xv-\xv_2|}\right)\right]
\eeq
which gives the squared field strength
\beq\label{e2}
\big[\nv A^0(\xv)\big]^2 = \kappa^2(\xv_1-\xv_2)^2+\frac{e\kappa}{2\pi}(\xv_1-\xv_2)\cdot\nv \left(\inv{|\xv-\xv_1|}-\inv{|\xv-\xv_2|}\right) +\morder{e^2}
\eeq
In order for the asymptotic ($|\xv|\to\infty$) field strength to be universal ($\la^2$)  we must choose
\beq\label{uni1}
\kappa = \frac{\Lambda^2}{|\xv_1-\xv_2|}
\eeq

The interaction term of the Hamiltonian \eq{ham2} is, neglecting the perturbative \order{e^2} contribution,
\beq\label{intham2}
H_{int}(t)= \frac{e\kappa}{2}\int d^3\xv\, d^3\yv\,\big[\psi^\dag(t,\xv)\psi(t,\xv)\big]\big[\psi^\dag(t,\yv)\psi(t,\yv)\big]\xv\cdot\yv +\morder{e^2}
\eeq
and generates a linear potential
\beq\label{genpot}
H_{int}(t)\,\bar\psi(t,\xv_1)\psi(t,\xv_2)\ket{0} =V(\xv_1-\xv_2)\bar\psi(t,\xv_1)\psi(t,\xv_2)\ket{0}
\eeq
where
\beq\label{linpot}
V(\xv_1-\xv_2)=\halft e\la^2\,|\xv_1-\xv_2|
\eeq

The linear potential obtained in this way is invariant under translations ($\xv_1 \to \xv_1+\bs{a},\ \xv_2 \to \xv_2+\bs{a}$) only for neutral states. If the fermion and antifermion had charges $e_1$ and $e_2$, respectively, the potential would be $V(\xv_1,\xv_2)\propto |e_1\xv_1-e_2\xv_2|$. Furthermore, multiplying the homogeneous solution in \eq{a0sol2} by any function of $|\yv|$ would destroy translation invariance even for neutral states. Thus the term $\propto \kappa$ in \eq{a0sol2} appears unique, and maintains translation invariance only for neutral states.

An $f\bar f$ bound state of momentum $\Pv$,
\beq\label{posstate2}
\ket{E,\Pv} = \int d^3\xv_1\,d^3\xv_2\,e^{i\Pv\cdot(\xv_1+\xv_2)/2}\,\bar\psi_\alpha(t,\xv_1)\Phi_{\alpha\beta}(\xv_1-\xv_2)\psi_{\beta}(t,\xv_2)\ket{0}
\eeq
is a Hamiltonian eigenstate of energy $E$ provided the wave function $\Phi$ satisfies
\beq\label{bse1}
i\nv\cdot\acom{\gamma^0\gv}{\Phi(\xv)}-\halft \Pv\cdot\com{\gamma^0\gv}{\Phi(\xv)}+m\com{\gamma^0}{\Phi(\xv)} = \big[E-V(\xv)\big]\Phi(\xv),
\eeq
with the potential $V(\xv)$ given by \eq{linpot}.

In the standard derivation of the QED Hamiltonian \eq{ham2} one assumes that the fields vanish at spatial infinity. A constant asymptotic field strength contributes an \order{\Lambda^2} term $\propto$ the volume of space from the field energy. Since we took $\Lambda$ to be the same for all states this common (infinite) contribution is irrelevant. The space integral of the \order{e\kappa} term in \eq{e2} may be converted to an integral over a surface at infinity. It gives a finite contribution of the same form as the linear potential \eq{linpot}, and thus only modifies the coefficient of $|\xv_1-\xv_2|$.

\subsection{Color algebra\label{3dcolor}}

We consider the straightforward generalization of the abelian ansatz \eq{a0sol2} for the homogeneous term,
\beq\label{a0sol3}
\hat A^0_a(t,\xv)= \kappa\int d^3\yv\,\psi_A^\dag(t,\yv) T_a^{AB}\psi_B(t,\yv)\, \xv\cdot\yv
\eeq 
where $T_a$ generates SU(3) transformations in the fundamental representation. Since $\nv^2{\hat A}^0_a=0$ the \order{\kappa} terms in Gauss' law are satisfied (in $\nv\cdot \Av_a=0$ gauge). Perturbative $A^\mu$ fields of \order{g} will couple to the \order{g^0} field ${\hat A}^0$ \eq{a0sol3}, giving \order{\as} effects which we neglect. 

We assume meson ($\mM$) and baryon ($\mB$) ``Born states'' of the form (Dirac indices are suppressed and repeated color indices summed over)
\beqa\label{mbstates}
\ket{\mM;E,\Pv} &=& \int d^3\xv_1\,d^3\xv_2\,e^{i\Pv\cdot(\xv_1+\xv_2)/2}\,\bar\psi_A(t,\xv_1)\Phi_\mM^{AB}(\xv_1-\xv_2)\psi_{B}(t,\xv_2)\ket{0} \label{meson}\\[2mm]
\ket{\mB;E,\Pv} &=& \int d^3\xv_1\,d^3\xv_2\,d^3\xv_3\,e^{i\Pv\cdot(\xv_1+\xv_2+\xv_3)/3}\,\psi_{A}^\dag(t,\xv_1)\psi_{B}^\dag(t,\xv_2)\psi_{C}^\dag(t,\xv_3)\Phi_{\mB}^{ABC}(\xv_1,\xv_2,\xv_3)\ket{0}\hspace{1cm} \label{baryon}
\eeqa
We look for a solution where the wave functions have the standard color structure,
\beqa
\Phi_\mM^{AB}(\xv_1-\xv_2) &=& \inv{\sqrt{N_C}}\delta^{AB}\Phi_\mM(\xv_1-\xv_2) \label{mcolor} \\[-1mm]
&&\hspace{6cm} (N_C=3) \nn\\[-1mm]
\Phi_\mB^{ABC}(\xv_1,\xv_2,\xv_3) &=& \inv{\sqrt{N_C!}}\epsilon^{ABC}\Phi_\mB(\xv_1,\xv_2,\xv_3) \label{bcolor}
\eeqa
The baryon wave function is a $4\times 4 \times 4$ matrix in Dirac indices and invariant under a common translation of $\xv_1,\xv_2$ and $\xv_3$. The color structure of the wave functions transforms analogously to \eq{wfgauge} under gauge transformations.

The ${\hat A}^0_a$ field \eq{a0sol3} transforms a component of the meson state as
\beq\label{a0meson}
\hat A_a^0(\xv)\,\bar\psi_C(\xv_1)\psi_{C}(\xv_2)\ket{0}=\kappa\,\xv\cdot\xv_1 \,\bar\psi_A(\xv_1)T_a^{AC}\psi_{C}(\xv_2)\ket{0}
-\kappa\,\xv\cdot\xv_2 \,\bar\psi_C(\xv_1)T_a^{CB}\psi_{B}(\xv_2)\ket{0}
\eeq
The gauge invariant square of the field strength for this component is
\beq
(\nv\hat A_a^0)\cdot(\nv\hat A_a^0)\,\bar\psi_C(\xv_1)\psi_{C}(\xv_2)\ket{0}=
C_F\kappa^2(\xv_1-\xv_2)^2\bar\psi_C(\xv_1)\psi_{C}(\xv_2)\ket{0}
\eeq
where we used
\beq\label{genid1}
T_a^{AC}T_a^{CB}= C_F\delta^{AB}
\eeq
with $C_F=4/3$. As in the abelian case \eq{uni1} we choose
\beq\label{mkappa}
\sqrt{C_F}\,\kappa = \frac{\Lambda^2}{|\xv_1-\xv_2|}
\eeq
to ensure the universal asymptotic field strength $\la^2$ for all meson components.

The gluon field \eq{a0sol3} contributes an interaction term to the QCD Hamiltonian,
\beq\label{intham3}
H_{int}= \frac{g\kappa}{2}\int d^3\xv\, d^3\yv\,\big[\psi^\dag_A(\xv)T_a^{AB}\psi_B(\xv)\big]\big[\psi^\dag_C(\yv)T_a^{CD}\psi_D(\yv)\big]\xv\cdot\yv
\eeq
Operating on a component of the meson state this gives
\beq\label{genpot2}
H_{int}\bar\psi_C(\xv_1)\psi_C(\xv_2)\ket{0} =V_\mM(\xv_1-\xv_2)\bar\psi_C(t,\xv_1)\psi_C(t,\xv_2)\ket{0}
\eeq
where $V_\mM$ is the linear potential
\beq\label{mpot}
V_\mM(\xv_1-\xv_2) = \halft C_Fg\kappa (\xv_1-\xv_2)^2=\halft\sqrt{C_F}\,g\la^2 |\xv_1-\xv_2|
\eeq
Since the free part of the QCD Hamiltonian is diagonal in color the meson state \eq{meson} is an eigenstate of the full Hamiltonian, $H\ket{\mM;E,\Pv}=E\ket{\mM;E,\Pv}$ when the meson wave function $\Phi_\mM$ satisfies the bound state equation \eq{bse1}, with $V(\xv)=V_\mM(\xv)$.

A component of the baryon state \eq{baryon} 
\beq
\ket{A,B,C} \equiv \psi^\dag_A(\xv_1)\psi^\dag_{B}(\xv_2)\psi^\dag_{C}(\xv_3)\ket{0}
\eeq
is transformed by the ${\hat A}^0_a$ field \eq{a0sol3} as
\beq\label{a0baryon}
\hat A_a^0(\xv)\ket{A,B,C}=\kappa\Big(T_a^{DA}\xv\cdot\xv_1\ket{D,B,C}+T_a^{DB}\xv\cdot\xv_2\ket{A,D,C}+T_a^{DC}\xv\cdot\xv_3\ket{A,B,D}\Big)
\eeq
Using the identity \eq{genid1} and \cite{MacFarlane:1968vc}
\beq\label{genid2}
T_a^{AB}T_a^{CD}= \halft\delta^{AD}\delta^{BC}-\sfrac{1}{6}\delta^{AB}\delta^{CD}
\eeq
the squared field strength of the color singlet component with quarks at $\xv_1,\xv_2,\xv_3$ is translation invariant,
\beq\label{bstrength}
(\nv\hat A_a^0)\cdot(\nv\hat A_a^0)\,\epsilon^{ABC}\ket{A,B,C}=\halft C_F\kappa^2\big[(\xv_1-\xv_2)^2+(\xv_2-\xv_3)^2+(\xv_3-\xv_1)^2\big]\epsilon^{ABC}\ket{A,B,C}
\eeq
A universal field strength requires for this component
\beq\label{bkappa}
\sqrt{\halft C_F}\,\kappa = \frac{\Lambda^2}{\sqrt{(\xv_1-\xv_2)^2+(\xv_2-\xv_3)^2+(\xv_3-\xv_1)^2}}
\eeq
The eigenvalue of the interaction Hamiltonian is similarly
\beq\label{bstrength2}
H_{int}\,\epsilon^{ABC}\ket{A,B,C}=V_\mB(\xv_1,\xv_2,\xv_3)\epsilon^{ABC}\ket{A,B,C}
\eeq
where the baryon potential is
\beq\label{bpot}
V_\mB(\xv_1,\xv_2,\xv_3)=\sfrac{1}{2\sqrt{2}} \sqrt{C_F}\, g\la^2\sqrt{(\xv_1-\xv_2)^2+(\xv_2-\xv_3)^2+(\xv_3-\xv_1)^2}
\eeq
When two quarks are at the same position, \eg, $\xv_2=\xv_3$, the baryon potential agrees with the meson one,
\beq\label{mbpot}
V_\mB(\xv_1,\xv_2,\xv_2) = V_\mM(\xv_1-\xv_2)
\eeq
The baryon state \eq{baryon} is an eigenstate of the full Hamiltonian when the wave function in \eq{bcolor} satisfies
\beq\label{bbse}
\sum_{j=1}^3 \gz_j(-i\nv_j\cdot\bs{\gamma}_j+\sfrac{1}{3}\Pv\cdot\gv_j+m)\Phi_\mB=(E-V_\mB)\Phi_\mB
\eeq
Here $\nv_j \equiv \partial/\partial\xv_j$ and the subscript $j$ on the Dirac matrices indicates which Dirac index on $\Phi_\mB$ it is contracted with.

\subsection{Properties of the bound state equations\label{3dproperties}}

The meson bound state equation \eq{bse1} is (in the rest frame, $\Pv=0$) a natural generalization of the Dirac equation. As such it was considered already in 1929 by Breit \cite{Breit:1929zz} (see also \cite{Krolikowski:1992fy}). The separation of variables for central potentials $V(|\xv|)$ in the rest frame, and the radial equations for states of any spin $J$, parity and charge conjugation, may be found in \cite{Geffen:1977bh}, together with a phenomenological study.

The radial wave functions given in \cite{Geffen:1977bh} share some of the properties of the $D=1+1$ rest frame wave functions. In particular, they are generally singular at $M=V(r)$, corresponding to $\sigma=0$ in \eq{sigdef}. Requiring local normalizability at this point imposes a discrete spectrum with asymptotically linear Regge trajectories, $J \propto M^2/8c$, where $c=\halft\sqrt{C_F}g\la^2$ is the coefficient of the linear potential in \eq{mpot}. In \cite{Geffen:1977bh} the wave functions were taken to vanish for $V(r) > M$. However, this violates the bound state equation and upsets orthogonality, Poincar\'e and gauge invariance. We discuss the issue of the non-normalizability of the wave functions in Section \ref{disc2}. 

In $D=1+1$ dimensions the equation \eq{wfframe} expresses the wave function in any frame in terms of that in the rest frame, and the bound state equation may be cast in the covariant form \eq{eom4}. In $D=3+1$ an analogous result was found \cite{Hoyer:1986ei} for the special configuration where $\xv \parallel \Pv$. Taking $\Pv=(0,0,P)$ along the $z$-axis it was shown that the bound state equation \eq{bse1} is for $\xv=(0,0,z)$ equivalent to
\beq\label{eom5}
-i\sigma\partial_\sigma \big\{\gz\gamma^z,\Phi^{(P)}(\bs{0}_\perp,\sigma)\big\} -i\sum_{j=x,y} \big\{\gamma^j\Pisl^\dag,\partial_j\Phi^{(P)}(\bs{0}_\perp,\sigma)\big\}+m\big[\Pisl^\dag,\Phi^{(P)}(\bs{0}_\perp,\sigma)\big]=\sigma\Phi^{(P)}(\bs{0}_\perp,\sigma)
\eeq
The quantities $\Pisl$ and $\sigma$ are defined as in \eq{sigdef}, with $V=V(z)$ being the linear potential \eq{linpot} or \eq{mpot}. The $P$-dependence of the wave functions that solve \eq{eom5} is given by \eq{wfframe}, with $\zeta$ as in \eq{zetadef}. Thus knowing $\Phi^{(P=0)}(\xv)$ in the rest frame, making use of the spherical symmetry, allows to determine the wave function at $\xv=(0,0,z)$ for all $P$ and $z$. The frame dependence \eq{wfframe} applies also to the first transverse derivatives $\partial_x\Phi^{(P)}(\bs{0}_\perp,\sigma),\,\partial_y\Phi^{(P)}(\bs{0}_\perp,\sigma)$. The bound state equation \eq{bse1} implicitly determines $\Phi^{(P)}(\xv)$ for all $\xv$ when it is known at $(0,0,z)$.

The electromagnetic form factor $F_{AB}^\mu(x)$ may be defined as in \eq{formfac} and expressed in terms of the $D=3+1$ wave functions of $A,\,B$ as in \eq{ff1}. The proof of gauge invariance, $\partial_\mu F_{AB}^\mu(x)=0$, can be carried through in $D=3+1$ similarly to the proof in Section \ref{ffpdf} \cite{Dietrich:2012un}. The covariant result indicates that the bound state equation \eq{eom1} has a hidden  Poincar\'e invariance for general $\xv$.

Wave functions of definite total spin $J$ can readily be found in the rest frame \cite{Geffen:1977bh}. Knowing their properties under boosts would give information about spin in the infinite momentum frame, corresponding to quantization on the light-front $x^+ = t+z=0$. The issue of spin for hadron LF wave functions is a topical problem \cite{Leader:2013jra}. 

\vspace{-.2cm}

\section{Discussion (Part 2)\label{disc2}}

\vspace{-.3cm}

We studied whether well-known techniques developed for QED bound states can be extended to QCD. Hadrons have relativistic, confined constituents and are in this sense quite different from atoms. Nevertheless, there are also similarities. Quarkonium is often referred to as the ``Positronium atom of QCD'' in view of the atomic-like spectra of $c\bar c$ and $b\bar b$ mesons. Hadron quantum numbers reflect the degrees of freedom only of their valence quarks ($q\bar q$ or $qqq$), despite their rich parton structure. There is no definite evidence of gluonic degrees of freedom (glueballs, hybrids), whereas the recent discoveries of ``molecular quarkonia" further reinforce the analogy to QED \cite{Lange:2013rxa}.

Atoms are described by the Schr\"odinger equation at lowest order in $\alpha$. In this approximation the interactions are mediated by the classical potential of the charged constituents, and are described by Born-level Feynman diagrams. Loop corrections allow fluctuations in the gauge and matter fields. Accurate calculations of the properties of atoms have been successfully compared to precision measurements \cite{Sapirstein,Kinoshita,Czarnecki:1999uk,Ishida:2013waa}. Atomic electrons are non-relativistic, moving with velocities $v/c \simeq \alpha \simeq 1/137$. For $\alpha$ of \order{1} the binding would become relativistic, but the perturbative (loop) expansion is then unreliable.

The Dirac equation describes a relativistic electron bound by a strong external field. This dynamics is also given by Born-level Feynman diagrams, in the limit where one particle (the source of the field) is very massive and has a large charge $eZ$ ($Z\gg 1$). Loop corrections on the electron propagator and vertices are of \order{\alpha} and may thus be neglected, compared to interactions of \order{\alpha Z} with the source particle. A relativistic electron can move backward in time, which corresponds to virtual $e^+e^-$ pair production and annihilation ($Z$-diagrams, \fig{pairfig}a). At any instant of time a Dirac state therefore has has multi-particle components. Yet the Dirac wave function has only a single electron degree of freedom. The ``original'' electron cannot be distinguished from those in the pairs, making the interpretation of the Dirac wave function more subtle than its Schr\"odinger equivalent.

The Schr\"odinger and Dirac examples suggest a Born-level approximation of hadrons in terms of a classical gauge field. This is also supported by the phenomenological success of the quark model. The challenge is to actually derive the potential from the classical equations of motion of the gauge field, and to maintain the Poincar\'e invariance of the QCD action. In order to describe hadron dynamics (electromagnetic form factors, scattering processes, \ldots) we must consider bound states in motion. 

Scattering amplitudes for processes like $e^+e^-\to e^+e^-$ are Poincar\'e invariant at each order in $\alpha$. Bound states are stationary in time (\ie, they are eigenstates of the Hamiltonian), and their wave functions involve all powers of $\alpha$. Boosts transform time and thus become dynamic, rather than explicit, symmetries of bound states. The Poincar\'e invariance of the action ensures that the bound state energy has a simple frame dependence ($E=\sqrt{M^2+\Pv^2}\,$), whereas equal-time wave functions transform in a more complex way. In Section \ref{motion} we saw that Positronium acquires an $\ket{e^+e^-\gamma}$ Fock component in frames with $\Pv\neq 0$. There are few (if any) cases where the boost dependence of a relativistic, equal-time wave function is explicitly known, except those discussed in these lectures.

We determined bound states as eigenstates of the Hamiltonian, with a gauge field that satisfies the field equations of motion at lowest order in the coupling $g$. Both the Hamiltonian and the equations of motion are determined by the gauge theory action, leaving little freedom. Nevertheless, there is a possibility to introduce an \order{g^0} contribution (for color neutral states) through the homogeneous solution \eq{a0sol3} of Gauss' law. Such a solution is usually discarded, as it introduces a non-vanishing field strength at large distances $|\xv|$.  The corresponding field energy is proportional to the volume of space, and must be a common (infinite) contribution to all components of the bound states. This leaves a single parameter $\Lambda$, and implies the linear potential \eq{mpot} for mesons and the analogous potential \eq{bpot} for baryons (plus perturbative, \order{\as} contributions).

Formulating the Hamiltonian Dirac dynamics as in \eq{hamdirstate} provided two lessons. 

({\it i}) It is necessary that the Dirac state is built on a vacuum that is itself an eigenstate of the Hamiltonian. A static external field does not transfer energy, preventing on-shell pair production. Consequently, the scattering amplitude is insensitive to the $\ieps$ prescription at the negative energy pole of the electron propagator. Using retarded (instead of Feynman) electron propagators implies a retarded vacuum satisfying $H\ket{0}_R=0$.

({\it ii}) It has been known since 1932 \cite{plesset} that the normalization integral $\int d^3\xv |\Psi(\xv)|^2$ of the Dirac wave function diverges for all polynomial potentials $V(|\xv|)$ and that the energy spectrum is continuous\footnote{The sole exception is the $V(r) \sim 1/r$ potential in $D=3+1$ dimensions, which is often found in textbooks.}. There is little awareness and understanding of this property of the Dirac bound states (see \cite{Giachetti:2007vq} for a recent discussion). With retarded boundary conditions $\psi^\dag\psi$ is the number operator of positive and negative energy fermions, and its expectation value in the Dirac state is $|\Psi(\xv)|^2$. \fig{Diracwf} supports the interpretation of $|\Psi(x)|^2$ as an {\it inclusive} particle density.

%
\begin{wrapfigure}[14]{r}{0.3\textwidth}
  \vspace{-25pt}
  \begin{center}\hspace{-.5cm}
    \includegraphics[width=0.25\textwidth]{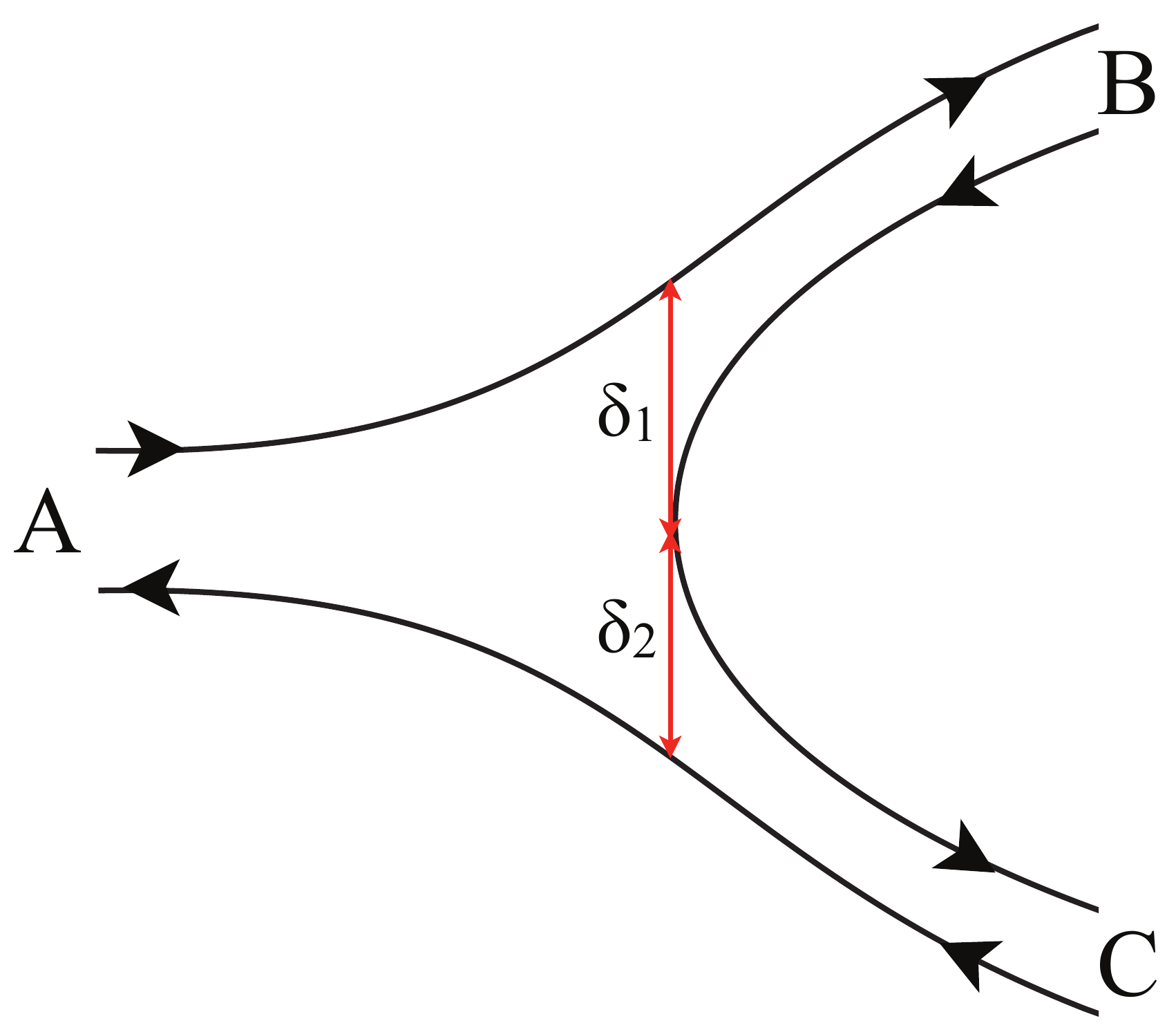}
  \end{center}
  \vspace{-15pt}
  \caption{The dual diagram for meson splitting $A\to B+C$, given by \eq{hadvert}. The $q\bar q$ pair is created at distance $\delv_1$ from the quark and $\delv_2$ from the antiquark of meson $A$.}\label{dualdiag}
\end{wrapfigure}

The $f\bar f$ bound states that we studied in Section \ref{ffstatesec} also need to be built on a vacuum that is an eigenstate of the Hamiltonian. This suggests an analogy to the $in$ and $out$ states used as asymptotic states of the perturbative $S$-matrix, which are eigenstates of the free Hamiltonian $H_0$. States defined at asymptotic times are on-shell and thus independent of the $\ieps$ prescription in their propagator. The $f\bar f$ states discussed here may be used as asymptotic states of the $S$-matrix, as in the electromagnetic form factor \eq{formfac}.

The time development from $t=\pm\infty$ to the (finite) scattering time is determined by the full Hamiltonian. The asymptotic states therefore develop into eigenstates of $H$ by the time of scattering. In addition to contributions from higher orders in $\as$, the bound states can split and merge as illustrated in the dual meson diagram of \fig{dualdiag}. The amplitude $\bra{B,C}A\rangle$ can be evaluated directly from the definition \eq{meson} of the meson states, using anticommutation relations for the quark fields according to \fig{dualdiag}. Suppressing Dirac and color indices,
\beqa\label{hadvert}
\bra{B,C}A\rangle &=& \inv{\sqrt{N_C}}\int\Big[\prod_{k=A,B,C}d\xv_1^kd\xv_2^k\Big]e^{i(\xv_1^A+\xv_2^A)\cdot \Pv_A/2-i(\xv_1^B+\xv_2^B)\cdot \Pv_B/2-i(\xv_1^C+\xv_2^C)\cdot \Pv_C/2}\nn\\
&\times& \bra{0}\big[\psi^\dag(\xv_2^B)\Phi_B^\dag\gz\psi(\xv_1^B)\big]
\big[\psi^\dag(\xv_2^C)\Phi_C^\dag\gz\psi(\xv_1^C)\big]
\big[\psi^\dag(\xv_1^A)\gz(\xv_1^A)\Phi_A\psi(\xv_2^A)\big]\ket{0} \nn\\[2mm]
&=& -\frac{(2\pi)^3}{\sqrt{N_C}}\delta^3(\Pv_A-\Pv_B-\Pv_C)\int d\delv_1 d\delv_2\,e^{i\delv_1\cdot\Pv_C/2-i\delv_2\cdot\Pv_B/2}\tr\big[\gz\Phi_B^\dag(\delv_1)\Phi_A(\delv_1+\delv_2)\Phi_C^\dag(\delv_2)\big]
\eeqa

If the $A\to B+C$ amplitude is combined with $B+C \to A$ we get a hadron loop correction to the propagation of $A$. The loop also induces mixing between hadrons, $A\to B+C \to D$. Thus the orthogonal basis of wave functions $\Phi(\xv)$ which satisfy the bound state equation \eq{bse1} needs to be rediagonalized when hadron loop corrections are considered. Similarly to the Dirac wave functions (see remark ($ii$) above) the original basis functions are not normalizable, as their norm $\Phi^\dag(\xv)\Phi(\xv)$ approaches a constant at large $|\xv|$. The mixing will likely redistribute the large $|\xv|$ components of low-lying states onto higher-lying states (which then decay into on-shell pairs, much like the pions produced in phenomenological string breaking). The states of the rediagonalized basis may thus become normalizable. The importance of the loop corrections for physical quantities depends on how sensitive measurables are to the large $|\xv|$ components of the wave functions. In $D=1+1$ both the parton distributions and duality relations were determined by low values of $x$, and should therefore be fairly insensitive to the mixing effects.  

There is an essential difference between the Dirac wave functions and the $f\bar f$ solutions of \eq{bse1}. The $f\bar f$ wave functions $\Phi(\xv)$ are (in the rest frame) generally singular at $M=V(|\xv|)$ \cite{Geffen:1977bh}. Regular (locally normalizable) solutions exist only for discrete bound state masses. The Dirac wave functions have no singularities, implying a continuous mass spectrum \cite{plesset,titchmarsh}.

The bound state equation \eq{bse1} appears to have a hidden boost invariance, which ensures the correct frame dependence for the energy eigenvalues, $E=\sqrt{M^2+\Pv^2}$. We investigated this in some detail in $D=1+1$ dimensions, where the $P$-dependence of the wave function is given by \eq{wfframe}. In $D=3+1$ a similar relation holds when $\xv || \Pv$, in which case the bound state equation can be cast in the covariant form \eq{eom5}. Whether the frame dependence of the wave function can be expressed analytically for general $\xv$ is an open question.

The Poincar\'e covariance makes it possible to consider dynamical processes involving bound states. We studied electromagnetic form factors and parton-hadron duality in $D=1+1$. Many more processes are of interest, including hadron-hadron scattering. The outcome of such studies, including the loop corrections mentioned above, will determine whether considering the \order{g^0} homogeneous solution \eq{a0sol3} of Gauss' law is physically viable.

\acknowledgments
I wish to thank Stan Brodsky, Dennis D.~Dietrich and Matti J\"arvinen for fruitful collaborations and valuable advice. Thanks are due also to Hanna Gr\"onqvist for reading and commenting on the first part of these notes. I am very grateful to Roman Zwicky and Luigi Del Debbio for their invitation to the {\it Mini-school on theoretical methods in particle physics} in Edinburgh. The present material is based on lectures presented at the school in October 2013. During the course of this work I have benefitted from a travel grant from the Magnus Ehrnrooth Foundation.


\appendix
\renewcommand{\theequation}{\thesection.\arabic{equation}}

\section{Bound states of scalar QED$_2$} \label{sqedapp}

\subsubsection*{1. Motivation}

An essential feature of the bound state equation (BSE) \eq{eom1} of QED$_2$ is that the energy eigenvalues have the correct dependence on the CM momentum of the state, $E=\sqrt{M^2+P^2}$. To some extent this was expected, since the equation was derived without breaking the Poincar\'e invariance of the action. The Lorentz covariance of an equal-time wave function is dynamical, since the concept of equal time is frame dependent (by contrast, the Bethe-Salpeter wave function \eq{BSwavef} is explicitly covariant since $x_1 - x_2$ transforms as a four-vector). The covariance of the BSE became manifest in \eq{eom3}: All dependence on $P$ and $E$ disappeared when the separation $x$ between the constituents was expressed in terms of $\sigma =(E-V(x))^2-P^2$. Gauss' law in $D=1+1$ requires the potential $V(x)$ to be linear in $x$, and the BSE \eq{eom1} is covariant only for linear potentials.

In Section \ref{boost} we verified that the frame dependence of the wave function found from the BSE agrees with that obtained in \eq{boost2} when the boost generator operates on the state. This ensures that quantities like the electromagnetic form factor \eq{formfac} are Lorentz covariant. A complete check of this would be fairly laborious, see Section \ref{formboost}.

The novel Lorentz covariance found for QED$_2$ bound states appears to hold also in $D=3+1$, again only for a linear $A^0$ potential (corresponding to a homogeneous solution of Gauss' equation)\footnote{As seen for Positronium in Section \ref{motion}, the boost covariance of perturbative contributions involves both the $A^0$ and $\Av$ potentials.}. The wave function $\Phi(\xv)$ is regular at $\sigma=0$ only when the $\Pv$-dependence of its $\xv\parallel \Pv$ component transforms similarly to the $D=1+1$ case \eq{wfframe}. Together with the BSE \eq{bse1} this implicitly determines $\Phi(\xv)$ for all $\xv$.

It is important to get a better understanding of the Lorentz covariance of the equal-time wave functions, and its dependence on the ``kinetic two-momentum'' $\Pi(x)=(P^0-eA^0,P^1)$ in $A^1=0$ gauge ($D=1+1$). In this Appendix we study the bound state equation of scalar QED$_2$. There are similarities with the fermion case: the wave function is again frame independent (apart from kinematic factors) when expressed as a function of $\sigma=\Pi^2$. This system may thus be helpful in providing further insight into the novel covariance.

\subsubsection*{2. Hamiltonian formulation\label{shamder}}

The Lagrangian of SQED$_2$ is
\beq\label{sqedlag}
\mathcal{L}=-\quart F_{\mu\nu}F^{\mu\nu}+\big[(i\partial_\mu-eA_\mu)\vphi\big]^\dag \big[(i\partial^\mu-eA^\mu)\vphi\big]-m^2\vphi^\dag\vphi
\eeq
where $-\quart F_{\mu\nu}F^{\mu\nu}=\halft (\partial_1 A^0)^2$ in $A^1=0$ gauge. We may express the complex scalar field $\vphi$ and its conjugate field $\pi$ in terms of their real and imaginary parts,
\beq\label{complfields}
\vphi = \frac{1}{\sqrt{2}}(\vphi_1 +i\vphi_2) \hspace{3cm}
\pi = \frac{1}{\sqrt{2}}(\pi_1 +i\pi_2)
\eeq
The real conjugate fields are then
\beqa\label{conjfields}
\pi_1 &=& \frac{\partial\mathcal{L}}{\partial(\partial_t\vphi_1)} = \partial_t\vphi_1-eA^0\vphi_2 \\
\pi_2 &=& \frac{\partial\mathcal{L}}{\partial(\partial_t\vphi_2)} = \partial_t\vphi_2+eA^0\vphi_1
\eeqa
where $\partial_t\equiv\partial/\partial x^0$ denotes differentiation wrt. time, and (below) $\partial_x\equiv\partial/\partial x^1$. The canonical commutation relations are,
\beq\label{cancomm}
\com{\vphi_j(t,x)}{\pi_k(t,y)} = i\delta_{jk}\delta(x-y)
\eeq

The equation of motion for the $\vphi_j$ ($\partial\mathcal{L}/\partial\vphi_j-\partial_\mu\big[\partial\mathcal{L}/\partial(\partial_\mu\vphi_j)\big]=0$, $j=1,2$) and for $A^0$ give
\beqa
\partial_t\pi_1&=& (\partial_x^2-m^2)\vphi_1+eA^0\pi_2 \\[2mm]
\partial_t\pi_2&=& (\partial_x^2-m^2)\vphi_2-eA^0\pi_1\\[2mm]
\partial_x^2 A^0 &=& e(\vphi_1\pi_2-\vphi_2\pi_1)
\eeqa
The above relations may be expressed in terms of the complex fields $\vphi$ and $\pi$ given in \eq{complfields},
\beqa\label{compexpr}
\com{\vphi^\dag(t,x)}{\pi(t,y)} &=& i\delta(x-y)\\[2mm]
\partial_t\vphi&=&\pi-ieA^0\vphi \\[2mm]
\partial_t\pi &=& (\partial_x^2-m^2)\vphi-ieA^0\pi \\[2mm]
\partial_x^2 A^0 &=& -ie(\vphi^\dag\pi-\pi^\dag\vphi) \label{gausslaw2}
\eeqa
The last equation (Gauss' law) allows to express $A^0$ in terms of the scalar fields,
\beq
A^0(t,x) = -\halft ie \int dy\big[\vphi^\dag(t,y)\pi(t,y)-\pi^\dag(t,y)\vphi(t,y)\big]\, |x-y|
\eeq
Using this in the Hamiltonian density $\mathcal{H} = \pi_1\partial_0\vphi_1+\pi_2\partial_0\vphi_2-\mathcal{L}$ gives the free and interacting parts of the Hamiltonian, $H=H_0+H_{int}$ with
\beqa\label{sqedham}
H_0&=&\int dx \big[\pi^\dag\pi + (\partial_1\vphi^\dag)(\partial_1\vphi) + m^2\vphi^\dag\vphi\big] \\[2mm]
H_{int}&=& \quart e^2\int dx\,dy \big[(\vphi^\dag\pi-\pi^\dag\vphi)(t,x)\big]\big[(\vphi^\dag\pi-\pi^\dag\vphi)(t,y)\big] |x-y| \label{intham4}
\eeqa

\subsubsection*{3. Bound state\label{sqedstate}}

In analogy to \eq{statedef} we express the bound state of two scalars with CM momentum $P$ at $t=0$ as
\beq\label{bound}
\ket{E,P}=\int dx_1 dx_2\, e^{iP(x_1+x_2)/2} \left[\phi_{aa}\vphi^\dag(x_1)\vphi(x_2) +\phi_{ab}\vphi^\dag(x_1) \pi(x_2) +\phi_{ba}\pi^\dag(x_1) \vphi(x_2) +\phi_{bb}\pi^\dag(x_1)\pi(x_2)\right]\ket{0}
\eeq
where the $c$-numbered wave functions $\phi_{aa},\,\phi_{ab},\,\phi_{ba}$ and $\phi_{bb}$ are functions of $x_1-x_2$. With $H\ket{0}=0$ the time dependence of the state is given by the commutators with the Hamiltonian (here $\partial_i \equiv \partial/\partial x_i$):
\beqa\label{h0comm}
\com{H_0}{\vphi^\dag(x_1)\vphi(x_2)} &=& \com{H_0}{\vphi^\dag(x_1)}\vphi(x_2)+ \vphi^\dag(x_1)\com{H_0}{\vphi(x_2)} = -i\big[\pi^\dag(x_1)\vphi(x_2)+\vphi^\dag(x_1)\pi(x_2)\big] \\[1mm]
\com{H_0}{\vphi^\dag(x_1)\pi(x_2)} &=& i\big[-\pi^\dag(x_1)\pi(x_2)+\vphi^\dag(x_1)(-\partial_2^2+m^2)\vphi(x_2)\big] \\[1mm]
\com{H_0}{\pi^\dag(x_1)\vphi(x_2)} &=& i\big[(-\partial_1^2+m^2)\vphi^\dag(x_1)\vphi(x_2)-\pi^\dag(x_1)\pi(x_2)\big] \\[1mm]
\com{H_0}{\pi^\dag(x_1)\pi(x_2)} &=& i\big[(-\partial_1^2+m^2)\vphi^\dag(x_1)\pi(x_2) + \pi^\dag(x_1)(-\partial_2^2+m^2)\vphi(x_2)\big]
\eeqa
All components of the state \eq{bound} are eigenstates of the interaction Hamiltonian \eq{intham4}, \eg,
\beq\label{hicomm}
H_{int}\vphi^\dag(x_1)\vphi(x_2)\ket{0}=  V(x_1-x_2)\vphi^\dag(x_1)\vphi(x_2)\ket{0}
\eeq
where
\beq\label{slinpot}
V(x_1-x_2)=\halft e^2 |x_1-x_2|
\eeq

The bound state condition
\beq\label{bsc}
H\ket{E,P} = E\ket{E,P} 
\eeq
imposes on the coefficients of the various field components, after partial integrations,
\beqa
\vphi^\dag(x_1)\vphi(x_2): && i\big(\quart P^2+m^2-\partial_x^2\big)\big[\phi_{ab}(x)+\phi_{ba}(x)\big] -P\partial_x \big[\phi_{ab}(x)-\phi_{ba}(x)\big] = (E-V)\phi_{aa}(x) \label{phiaa} \\[1mm]
\vphi^\dag(x_1)\pi(x_2): && -i\phi_{aa}+i\big(\quart P^2+m^2-\partial_x^2-iP\partial_x\big)\phi_{bb}(x) = (E-V)\phi_{ab}(x) \label{phiab} \\[1mm]
\pi^\dag(x_1)\vphi(x_2): && -i\phi_{aa}+i\big(\quart P^2+m^2-\partial_x^2+iP\partial_x\big)\phi_{bb}(x) = (E-V)\phi_{ba}(x) \label{phiba} \\[1mm]
\pi^\dag(x_1)\pi(x_2): && -i\big[\phi_{ab}(x)+\phi_{ba}(x)\big] = (E-V)\phi_{bb}  \label{phibb}
\eeqa
Equations \eq{phiaa} - \eq{phibb} are consistent with $\phi_{aa}, \phi_{bb}$ and $\phi_{ab}+\phi_{ba}$ being even and $\phi_{ab}-\phi_{ba}$ being odd in $x$, or {\it vice versa}. We may therefore solve the equations for $x \ge 0$. The wave functions for $x<0$ are given by parity, with a continuity condition at $x=0$. 
The equations allow to express $\phi_{aa}$ and $\phi_{ab}\pm\phi_{ba}$ in terms of $\phi_{bb}$ and its derivatives. Introducing the same variable $\sigma$ as in the fermion case \eq{sigdef},
\beq\label{sigmadef}
\sigma = (E-V)^2 - P^2 \hspace{2cm} \partial_x = -e^2(E-V)\partial_\sigma \hspace{2cm} (x>0)
\eeq
the differential equation for $\phi_{bb}$ has no explicit dependence on $P$ or $E$,
\beq\label{bsebb}
e^4 \partial_\sigma^2{\phi}_{bb} + \frac{e^4}{\sigma} \partial_\sigma{\phi}_{bb} + \Big(\inv{4}-\frac{m^2}{\sigma}\Big)\phi_{bb}=0
\eeq
Thus $\phi_{bb}(\sigma)$ is frame independent as a function of $\sigma$. Since $\sigma$ is a $P$-dependent function of $x$ the wave function viewed as a function of $x$ is frame dependent.
The remaining wave functions can be expressed in terms of $\phi_{bb}$ and $\partial_\sigma\phi_{bb}$ as follows,
\beqa
\phi_{aa} &=& -\quart\sigma\phi_{bb}+\halft e^4\partial_\sigma{\phi}_{bb}-\frac{P^2}{\sigma}(m^2\phi_{bb}-e^4\partial_\sigma{\phi}_{bb})  \label{phiaaexpr} \\
\phi_{ab}-\phi_{ba} &=& -2Pe^2 \partial_\sigma\phi_{bb} \label{phiabmexpr} \\[1mm]
\phi_{ab}+\phi_{ba} &=& i(E-V)\phi_{bb} \label{phiabpexpr}  \\
\eeqa

The differential equation \eq{bsebb} has the analytic solution
\beq\label{phibbsol}
\phi_{bb}(\sigma)= e^{-i\sigma/2}\big[c_1\,{_1F_1}(\halft-im^2,1,i\sigma) + c_2\,U(\halft-im^2,1,i\sigma)\big]
\eeq 
where we set $e=1$ and $c_1,c_2$ are arbitrary constants. For $\sigma \to 0$ this solution behaves as
\beq\label{bbsiglim}
\phi_{bb}(\sigma)= c_1\big[1+m^2\sigma+\morder{\sigma^2}\big] -  c_2\big[\log(\sigma)/\Gamma(\halft-im^2)+\morder{\sigma^0}\big]
\eeq
The coefficient of the $\sigma^{-1}$ term in $\phi_{aa}$ \eq{phiaaexpr} vanishes at $\sigma=0$ only provided $c_2=0$, which is then the requirement for a physical solution. The continuity condition at $x=0$ determines the bound state masses as in Section \ref{ansol}. Unlike in the fermion case \eq{bse4}, the differential equation \eq{bsebb} remains singular at $\sigma=0$ for $m=0$, so the spectrum is discrete even for scalars of zero mass (in which case the Hypergeometric functions reduce to Bessel functions). This is also seen from \eq{bbsiglim}.

Similarly to the fermion wave functions \eq{aschi}, the physical solution for $\phi_{bb}$ oscillates in the $\sigma \to \infty$ limit,
\beq\label{phibbas}
\phi_{bb}(\sigma\to\infty) =c_1 \sqrt{\frac{2}{\pi\sigma}\big(e^{2\pi m^2}+1\big)}\cos\Big[\halft\sigma-m^2\log\sigma-\quart\pi+\arg\Gamma(\halft+im^2)\Big]\big[1+\morder{\sigma^{-1}}\big]
\eeq
The power of $\sigma$ modulating the oscillations is different for $\phi_{aa}$ and $\phi_{ab}\pm \phi_{ba}$.

\break

\section{Non-relativistic limits} \label{NRlimit}

\subsubsection*{1. The NR limit of the Dirac solution}

The non-relativistic limit of the Dirac equation is usually obtained by writing the energy $M$ in terms of the constituent mass $m$ and the binding energy $E_b$ as $M=m+E_b$, and assuming that the potential and binding energy may be neglected compared to the constituent mass, $V\sim E_b \ll m$. It is readily seen that the equation \eq{phieq} for the upper component of the Dirac wave function reduces to the Schr\"odinger equation in this limit. 

This procedure necessarily fails for a linear potential when $|x|$ is sufficiently large. The behaviour of the wave function at large $|x|$ is indeed quite different from that of the Schr\"odinger wave function, as seen in \fig{Diracwf}. The normalization integral of the Dirac wave function diverges, implying a continuous energy spectrum for the Dirac solutions \cite{plesset,titchmarsh}. How can the spectrum become discrete in the NR limit?

Here we derive the NR limit of the Dirac wave function \eq{phisol}. Recalling that in \eq{qedtwopot} we normalized the charges to unity, $V(x)= \halft |x|$, we need to consider $m\to\infty$. In the Schr\"odinger equation $\partial_x^2/2m \sim V(x)$ implies that 
\beq\label{dirscaling}
x \sim E_b \sim m^{-1/3} \hspace{2cm} \sigma=(m+E_b-V)^2 \simeq m^2+m(2E_b-|x|)
\eeq
The first solution in \eq{phisol} has the integral representation
\beq\label{dir1}
e^{-i\sigma}{_1}F_1\Big(-\frac{im^2}{2},\inv{2},2i\dsi\Big)=\frac{\sqrt{\pi}}{\Gamma[(1+im^2)/2]\Gamma(-im^2/2)}\,I
\eeq
where
\beq\label{int}
I = \int_0^1du\, e^{i\sigma(2u-1)}u^{-(im^2+2)/2}(1-u)^{(im^2-1)/2} \equiv \int_0^1du\, e^{im^2f(u)}
\eeq
For $m\to\infty$ the phase of the integrand oscillates rapidly so we may use the stationary phase approximation. Expanding $f(u)$ around its stationary point $u=\halft$ we get
\beq
m^2f(u)\simeq -\frac{3i}{2}\log 2 -\frac{8m^2}{3}(u-\halft)^3+2(\sigma-m^2)(u-\halft)
\eeq
Changing the integration variable to $v=2m^{2/3}(u-\halft)$ the large $m$ limit of \eq{dir1} becomes
\beqa\label{nrdir1}
e^{-i\sigma}{_1}F_1\Big(-\frac{im^2}{2},\inv{2},2i\dsi\Big)&\simeq& \frac{\sqrt{2\pi}\,m^{-2/3}}{\Gamma[(1+im^2)/2]\Gamma(-im^2/2)}\int_{-\infty}^\infty dv\,\exp\Big[-\frac{i}{3}v^3+i\frac{\sigma-m^2}{m^{2/3}}v\Big]\nn\\[2mm]
&\simeq& \sqrt{\pi}\,m^{1/3}e^{\pi(2m^2-i)/4}{\rm Ai}\big[m^{1/3}(|x|-2E_b)\big]
\eeqa
where in the second expression we used the large $m$ limit of the $\Gamma$-functions,
\beq\label{gamlimit}
\inv{\Gamma[(1+im^2)/2]\Gamma(-im^2/2)} \simeq \frac{m}{2\sqrt{2}\,\pi}e^{\pi m^2/2-i\pi/4}
\eeq

The NR limit of the second solution in \eq{phisol} is found using the same method and gives an identical result,
\beq\label{nrdir2}
2m\,(M-V)\,e^{-i\sigma}{_1}F_1\Big(\frac{1-im^2}{2},\frac{3}{2},2i\dsi\Big)
\simeq \sqrt{\pi}\,m^{1/3}e^{\pi(2m^2-i)/4}{\rm Ai}\big[m^{1/3}(|x|-2E_b)\big]
\eeq
Thus both independent solutions of the Dirac equation reduce to the same, normalizable Airy wave function of the Schr\"odinger equation with a linear potential. Using the expressions \eq{nrdir1} and \eq{nrdir2} in \eq{phisol} gives \eq{dirnrlimit}.

Numerically we find that the Dirac eigenvalue $M$ determined by the continuity condition at $x=0$ becomes independent of the ratio $a/b$ at large $m$, except in a narrow interval around $a/b=-1$. The continuous range of the Dirac spectrum is thus, in the $m\to\infty$ limit, restricted to parameters satisfying $a+b=0$. For $a/b=-1$ the NR limit \eq{dirnrlimit} vanishes, and the derivation above is incomplete.

\subsubsection*{2. The NR limit of the $f\bar f$ wave function}

The general solution \eq{Phi1sol} of the $f\bar f$ equation \eq{bse4} involves Confluent Hypergeometric functions of both the first (${_1}F_1$) and second ($U$) kind. The wave function is regular at $\sigma=0$ (with $\sigma(x)$ given in \eq{sigdef}) only for the ${_1}F_1$ solution.

The NR limit of the regular solution in \eq{Phi1sol} may be derived as for the Dirac wave function. In the rest frame,
\beq\label{ffscaling}
\sigma(x)=(2m+E_b-\halft |x|)^2 \simeq 4m^2+2m(2E_b-|x|) \hspace{2cm} (m\to\infty)
\eeq
The result, with $x\sim E_b \sim m^{-1/3}$,
\beq\label{nr1}
\dsi\, e^{-i\dsi/2}\kum(1-im^2,2,i\dsi) \simeq \frac{e^{\pi m^2}}{(m/2)^{2/3}}{\rm Ai}\big[((m/2)^{1/3})(|x|-2E_b)\big]
\eeq
involves only the normalizable Airy function.

%
\begin{wrapfigure}[9]{r}{0.3\textwidth}
  \vspace{-30pt}
  \begin{center}\hspace{-.5cm}
    \includegraphics[width=0.25\textwidth]{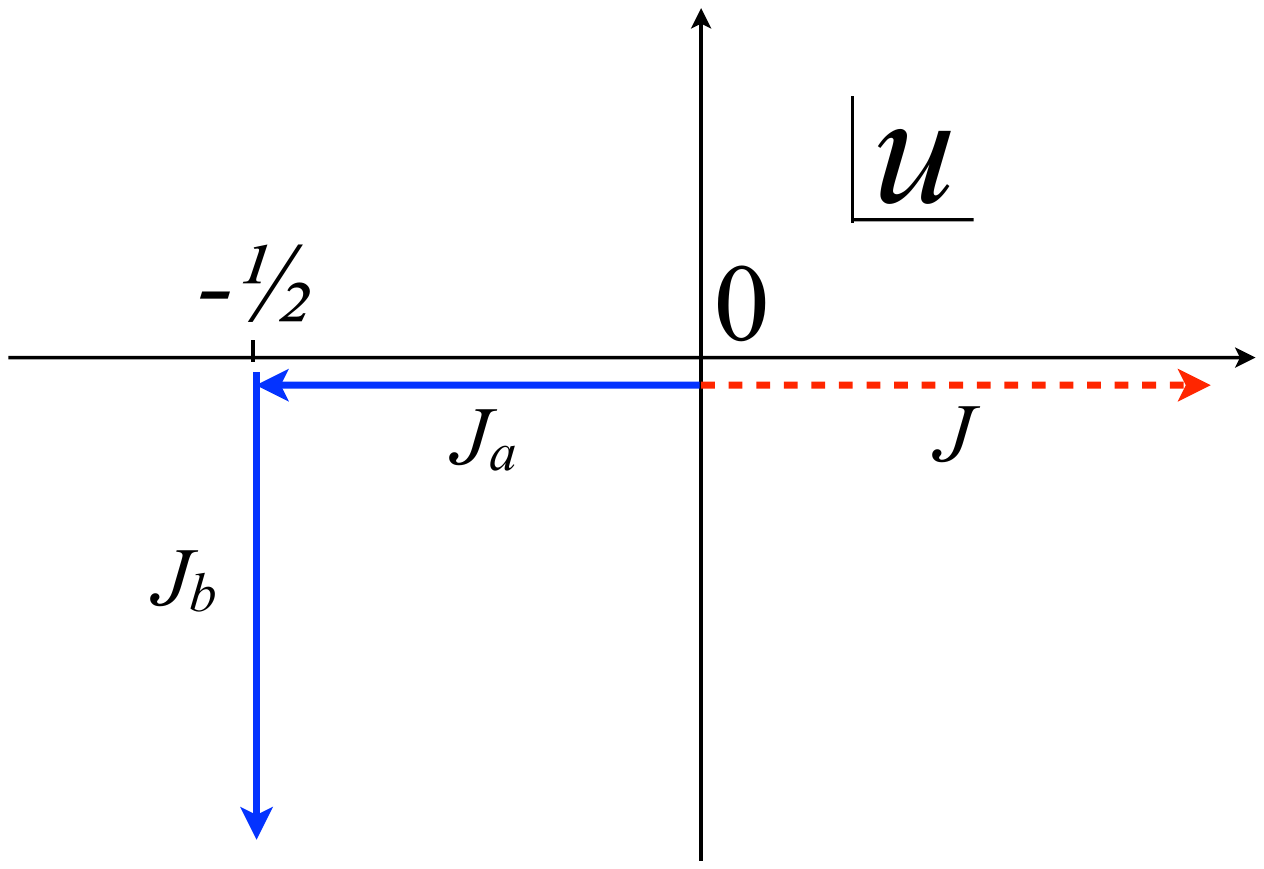}
  \end{center}
  \vspace{-20pt}
  \caption{The integration path of $J$ in \eq{Jdef} (dashed red line) may be rotated by $-\pi/2$ and shifted by $-\halft$, giving the paths of $J_a$ in \eq{Jaint} and $J_b$ in \eq{Jbint} (solid blue lines).}\label{intpath}
\end{wrapfigure}

The second solution in \eq{Phi1sol} has the integral representation
\beqa
\dsi\, e^{-i\dsi/2}U(1-im^2,2,i\dsi) &=&\frac{\dsi}{\Gamma(1-im^2)}\,J \label{uint}\\[2mm]
&&\hspace{-2cm}J=\int_0^\infty du\,e^{-i\dsi(2u+1)/2}\, u^{-im^2}(1+u)^{im^2} \label{Jdef}
\eeqa
The phase of the $J$-integrand rotates rapidly for large $m$, but now the stationary point is at $u=-\halft$, outside the integration path. In order to bring this point onto the path we may rotate the contour by $-\pi/2$, so that it extends from $u=0$ to $u=-i\infty$, and then shift it by $-\halft$. As indicated in \fig{intpath} the new path has two segments, one ($J_a$) from $u=0$ to $u=-\halft$ and another ($J_b$) from $u=-\halft$ to $u=-\halft-i\infty$ (the asymptotic contour in the lower half $u$-plane may be neglected). The first may be written ($u=w-\halft$)
\beqa
J_a&=&-e^{-\pi m^2}\int_0^\halft dw\,e^{-i\dsi w}(\halft-w)^{-im^2}(\halft +w)^{im^2}\nn\\[2mm]
&\simeq& -e^{-\pi m^2} \int_0^\infty dw\exp\Big[i\,\sfrac{16}{3}m^2w^3-i\,2m(2E_b-|x|)w\Big]\nn\\[2mm]
&=& -\frac{\pi\,e^{-\pi m^2}}{(16m^2)^{1/3}}\left\{{\rm Ai}\big[\big(\halft m\big)^{1/3}(|x|-2E_b)\big]+\frac{i}{\pi}\int_0^\infty dv\,\sin\big[\sfrac{1}{3}v^3+\big(\halft m\big)^{1/3}(|x|-2E_b)\big]\right\}\label{Jaint}
\eeqa
where we used the stationary phase approximation around $w=0$.

For the contour from $u=-\halft$ to $u=-\halft-i\infty$ we have ($u=-iw-\halft$)
\beqa
J_b&=&-i\,e^{-\pi m^2}\int_0^\infty dw\,e^{-\dsi w}\big(\halft+iw\big)^{-im^2}\big(\halft-iw\big)^{im^2}\nn\\[2mm]
&\simeq& -\frac{i\,e^{-\pi m^2}}{(16m^2)^{1/3}}\int_0^\infty dv\,\exp\big[-\sfrac{1}{3}v^3+\big(\halft m\big)^{1/3}(|x|-2E_b)v\big] \nn\\[2mm]
&=& -\frac{\pi\,e^{-\pi m^2}}{(16m^2)^{1/3}}\left\{ i\,{\rm Bi}\big[\big(\halft m\big)^{1/3}(|x|-2E_b)\big]-\frac{i}{\pi}\int_0^\infty dv\,\sin\big[\sfrac{1}{3}v^3+\big(\halft m\big)^{1/3}(|x|-2E_b)\big]\right\}\label{Jbint}
\eeqa
where we again used the stationary phase approximation around $w=0$. Substituting $J=J_a+J_b$ in \eq{uint} we find the NR limit of the second solution of the $f\bar f$ equation \eq{bse4},
\beq\label{nr2}
\dsi\, e^{-i\dsi/2}U(1-im^2,2,i\dsi)\simeq-(2m^2)^{2/3}\frac{\pi\,e^{-\pi m^2}}{\Gamma(1-im^2)}\Big\{{\rm Ai}\big[\big(\halft m\big)^{1/3}(|x|-2E_b)\big]+i\,{\rm Bi}\big[\big(\halft m\big)^{1/3}(|x|-2E_b)\big]\Big\}
\eeq
Thus the $U$-function reduces at large $m$ to a combination of the Airy functions Ai and Bi. The latter grows exponentially with $|x|$ and is thus an unphysical solution of the Schr\"odinger equation. The criterion of local normalizability of the relativistic $f\bar f$ wave function is consistent with the usual normalizability requirement in the non-relativistic limit.

\end{document}